\documentclass[twocolumn,tighten,iop,twocolappendix]{aastex702}

\usepackage{grffile}

\received{}
\revised{}
\accepted{}

\DeclareRobustCommand{\VAN}[3]{#2}
\let\VANthebibliography\thebibliography
\def\thebibliography{\DeclareRobustCommand{\VAN}[3]{##3}\VANthebibliography}

\graphicspath{{./}{figures/}}

\usepackage{graphicx}
\graphicspath{{Plots/}}
\usepackage{amsmath}
\usepackage{ion-defs, farhan-defs}

\usepackage[super]{nth}

\usepackage[T1]{fontenc}

\usepackage{subfigure}
\usepackage{amsmath}
\usepackage{mathrsfs}
\usepackage{esvect}

\usepackage{bigints}
\usepackage[fulladjust]{marginnote}

\usepackage{appendix}

\usepackage{CJK}

\hypersetup{colorlinks=true, citecolor=blue, filecolor=magenta,urlcolor=blue, linkcolor=blue}

\begin{document}

\title{The Role of Star Formation and Morphology in the Mass-Metallicity Relation of Galaxies at Cosmic Noon}

\correspondingauthor{Farhanul~Hasan}
\email{fhasan@stsci.edu}

\author[0000-0002-0072-0281]{Farhanul Hasan}
\affiliation{Space Telescope Science Institute, 3700 San Martin Drive, Baltimore, MD 21218, USA}
\email{fhasan@stsci.edu}

\author[0000-0002-9946-4731]{Marc Rafelski}
\affiliation{Space Telescope Science Institute, 3700 San Martin Drive, Baltimore, MD 21218, USA}
\affiliation{Department of Physics and Astronomy, Johns Hopkins University, 3400 North Charles Street, Baltimore, MD 21218, USA}
\email{mrafelski@stsci.edu}

\author[0000-0001-5294-8002]{Kalina~V.~Nedkova}
\affiliation{IPAC, California Institute of Technology, 1200 E.~California Blvd, Pasadena, CA 91125, USA}
\email{knedkova@ipac.caltech.edu}

\author[0000-0003-3108-0624]{Peter J. Watson}
\affiliation{INAF -- Osservatorio Astronomico di Padova, Vicolo Osservatorio 5, 35122 Padova, Italy}
\affiliation{Universit\'e C\^ote d'Azur, Observatoire de la Côte d’Azur, CNRS, Laboratoire Lagrange, 06000 Nice, France}
\email{peter.watson@oca.eu}

\author[0000-0002-1392-3520]{Debra M. Elmegreen}
\affiliation{Department of Physics and Astronomy, Vassar College, Poughkeepsie, NY 12604 USA}
\email{elmegreen@vassar.edu}

\author[0000-0002-1723-6330]{Bruce G. Elmegreen}
\affiliation{Retired, Katonah, NY 10536 USA}
\email{belmegreen@gmail.com}

\author[0000-0002-0364-1159]{Zahra Sattari}
\affiliation{IPAC, California Institute of Technology, 1200 E.~California Blvd, Pasadena, CA 91125, USA}
\email{zsattari@ipac.caltech.edu}

\author[0000-0001-7166-6035]{Vihang Mehta}
\affiliation{IPAC, California Institute of Technology, 1200 E.~California Blvd, Pasadena, CA 91125, USA}
\email{vmehta@ipac.caltech.edu}

\author[0000-0002-6586-4446]{Alaina Henry}
\affiliation{Space Telescope Science Institute, 3700 San Martin Drive, Baltimore, MD 21218, USA}
\email{ahenry@stsci.edu}

\author[0000-0002-4917-7873]{Mitchell Revalski}
\affiliation{Space Telescope Science Institute, 3700 San Martin Drive, Baltimore, MD 21218, USA}
\email{mrevalski@stsci.edu}

\author[0000-0003-4804-7142]{Ayan Acharyya}
\affiliation{INAF -- Osservatorio Astronomico di Padova, Vicolo Osservatorio 5, 35122 Padova, Italy}
\email{ayan.acharyya@inaf.it}


\author[0000-0003-0980-1499]{Benedetta Vulcani}
\affiliation{INAF -- Osservatorio Astronomico di Padova, Vicolo Osservatorio 5, 35122 Padova, Italy}
\email{benedetta.vulcani@inaf.it}

\author[0000-0003-0028-4130]{Grecco A. Oyarz\'un}
\affiliation{Homer L. Dodge Department of Physics and Astronomy, The University of Oklahoma, 440 W. Brooks St., Norman, OK 73019, USA}
\affiliation{Department of Physics and Astronomy, Johns Hopkins University, 3400 North Charles Street, Baltimore, MD 21218, USA}
\email{grecco.oyarzun@ou.edu}

\author[0000-0002-7928-416X]{Y. Sophia Dai}
\affiliation{Chinese Academy of Sciences South America Center for Astronomy (CASSACA), National Astronomical Observatories(NAOC),
20A Datun Road, Beijing 100012, China}
\email{ydai@nao.cas.cn}

\author[0000-0002-9373-3865]{Xin Wang}
\affiliation{School of Astronomy and Space Science, University of Chinese Academy of Sciences (UCAS), Beijing 100049, China}
\affiliation{National Astronomical Observatories, Chinese Academy of Sciences, Beijing 100101, China}
\affiliation{Institute for Frontiers in Astronomy and Astrophysics, Beijing Normal University, Beijing 102206, China}
\email{xwang@ucas.ac.cn}

\author[0000-0001-6919-1237]{Matthew A.~Malkan}
\affiliation{Department of Physics and Astronomy, University of California, Los Angeles, 475 Portola Plaza, Los Angeles, CA 90095, USA}
\email{malkan@astro.ucla.edu}

\author[0000-0002-9136-8876]{Claudia~M.~Scarlata}
\affiliation{Minnesota Institute for Astrophysics, University of Minnesota, 116 Church Street SE, Minneapolis, MN 55455, USA}
\email{mscarlat@umn.edu}

\author[0000-0003-1767-6421]{Alexandra Le Reste}
\affiliation{Minnesota Institute for Astrophysics, University of Minnesota, 116 Church Street SE, Minneapolis, MN 55455, USA}
\email{alereste@umn.edu}

\author[0000-0001-8587-218X]{Matthew~J.~Hayes}
\affiliation{Stockholm University, Department of Astronomy and Oskar Klein Centre for Cosmoparticle Physics, AlbaNova University Centre, SE-10691, Stockholm, Sweden}
\email{matthew@astro.su.se}

\author[0000-0003-4569-2285]{Andrew J.~Battisti}
\affiliation{International Centre for Radio Astronomy Research, University of Western Australia, 7 Fairway, Crawley, WA 6009, Australia}
\affiliation{Research School of Astronomy and Astrophysics, Australian National University, Cotter Road, Weston Creek, ACT 2611, Australia}
\email{andrew.battisti@anu.edu.au}

\author[0000-0002-8630-6435]{Anahita Alavi}
\affiliation{IPAC, California Institute of Technology, 1200 E.~California Blvd, Pasadena, CA 91125, USA}
\email{anahita@ipac.caltech.edu}

\author[0000-0001-6482-3020]{James Colbert}
\affiliation{IPAC, California Institute of Technology, 1200 E.~California Blvd, Pasadena, CA 91125, USA}
\email{colbert@ipac.caltech.edu}

\author[0000-0002-7570-0824]{Hakim Atek}
\affiliation{CNRS, Institut d'Astrophysique de Paris, 98 bis Boulevard Arago, 75014 Paris, France}
\email{atek@iap.fr}

\author[0000-0001-7016-5220]{Michael Rutkowski}
\affiliation{Department of Astronomy, University of Florida, Gainesville, FL, 32611, USA}
\email{m.rutkowski@ufl.edu}

\author[0000-0001-5860-3419]{Tucker Jones}
\affiliation{Department of Physics and Astronomy, University of California Davis, 1 Shields Avenue, Davis, CA 95616, USA}
\email{tdjones@ucdavis.edu}

\author[0009-0007-8470-5946]{Ilias Goovaerts}
\affiliation{Space Telescope Science Institute, 3700 San Martin Drive, Baltimore, MD 21218, USA}
\email{igoovaerts@stsci.edu}

\author[0000-0001-7396-3578]{Alexander Beckett}
\affiliation{Aix Marseille Université, CNRS, CNES, LAM (Laboratoire d’Astrophysique de Marseille), 13388 Marseille, France}
\email[]{alexander.beckett@lam.fr}

\author[0000-0002-8651-9879]{Andrew~J.~Bunker}
\affiliation{Department of Physics, University of Oxford, Denys Wilkinson
Building, Keble Road, Oxford OX1 3RH, UK}
\email{andy.bunker@physics.ox.ac.uk}

\author[0009-0002-9932-4461]{Mason~S.~Huberty}
\affiliation{Minnesota Institute for Astrophysics, University of Minnesota, 116 Church Street SE, Minneapolis, MN 55455, USA}
\email{huber458@umn.edu}






\begin{abstract}

We investigate the dependence of the mass–metallicity relation (MZR) on star formation and galaxy morphology at $1.7\lesssim z\lesssim3.5$, using JWST/NIRISS wide-field slitless spectroscopy of $>$ 1000 galaxies. The observations span $>$ 85 arcmin$^2$ from the pure-parallel PASSAGE and OutThere surveys. We derive gas-phase metallicities using a Bayesian framework that combines multiple strong emission-line ratios calibrated against direct auroral-line metallicity measurements from JWST. We constrain stellar masses, star formation rates (SFRs), and morphological properties using ancillary HST and JWST imaging. We stack spectra by stellar mass to constrain the MZR over $\log(M_\star/M_\odot)\sim6.5$–11, and find that it is well described by a second-order polynomial, with a steep low-mass slope and metallicities $<$ 10\% solar at $\log(M_\star/M_\odot)<7.5$. Additional stacking by SFR and morphology reveals strong evidence for a fundamental metallicity relation (FMR). Specifically, at $\log(M_\star/M_\odot)\lesssim9.5$, galaxies with lower SFRs are more metal-rich than those with higher SFRs at fixed mass.
At $\log(M_\star/M_\odot)\lesssim9.5$, we find that larger galaxies, and those having lower SFR surface densities, bulges, and/or spiral features are significantly metal rich relative to their smaller, higher SFR surface density, bulge-less, and/or non-spiral counterparts.  
These results suggest the secondary dependencies of the MZR reflect variations in the baryon cycle, such that recent metal-poor gas accretion drives concentrated star formation and metallicity dilution, while structurally developed galaxies retain more chemically enriched gas reservoirs.
\end{abstract}

\vspace{-40pt}

\keywords{ Galaxy evolution (594) --- Galaxy abundances (574) --- Galaxy chemical evolution (580)  --- High-redshift galaxies (734) --- Metallicity (1031) --- Star formation (1569) --- Galaxy structure (622)}

\section{Introduction}\label{sec:intro}

The modern picture of galaxy evolution places enormous emphasis on the cycling of baryonic matter through galaxies. In this so-called ``baryon cycle'' \citep[see, e.g., reviews by][]{Tumlinson2017ARA&A,PH20}, gas that accretes from the cosmic web helps fuel the formation of stars and feed central supermassive black holes of galaxies. Subsequent internal phenomena such as outflows from supernovae (SNe) and active galactic nuclei (AGN) expel gas from the interstellar medium (ISM) into the circumgalactic medium (CGM) and the large-scale cosmic web.

One of the most fundamental tracers of the baryon cycle is the chemical enrichment of galaxies. Inflows of gas via cosmic web filaments are expected to bring in mostly pristine material with low metallicity \citep[e.g.,][]{Keres05,DB06,Dekel09,Mandelker18}. Different generations of star formation produce metals that enrich the ISM, but stellar and AGN feedback can eject these metals and lower the metallicity of the ISM \citep[e.g.,][]{FD08,Dave11,Hopkins12,Chisholm18}. Furthermore, metals may be recycled back into the ISM from the CGM \citep[e.g.,][]{Oppenheimer10,AA17,Muratov17}. These physical processes result in the well-established relationship between stellar mass and gas-phase ISM metallicity -- the ``mass-metallicity relation,'' \citep[MZR; e.g.,][]{Tremonti2004ApJ,Curti2017MNRAS,Revalski2024ApJ}.

Being one of the key statistical descriptions of the baryon cycle, the MZR is one of the most well-studied scaling relations of galaxies across cosmic time. At low redshifts ($z\lesssim0.5$), the MZR has been well-characterized over many orders of magnitude in stellar mass and gas-phase metallicity with several hundreds of thousands of galaxies \citep[e.g.,][]{Tremonti2004ApJ,AM13,Curti2017MNRAS}. 
Wide-field spectroscopic surveys such as the Sloan Digital Sky Survey \citep{York00} have been tremendously prolific in measuring stellar masses and gas-phase metallicities in the nearby universe.

At higher redshifts, the MZR is less well-constrained, owing to the difficulty of obtaining high signal-to-noise (S/N) photometry and spectroscopy for large numbers of galaxies. The dramatic surface brightness dimming effect with redshift, the bias of observing brighter (therefore more massive) galaxies at higher redshifts, and the redshifting of common strong rest-frame optical lines such as {\OIIIwave}, {\Hbwave}, and {\Hawave} into IR wavelengths -- thus strongly benefiting from space-based observations -- all play a role.

Despite these challenges, the last decade or so has seen incredible progress in characterizing the MZR at higher redshifts, especially cosmic noon -- $z\sim1-3$ -- when cosmic star formation was at its peak \citep[e.g.,][]{MD14,FS20}. Large spectroscopic observing campaigns have now assembled statistically significant samples with which the MZR has been characterized at these epochs \citep[e.g.,][]{Cullen14,Sanders15,Kacprzak16,Henry2021ApJ,Sanders2021ApJ,Topping21,Khostovan26}. However, most of these constraints have been limited to stellar masses ${\logms}\gtrsim8.5$ \citep[with a few exceptions, such as][]{Revalski2024ApJ}, primarily underscoring the difficulty of observing emission lines required to measure metallicities of intrinsically faint low-mass galaxies. 
With the superior sensitivity of the James Webb Space Telescope (JWST), a handful of studies have recently characterized the MZR at lower masses at cosmic noon and beyond \citep{Li23,Chemerynska24,Sarkar2025ApJ,He26,Isobe26}.
However, these studies have painted a mixed picture at this mass regime, in terms of both the normalization and the slope of the MZR.

It is now well-established that in the local universe, the MZR has an important secondary dependence on star formation rate (SFR), thus constraining galaxies on a three-dimensional plane of stellar mass, metallicity, and SFR. This so-called ``fundamental metallicity relation'' (FMR) results in galaxies with higher SFR being more metal-poor on average compared to those with lower SFR \citep[e.g.,][]{Ellison08,Mannucci10,AM13,LL13,Curti20,Sui26}. This may be due to both pristine gas accretion and stellar outflows lowering the chemical enrichment of the ISM.

The existence and quantitative description of the FMR at higher redshifts are open questions in the literature. Several authors have reported either strong or tentative evidence for an SFR-dependence of the MZR out to at least $z\sim4$ \citep[e.g.,][]{Sanders18,Sanders2021ApJ,Nakajima2023ApJS,Curti2024A&A,Revalski2024ApJ,Lam26,Stanton26}. However, the exact nature of this dependence remains debated, owing to systematics associated with measurements of metallicities, SFRs, and sample selection biases. The largest samples to date investigating the FMR at $z\gtrsim2$ have been on the order of a few hundred galaxies, limited to ${\logms}\gtrsim8.5$ \citep[][]{Sanders2021ApJ,Lam26}, indicating the need for larger samples.
Some studies have even suggested a lack of an FMR at both low and high-redshifts \citep[e.g.,][]{Sanders15,Sanchez17,KC25,Laseter26}.

Although stellar mass and SFR are likely to be the primary and secondary predictors of gas-phase metallicity, the MZR may also depend on galaxy structure and morphology, given that all of these properties are shaped in some way by the baryon cycle. 
In the nearby Universe, the MZR has often been reported to depend on galaxy size such that larger galaxies are metal-poor compared to smaller galaxies at fixed mass \citep{Ellison08,BH12,DEugenio18,Ma24}.
Further evidence for an anti-correlation between size and metallicity at fixed mass has been found at $z\sim1-7$, albeit with fairly small samples ($\lesssim$100) \citep{Yabe12,Gillman21,Jia25}. 
This trend can be explained either by the late accretion of metal-poor gas preferentially forming stars at large distances, thus increasing the stellar size while keeping the ISM chemically diluted \citep[][]{SD18,Boardman26,Wang26}, or by the strength of the gravitational potential wells of compact galaxies in retaining metals \citep{DEugenio18,Baker23}. 
In contrast, \citet{Hoopes07} and \cite{LH23} found evidence for compact galaxies to be metal-poor relative to more extended galaxies in the nearby and high-redshift Universe, respectively. 
Most of the observational studies are based on small samples of massive galaxies {${\logms}\gtrsim10$}, typically at $z \lesssim 2$.


A growing body of evidence now suggests that the secondary dependence of metallicity on star formation may be more fundamentally linked to the surface density of SFR than to the global SFR alone. Spatially resolved analyses of nearby galaxies find that, at fixed stellar-mass surface density, regions with higher SFR surface density tend to exhibit lower gas-phase metallicities, which can be interpreted as a resolved analogue of the FMR \citep{Teklu20,Baker23}.  
At cosmic noon, direct measurements of the link between global metallicity and SFR surface density are comparatively limited, although spatially resolved observations at $z\sim2-3$ suggest that intensely star-forming regions can be relatively metal poor
\citep[e.g.,][]{Cresci10, XinWang2020ApJ, EC25}.
This phenomenological connection is qualitatively consistent with scenarios in which enhanced gas accretion both dilutes the ISM metallicity and increases the local star formation activity \citep[e.g.,][]{Lilly2013ApJ,DM14,Torrey18}.

Clumpy, irregular, and disturbed morphologies have also been linked to chemical enrichment. At $z\gtrsim1$, massive star-forming clumps are found in the majority of lower-mass galaxies \citep[e.g.,][]{ElmegreenD2007ApJ,Guo15,Sattari2023ApJ} and clumpy sub-populations such as ``chain'' and ``tadpole'' galaxies are commonplace \citep[e.g.,][]{Elmegreen04,Elmegreen09,EE10}.
It has been suggested that these clumps originate from the accretion of relatively pristine gas from the cosmic web \citep[e.g.,][]{Agertz09,Dekel2009ApJ,Ceverino16}. 
This picture is supported by observations showing clumps associated with metal-deficiency, at both low redshift \citep[e.g.,][]{SA2013ApJ,SA2014ApJ} and high redshift \citep{EC25,Sok25,Sok26}.
In addition, both lopsided/irregular morphologies \citep{Reichard09,SA14,Gillman21} and signs of interaction \citep{Ellison2008AJ,Horstman21} have been linked to reduced metallicity.

Wide field slitless spectroscopy (WFSS) surveys, especially in the era of JWST, are uniquely powerful for statistically robust investigations into the MZR and its dependence on star formation and morphological properties at cosmic noon.
In contrast to conventional slit-based spectroscopy, WFSS removes photometric target selection biases and enables, in principle, an unbiased spectroscopic search for galaxies \citep{Willott22}. This, combined with JWST's unparalleled sensitivity and NIR wavelength coverage, is now making it possible to constrain the MZR down to unprecedentedly low stellar masses of ${\logms}\sim6-7$ at cosmic noon and beyond \citep[e.g.,][]{Li23,He26}.

Large samples, combining deep spectroscopy and additional high-quality multi-band imaging, are key to 
statistically determining the effect of 
SFR and various morphological indicators on metal enrichment. HST and JWST photometry provides the broadband spectral energy distributions (SEDs) required to precisely estimate masses, and SFRs, while their high spatial resolution imaging allows morphological characterizations in terms of size, S{\'e}rsic index, shape (axis ratio), concentration, clumpiness, and merger signatures. 
Large legacy imaging surveys such as CANDELS \citep{Koekemoer2011ApJS,Grogin2011ApJS..197...35G}, COSMOS \citep{Scoville07,Weaver22}, and COSMOS-Web \citep{Casey23,Shuntov25} have established the value of such deep HST/JWST imaging for constraining these physical properties.
WFSS is a highly efficient method of building such large samples.


\begin{deluxetable*}{cccccccccc}
\tabletypesize{\scriptsize}
\tablewidth{0pt}
\tablecaption{Summary of Observations, indicating both NIRISS and ancillary data available
\label{tab:fields}}

\tablehead{
\colhead{Field} &
\colhead{PID} &
\colhead{RA} &
\colhead{Dec} &
\multicolumn{3}{c}{Grism Coverage} &
\colhead{$t_{\rm exp}$} &
\colhead{NIRISS Imaging} &
\colhead{Ancillary Imaging} \\
\cline{5-7}
\colhead{} &
\colhead{} &
\colhead{(deg)} &
\colhead{(deg)} &
\colhead{F115W} &
\colhead{F150W} &
\colhead{F200W} &
\colhead{(hr)} &
\colhead{} &
\colhead{}
}

\startdata
Par005 & GO-1571 & 150.653698  &   2.085580   & R/C & R/C &     &  5.58 & F115W, F150W        & HST (COSMOS) \\
Par006 & GO-1571 & 150.676216  &   2.129305   & R/C & R/C &     &  5.58 & F115W, F150W        & HST (COSMOS) \\
Par016 & GO-1571 & 152.659415  &  $-$4.750800   & R   & R   & R   &  2.48 & F150W, F200W        & HST (GO-17461) \\
Par019 & GO-1571 & 130.365070  &  48.603346   & R   & R   &     &  1.72 & F115W, F150W        & HST (GO-17461) \\
Par020 & GO-1571 & 149.622923  &   2.225106   & R   & R   &     &  5.92 & F150W               & HST (COSMOS) \\
Par021 & GO-1571 & 161.163517  &  33.929461   & R   & R   & R   &  1.91 & F200W               & HST (GO-17461) \\
Par023 & GO-1571 & 150.065529  &   2.551921   & R   & R   &     &  5.34 & F150W               & HST, JWST (COSMOS-Web) \\
Par024 & GO-1571 & 150.131524  &   2.499660   & R   & R   &     &  5.92 & F150W               & HST, JWST (COSMOS-Web) \\
Par025 & GO-1571 & 150.245773  &   1.977474   & R   & R   &     &  5.92 & F150W               & HST, JWST (COSMOS-Web) \\
Par026 & GO-1571 & 150.122717  &   2.439877   & R   & R   &     &  5.92 & F150W               & HST, JWST (COSMOS-Web) \\
Par028 & GO-1571 & 150.087407  &   2.418531   & R/C & R/C & R/C & 33.7  & F115W, F150W, F200W & HST, JWST (COSMOS-Web) \\
Par029 & GO-1571 & 150.252287  &   1.728557   & R   & R   &     &  5.92 & F150W               & HST (COSMOS) \\
Par034 & GO-1571 & 190.403103  &  22.499539   & R/C & R   & R   &  8.68 & F150W, F200W        & HST (GO-17461) \\
Par040 & GO-1571 & 181.362901  &  $-$7.536305   & R   & R   & R   &  2.85 & F200W               & HST (GO-17461) \\
Par042 & GO-1571 & 195.184632  &  12.534135   & R   & R   & R   &  4.58 & F150W, F200W        & HST (GO-17461) \\
Par044 & GO-1571 & 265.744237  &  67.090205   & R   & R   & R   &  3.19 & F200W               & HST (GO-17461) \\
Par052 & GO-1571 & 150.155221  &   2.041839   & R/C & R   & R   & 19.9  & F150W, F200W        & HST, JWST (COSMOS-Web) \\
Par682 & GO-3383 & 342.076592 & $-$44.555687 & R/C & R/C & R/C & 63.4  & F115W, F150W, F200W & None \\
\enddata

\tablecomments{
R and C indicate observations obtained with the
NIRISS GR150R and GR150C grisms, respectively.
The exposure time ($t_{\rm exp}$) represents the total
grism integration time summed over all available
grism filters and orientations.
}
\vspace{-15pt}
\end{deluxetable*}


One of the largest JWST WFSS surveys to date is PASSAGE (GO-1571; PI: Malkan), which observed 63 independent fields covering an area of $>$300 sq. arcmin. for a total of $\sim$590~hr with JWST/NIRISS in pure-parallel mode in Cycle 1 \citep{Malkan25}. 
This has already resulted in the determination of robust spectroscopic redshifts and masses for $>$~1000 galaxies at $z\sim0-5$ \citep{Huberty26}, identification of several {\Lya} emitters at the epoch of reionization \citep{Runnholm25}, and spatially-resolved metallicity gradients at cosmic noon \citep{Acharyya26}. 
In this study, we combine 17 multiple-grism filter fields in PASSAGE with one in the JWST Wide Area 3D Parallel Survey OutThere survey \citep[GO-3383; PIs: Glazebrook \& Brammer;][]{Jaiswar26},
to investigate the MZR and its dependence on SFR and morphology at redshifts $1.7\lesssim z \lesssim 3.5$.

This paper is organized as follows. 
In \S~\ref{sec:data}, we present the data used in our study. In \S~\ref{sec:measure}, we describe our sample selection and measurements.
In \S~\ref{sec:analysis}, we present our analysis of the MZR and investigate its dependence on SFR and morphology. We discuss our results in \S~\ref{sec:discuss}, and conclude in \S~\ref{sec:conc}.
We assume \citet{Planck20} cosmology with $\Omega_m=0.315$, $\Omega_{\Lambda}=0.685$, and $H_0=67.4~{\kmsmpc}$.

\section{Data} \label{sec:data}

\subsection{NIRISS Grism Spectra and Imaging}

This study is primarily based on PASSAGE \citep{Malkan25},
where majority of the fields have grism spectroscopy and imaging in more than one of the NIRISS filters F115W, F150W, and F200W ($\sim$1-2.3~{\micron}). Several of these also have both orthogonal R and C grism orientations to help mitigate contamination, from, e.g., source confusion.

Our sample for this study consists of 17 multi-filter fields from PASSAGE and one from OutThere \citep{Jaiswar26}. Table~\ref{tab:fields} lists details of all of these fields, including the available grism filters and orientations, total exposure times, and ancillary imaging (see below).
Given the NIRISS field-of-view (FOV) of 2.2{\arcmin}$\times$2.2{\arcmin}, our sample covers a total non-contiguous area of $\gtrsim$~87 sq. arcmin. 
Thirteen of our 18 fields have $\geq1$ hr exposures in each of the observed filters. 
The deepest PASSAGE field, Par028, has $>$33 hrs of grism observations across three filters with both orientations in each. 
Par682, which is the lone field from the OutThere survey, has $>$20-hr exposures in each of the three filters, totaling $\sim$63 hrs of NIRISS grism observations, rivaling that of NGDEEP \citep{Bagley24}.

All our NIRISS/WFSS data are reduced with the Grism Redshift \&\ Line analysis \citep[\textsc{Grizli;}][]
{Brammer_grizli_2021zndo} software using the NIRISS calibrations for the NGDEEP survey from \cite{Pirzkal2024ApJ}. Further details of the survey data reduction are presented in \cite{Malkan25}, \cite{Huberty26}, and Watson et al. (in prep.).
For each field, a detection image is made by taking the weighted mean of all direct images for all filters available for that field. All sources in the detection image are identified using the Python Library for Source Extraction and Photometry \citep[SEP;][]{barbary2016}. 
{\sc Grizli} is then used to extract the 1D and 2D spectra for each identified source.

\begin{figure*}[htbp]
\vspace{-10pt}
\centering
\gridline{
\fig{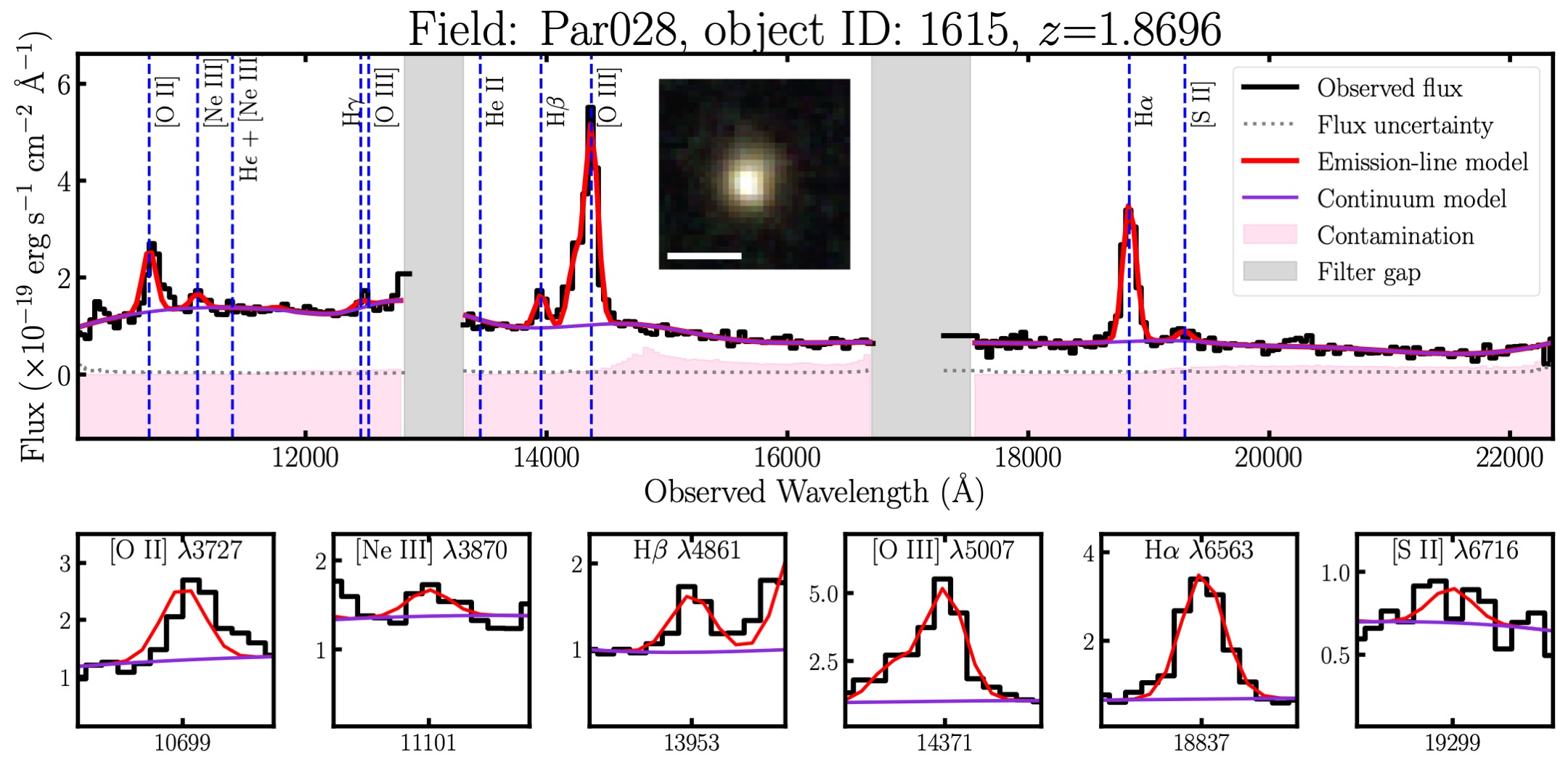}{\textwidth}{}
}
\vspace{-20pt}
\caption{Example of a combined 1D grism spectrum and RGB image cutout of a galaxy. 
The top panel shows the observed spectrum (solid black), uncertainty on the flux (dotted grey), emission-line model (solid red), continuum model (solid purple), selected emission lines (dashed blue lines), contamination model (shaded pink), masked filter gaps (shaded grey). 
The bottom panels show zoom-ins of selected emission lines.
In the RGB image cutout (combining NIRCam F115W, F150W, F277W, F444W) the solid white horizontal line represents a 0\farcs5 spatial scale.
}
\label{fig:spec+img_eg}
\end{figure*}

\subsection{Ancillary Imaging} \label{sec:ancimg}

Other than Par682, all of our fields have rich ancillary imaging. Six of the fields overlap with COSMOS-Web \citep{Casey23}, giving us access to high-resolution ancillary imaging with JWST/NIRCam in four filters, F115W, F150W, F277W, and F444W, and JWST/MIRI in F770W in some cases as well. We obtained science-ready mosaics in all four filters from the COSMOS2025 data release\footnote{\url{https://cosmos2025.iap.fr/}}. The data processing and reduction is described in detail in \citet{Franco26}. We use the images and weight maps at 0\farcs03 pix$^{-1}$ scale. 
We also obtained value-added catalogs from COSMOS2025 DR1, which included photometry, morphological measurements and spectral energy distribution (SED)-based physical parameters such as 
photometric redshifts \citep{Shuntov25}.

Four other fields overlap with the wider COSMOS area \citep{Scoville07}. For these, we obtained science-ready HST/ACS F814W mosaics processed with the v2.0 COSMOS data processing pipeline \citep{Koekemer07}. We use the 0\farcs03 pix$^{-1}$ images and weight maps. Further details of the reduction process are provided in \citet{Koekemer07} and \citet{Massey10}. 
We further obtained the COSMOS2020 value-added catalogs containing photometry, morphology, and physical parameter measurements \citep{Weaver22}.

Seven of our fields have ancillary imaging from a Cycle 31 HST follow-up program (GO-17461; PI: Mehta). These each have imaging in four HST filters, WFC3/UVIS F475W, F625W, and ACS/WFC F814W. All of the imaging data products are at 0\farcs03 pix$^{-1}$ scale. The data reduction will be described in detail in a follow-up paper (Mehta et al., in prep).

\section{Measurements} \label{sec:measure}

\begin{figure*}[htbp]
    \centering
\gridline{
\fig{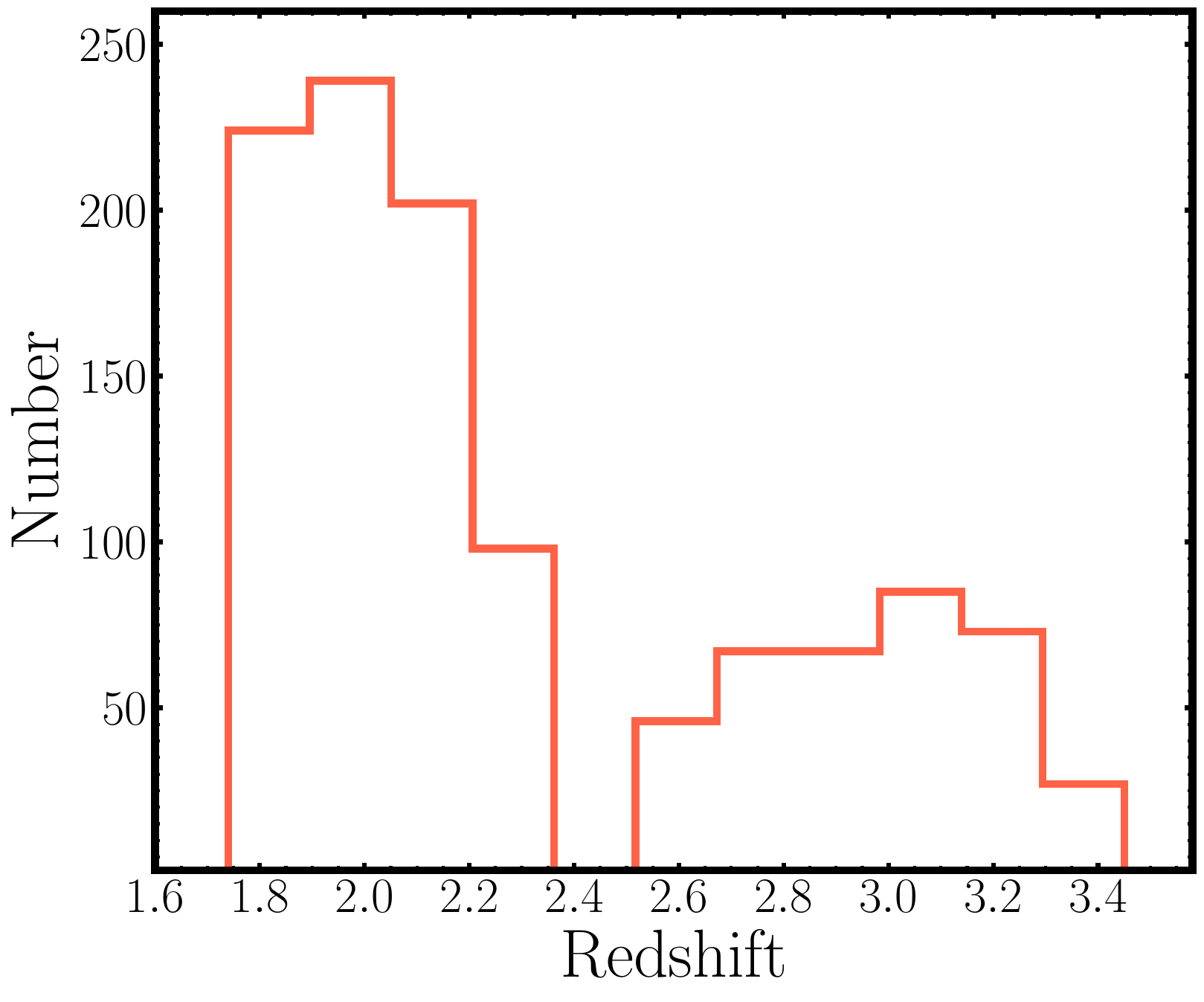}{0.25\textwidth}{}
\fig{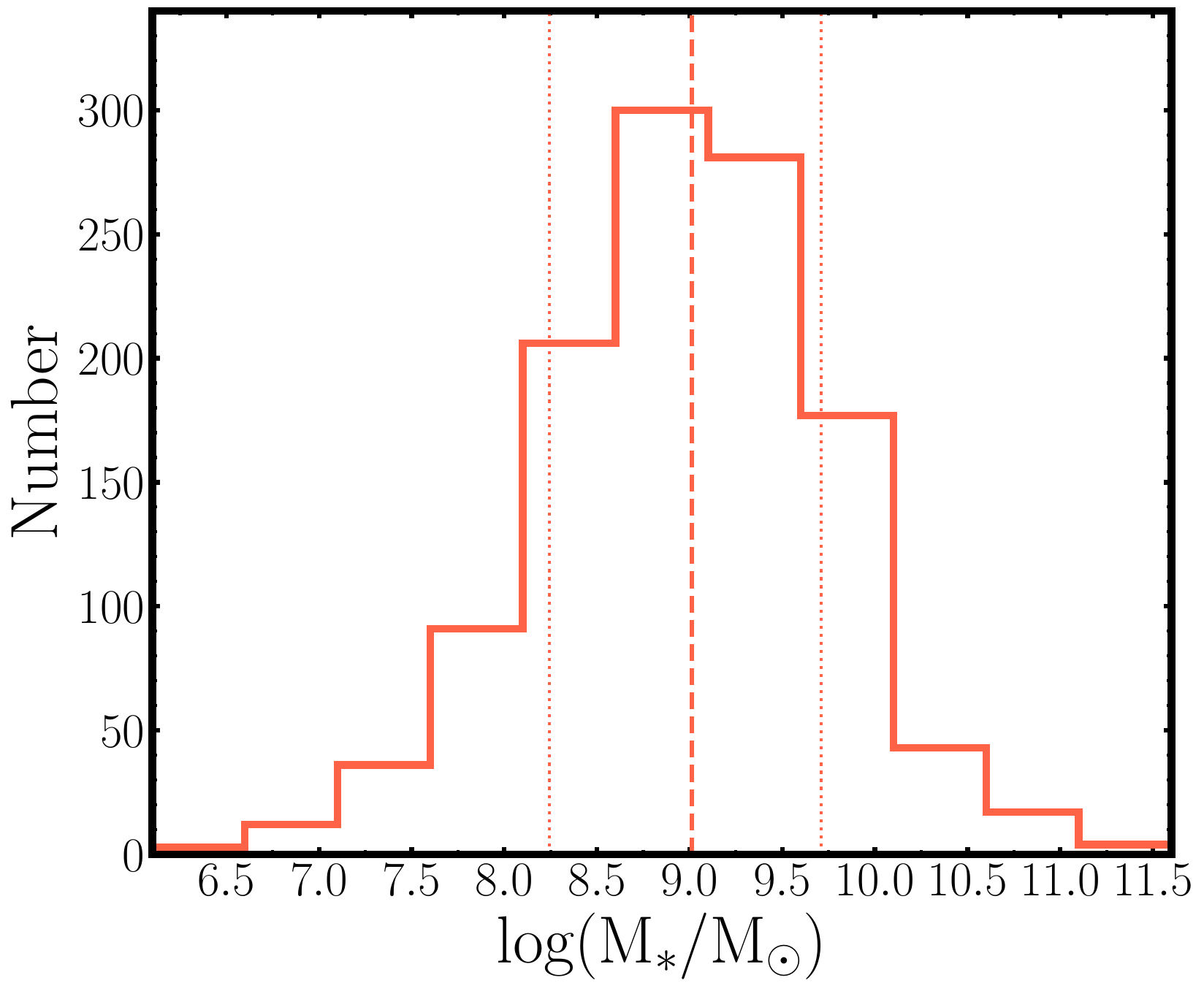}{0.25\textwidth}{}
\fig{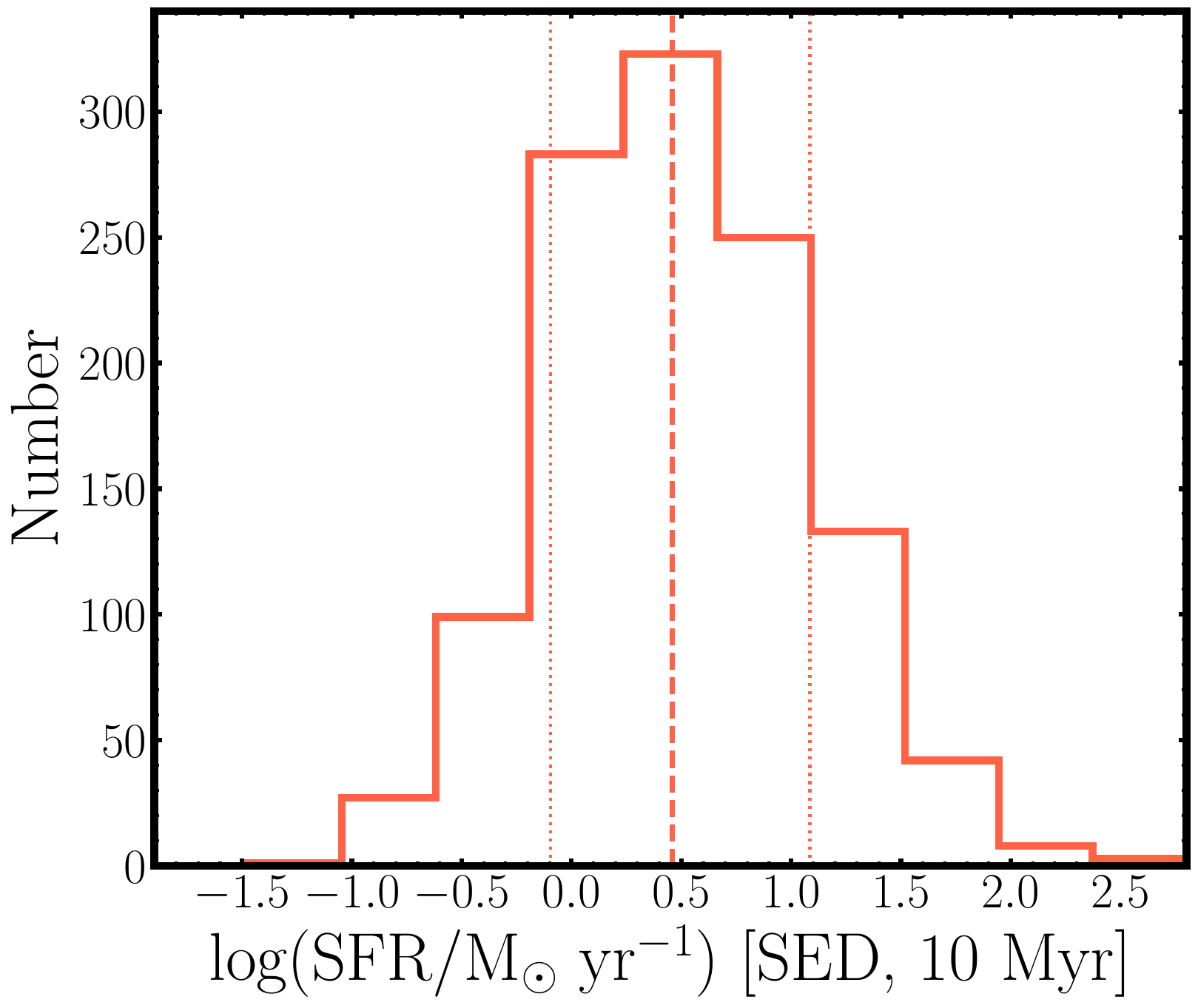}{0.25\textwidth}{}
\fig{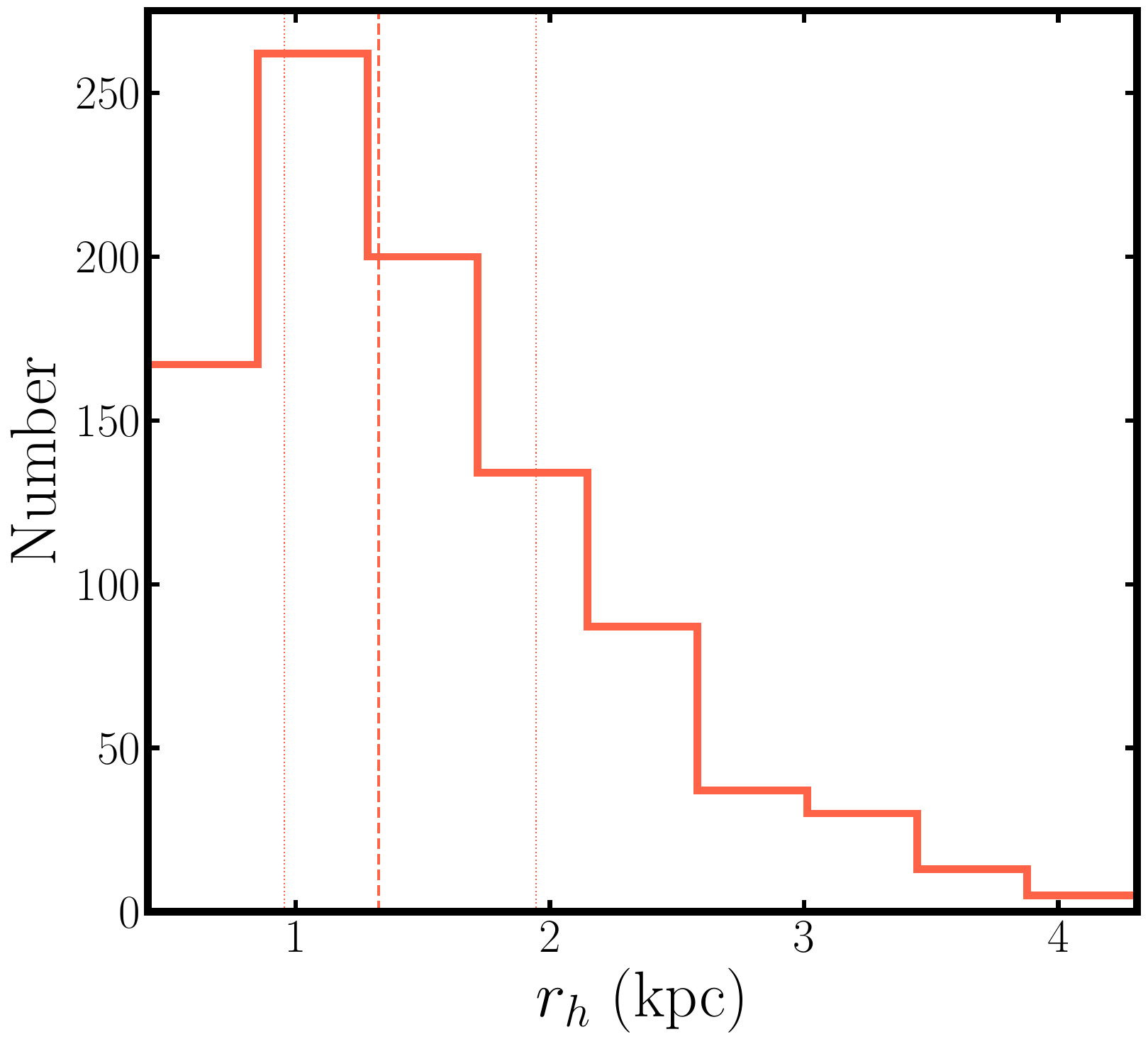}{0.23\textwidth}{}
}   
\vspace{-25pt}
\caption{Histograms showing properties of our final sample: {\it from left to right}: redshift, stellar mass, SFR, half-light radius.
In all but the left panel, the dashed and dotted lines indicate the median and middle 50\% of the distributions, respectively.
}
    \label{fig:hists_sample}
\end{figure*}

We present an example of the combined 1D grism spectrum and RGB composite image in Fig.~\ref{fig:spec+img_eg}. 
We also present histograms illustrating properties of our final sample, which we will describe in \S~\ref{sec:finalsamp}, in Fig.~\ref{fig:hists_sample}. The measurements we make to compile this sample are outlined below.

\subsection{Spectral Line Analysis and Metallicities}
\label{sec:spec}

We adopt a multi-step approach to identify and model emission lines from the NIRISS grism spectra. This includes 
(1) obtaining emission-line galaxy candidates using an automated line detection technique \citep{Colbert13,Henry2021ApJ}, 
(2) visually inspecting spectra and images to correctly identity emission lines, and 
(3) carefully modeling the continuum and emission lines for each spectrum.
We perform the spectral analysis using our team's custom emission line-finding and analysis software designed for JWST/NIRISS grism spectra\footnote{\url{https://github.com/jwstwfss/line-finding}} \citep{LineFindingCode2026zndo}.
This software, originally developed for HST/WFC3 grism data \citep{Henry2021ApJ, Revalski2024ApJ,Battisti2024MNRAS}, was publicly released in \citet{Huberty26}.
We performed careful manual continuum-fitting to estimate fluxes, equivalent widths, and other spectral quantities for each galaxy.
Details of this approach are described in other papers \citep[][Nedkova et al., in prep.]{Huberty26}.

While the gold standard for gas-phase metallicity measurements relies on electron temperature ($T_e$)-sensitive auroral lines such as {\OIIIauwave} \citep[i.e., ``direct'' metallicities; e.g.,][]{Kewley19,Curti25_review},
these lines are very difficult to detect and resolve.
The alternative, empirical calibrations to readily observable strong-line ratios such as {\OIIIf}/{\Hb} and {\OIIIf}/{\OIIf}, are traditionally derived from local analogs of high-$z$ star-forming galaxies \citep[e.g.,][]{Curti2017MNRAS,Bian2018ApJ,Nakajima2022ApJS}, which may introduce systematic biases due to evolving ISM conditions \citep[e.g.,][]{Strom2018ApJ,Kashino2022ApJ,Stanton2024MNRAS}.
Recent JWST observations of auroral lines in galaxies at $z\gtrsim1$ provide high-$z$ strong-line calibrations better suited for metallicity measurements at cosmic noon \citep[e.g.,][]{Sanders2024ApJ,Cataldi25,Chakraborty2025ApJ,Scholte2025MNRAS,Isobe26,Sanders26}.

We measure metallicities from our identified emission lines using a Bayesian approach that marginalizes over poorly constrained parameters and derive realistic uncertainties on the metallicity estimates, a method adopted from \citet{Henry2021ApJ}. 
This forward-modeling likelihood approach jointly infers gas-phase metallicity, dust extinction $E(B-V)$, stellar absorption in the Balmer lines, and contamination of {\Hg} by {\OIIIauwave}.
The methodology is described in detail in Appendix C of \citet{Henry2021ApJ}.
The primary change we make with respect to \citet{Henry2021ApJ} and \citet{Revalski2024ApJ} is the use of newer strong-line calibrations of \citet{Cataldi25} based on JWST measurements of cosmic noon galaxies, as opposed to the \citet{Curti2017MNRAS} calibrations based on local analogs of high-$z$ galaxies. 
This has the additional advantage of being able to extend the metallicity measurements down to ${\logoh}=7.0$, which is $>$~0.5 dex lower than the \citet{Curti2017MNRAS} calibrations.
We use the following features for metallicity measurements: {\NeIIIwave}, {\OIIwave}, {\Hbwave}, and {\OIIIwave}. 
The metallicities are measured in stacked, rather than individual, spectra (see~\S~\ref{sec:stack}).

We restrict our analysis to galaxies with ${1.74 \leq z \leq 3.45}$, as this is the redshift range where {\OIIwave}, {\Hbwave}, and {\OIIIwave} can all be detected with the NIRISS/WFSS F115W, F150W, and F200W filters \citep[see, e.g.,][]{Malkan25}. These lines are essential to obtaining robust metallicity measurements (described below). We require a $S/N({\OIIIf})\geq5$ for inclusion in our final sample as well, to ensure that galaxies with robust redshifts are selected and they will contribute meaningfully to stacked spectra (see below), with low noise levels.
This yields an initial sample of 1208 galaxies at ${1.74 \leq z \leq 3.45}$ with in this redshift range and with $S/N({\OIIIf})\geq5$. Our redshift measurements are very secure, with a median error of $\Delta z\sim0.0004$. 
The redshifts of our final sample are shown in the left panel of Fig.~\ref{fig:hists_sample}.
The median redshift is about $z\sim2.3$, 
and the gap at $z\sim2.3-2.5$ is from {\OIIf}, {\Hb}, and {\OIIIf} falling in the NIRISS filter gaps so we do not have access to lines we require for metallicity measurements in this window (see also Nedkova et al., in prep).

\subsection{Masses and SFRs from SED-fitting}
\label{sec:sed}

\begin{figure*}[htbp]
\vspace{-10pt}
\centering
\gridline{
\fig{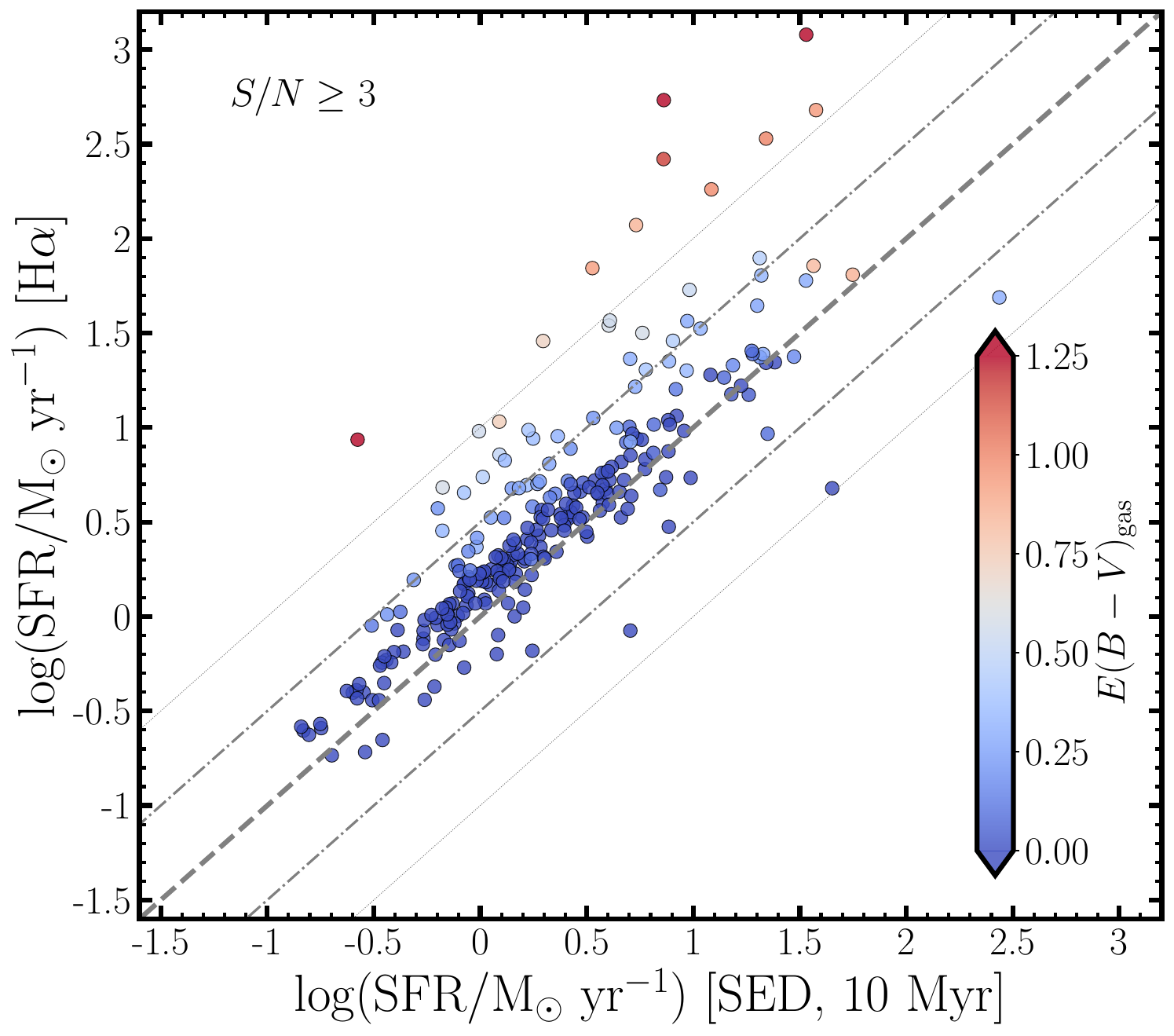}{0.325\textwidth}{}
\fig{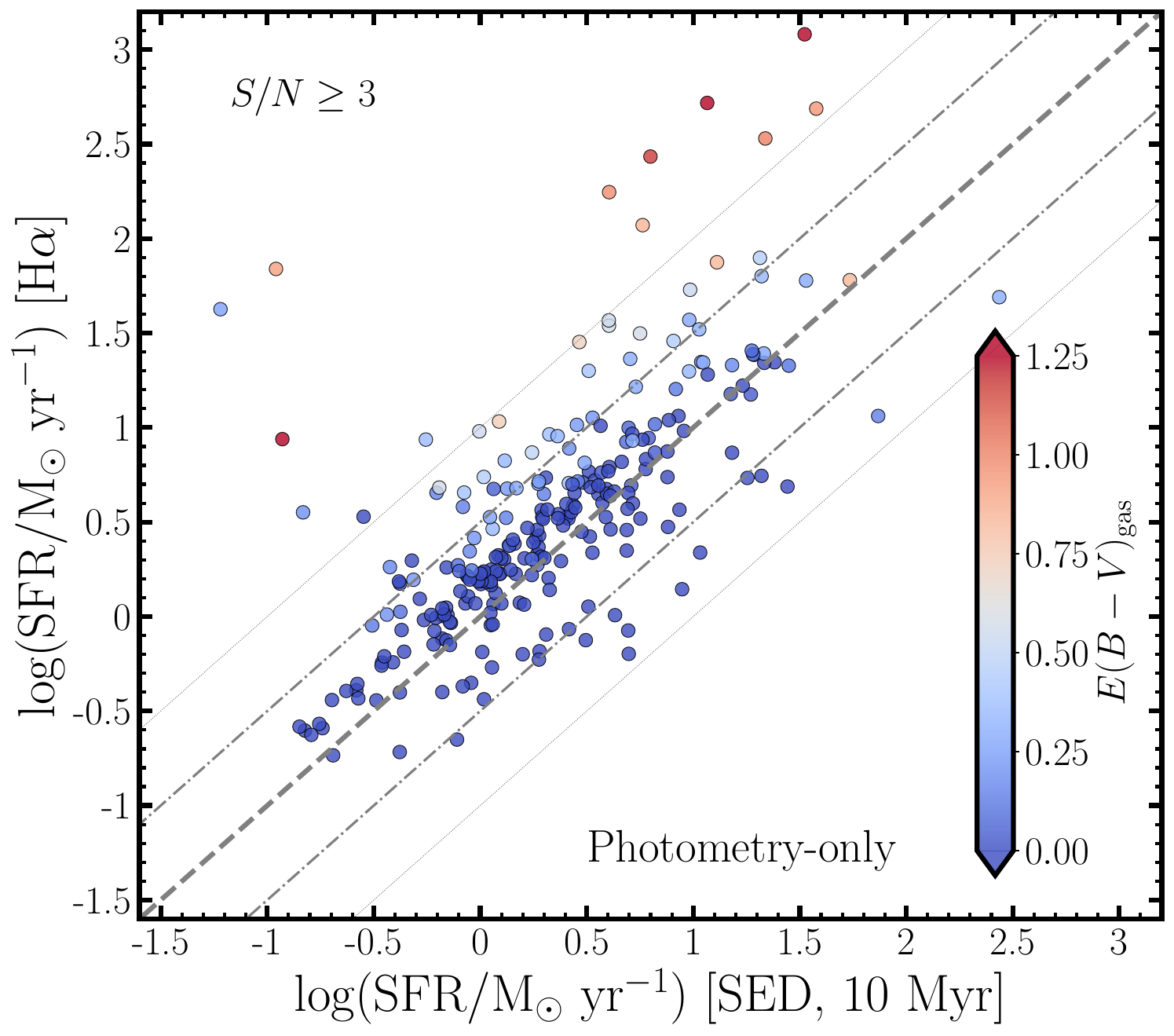}{0.325\textwidth}{}
\fig{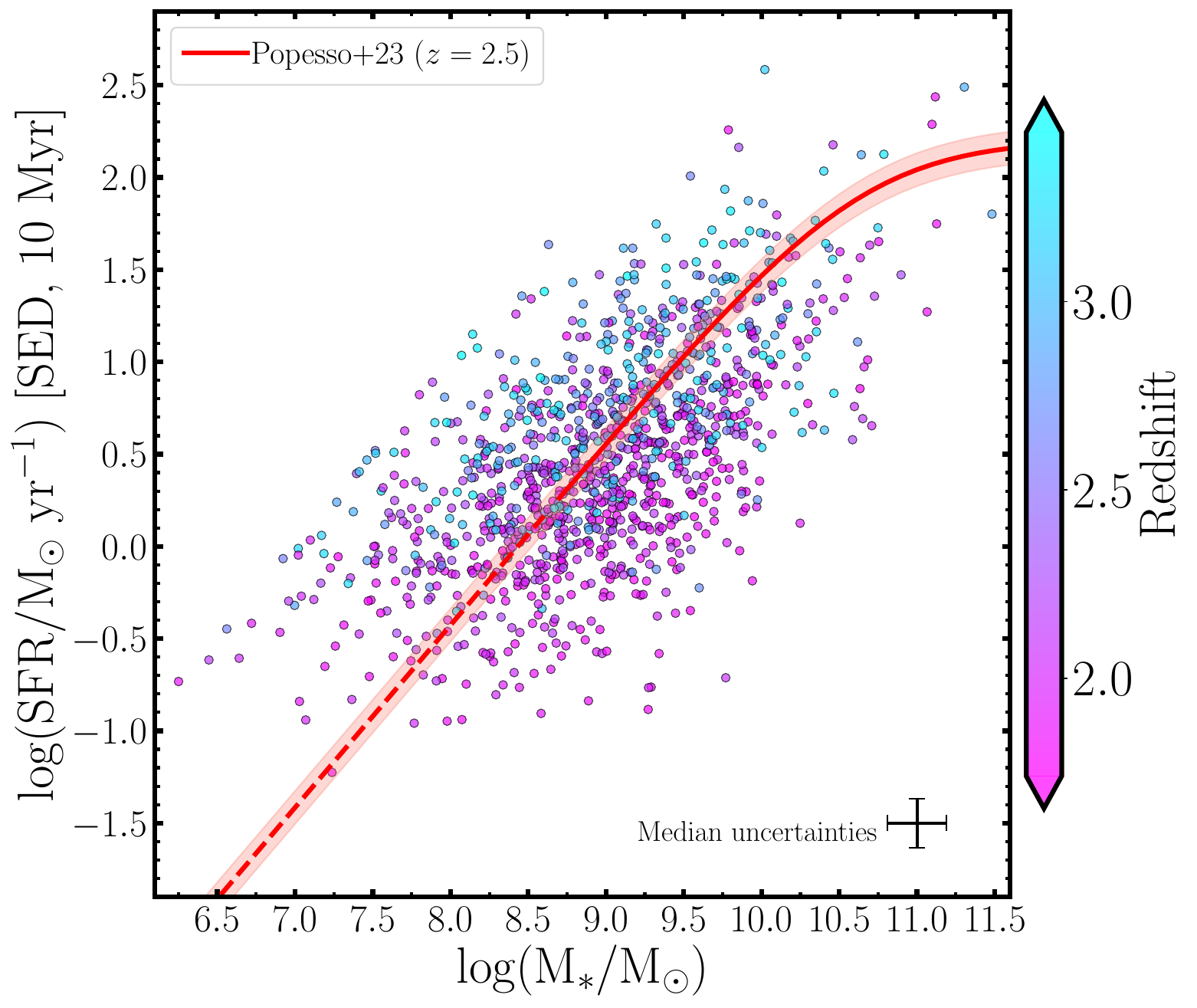}{0.345\textwidth}{}
}
\vspace{-25pt}
\caption{
Statistics from SED-derived quantities.
{\it Left}: The SFR estimated from dust-corrected {\Ha} emission with the SED-derived SFR averaged over a 10~Myr-timescale for our fiducial SED-fitting approach leveraging both the photometry and emission lines. 
{\it Middle}: SFR({\Ha}) vs. SFR(10~Myr) when SED-fitting only uses photometry.
In both these panels, the points are color-coded by the nebular color excess $E(B-V)_{\mathrm{gas}}$. Dashed line represents the 1:1 correspondence, while the dot-dashed and dotted lines represent $\pm$0.5 dex and $\pm$1 dex above and below 1:1 correspondence, respectively.
{\it Right}: SED-based SFR over 10~Myr timescale vs. {\mstar} colored by redshift. The best-fit SFMS at $z=2.5$ parameterized by \citet{Popesso23} and its $\pm1\sigma$ uncertainty interval are shown in magenta. Median uncertainties on {\logms} and {\logsfr} are indicated on the lower right.
10~Myr SFRs are generally in very good agreement with {\Ha} SFRs and our SFMS is in good agreement with the literature.
}
\label{fig:sfr-comp}
\end{figure*}

Constraining the MZR and its dependence on star formation and morphology depend critically on robust estimates of stellar masses and SFRs.
This is achieved via SED-fitting on a wealth of multi-band photometry at our disposal, spanning rest-frame UV ($\sim0.13~{\micron}$) to MIR ($\sim8.7~{\micron}$) 

In fields falling within COSMOS, the number of filters available for SED-fitting is over 30, with COSMOS-Web providing NIR and MIR coverage as well as optical and UV. Following \cite{Huberty26}, we match sources within 0\farcs3 of COSMOS-Web \citep{Shuntov25} sources (for the COSMOS-Web overlapping fields) or COSMOS-2020 \citep{Weaver22} sources (for the COSMOS overlapping fields). The photometry is described in extensive detail in these aforementioned papers.

The seven HST follow-up fields have three HST optical filters in addition to at least one NIRISS filter.
The HST fluxes are measured on mosaics drizzled to the JWST/NIRISS pixel grid, detected on a $\chi^2$ stack of the three bands, and matched to JWST sources by segmentation overlap. Therefore, the HST photometry is obtained from the same physical apertures and objects as the JWST photometry (Mehta et al., in prep).

We use the {\sc Bagpipes} SED-fitting package \citep{Carnall2018MNRAS} to fit all the available ancillary photometry and emission line fluxes. 
We adopt a non-parametric star formation history (SFH) with continuity prior \citep{Leja19a}, the 2016 update of the \cite{BruzualCharlot2003} stellar population models, and assume a \cite{kroupa2001} initial mass function (IMF), and a \citet{Calzetti2000} dust attenuation curve. 
Details of the SED-fitting process are described in 
Nedkova et al., (in prep.).

Since we wish to explore the physical connection between gas-phase metallicity and recent star formation activity, we carefully consider the timescales associated with different measures of the SFR.
Emission lines in nebular {\HII} regions surrounding massive/short-lived stars are used to measure metallicity. These regions reveal the current (observed) state of the star-forming ISM.
Accordingly, many previous studies adopted the SFR measured from nebular recombination lines such as {\Ha} to understand the dependence of SFR on the MZR \citep[e.g.,][]{Curti20,Sanders2021ApJ}.

However, we are severely limited in our ability to compute SFR({\Ha}). With NIRISS, {\Ha} can only be detected at $z\lesssim2.4$, and even then the {\Hawave} is blended with the {\NIIwave} doublet at its resolution (${R\sim150}$).
Furthermore, the {\Ha} emission is also subject to dust extinction in nebular regions, which we require much weaker lines such as {\Hb} to correct for. 
Only 272 galaxies -- $\sim$20\% of our initial sample -- show {\Ha} and {\Hb} emission at $S/N\geq3$.

Instead, we have access to SFRs derived from SED-modeling for all of our galaxies. However, the nebular-line and SED SFRs probe different timescales and depend on different modeling assumptions. Nebular lines such as {\Ha} trace the ionizing radiation of massive (O and B) stars and are therefore sensitive to star formation over the last few to 10 Myr, whereas SED-based SFRs reflect the integrated stellar continuum averaged over some timescale \citep[e.g.,][]{KE12}. Differences between the two types of SFRs may arise due to recent SFH, including bursts or rapid declines, as well as from systematic effects such as dust attenuation, nebular-to-stellar reddening, stellar population models, and the assumed form of the SFH \citep[e.g.,][]{Reddy15,Shivaei15}.

We estimate SFR({\Ha}) using the canonical \citet{KE12} prescription with a \citet{kroupa2001} IMF and \citet{Calzetti2000} dust extinction law to be consistent with the SED-based measurements. We apply a mass- and redshift-dependent correction to remove {\NIIf} from {\Ha} following \citet{Faisst18}. Note that the {\NIIf}/{\Ha} ratio is $<$~10\% for galaxies with ${\logms}\leq9.5$ at $z\sim1.7-2.7$, and rises strongly only at high masses (${\logms}\gtrsim10.5$).
We then measure the Balmer decrement ${\Ha}/{\Hb}$ and estimate the color excess $E(B-V)_{\mathrm{gas}}$  \citep[assuming an intrinsic Case B recombination line ratio of 2.86; e.g.,][]{HB87}.

The left panel of Fig.~\ref{fig:sfr-comp} compares the dust-corrected SFR({\Ha}) with the 10~Myr-averaged SED-based SFR given our fiducial SED setup using both photometry and emission lines. The middle panel presents the comparison where SFR(10~Myr) is from SEDs which only utilize photometry. Individual points are colored by $E(B-V)_{\mathrm{gas}}$.
For our fiducial SED models, SFR(SED, 10~Myr) agrees with SFR({\Ha}) to within $\approx$0.3~dex for most of our {\Ha}- and {\Hb}-detected galaxies.
This is consistent with the offset observed with the JADES survey at $z\geq3$ \citep{Curti2024A&A}. This level of scatter is also equivalent to the average uncertainties on SFR(SED, 10~Myr). A small fraction of the galaxies have SFR(SED, 10~Myr) offset from SFR({\Ha) by $>$~0.5 dex, due to high levels of dust extinction. The negligible fraction ($\lesssim$4\%) of galaxies with SFR({\Ha) $\geq$1 dex higher than SFR(SED, 10~Myr) have very high dust extinction, with $E(B-V)_{\mathrm{gas}}\gtrsim1$.

The inclusion of emission-line fluxes to the SED-fitting significantly improves the agreement between SFR({\Ha) and SFR(SED, 10~Myr). With photometry-only SED-fitting, the average scatter is larger, $\sim$0.5 dex, including several $>$1 dex outliers at lower SFRs (${\logsfr}<0$). This is not surprising as the constraints of emission lines -- which trace very recent star formation -- helps the SED fits better approximate short-timescale SFR. Ignoring the outliers, the photometry-only SFR (10~Myr) still correlate with the SFR({\Ha}) to a reasonably good degree.

We note that the SFRs averaged over 100~Myr from our fiducial SED fits are far less well-correlated with SFR({\Ha}) than SFR(10~Myr). The average disagreement is $>$0.5~dex, including many $>1\sigma$ outliers. The median $\pm1\sigma$ uncertainties on SFR(SED, 100~Myr) -- $\approx$0.45 dex -- are also much larger than those on SFR(SED, 10~Myr) -- $\approx$0.25 dex.

The fairly tight correlation of the 10 Myr-SFR with {\Ha} SFR indicates that the former traces the current state of the ISM fairly well. This supports our choice of SED-based SFR averaged over a 10~Myr-timescale in the remaining analysis. We note that the uncertainties on the individual SFRs ($\sim$0.25 dex on average) are not negligible and that the lack of rest-frame MIR or FIR photometry in the SEDs can bias the SFRs significantly \citep[e.g.,][]{Battisti15}.
However, these factors do not affect our scientific conclusions as we only bin galaxies near the top and bottom of the SFR distribution when investigating the SFR-dependence of the MZR and therefore individual uncertainties do not significantly affect our results (see \S~\ref{sec:analysis}).

The right panel of Figure~\ref{fig:sfr-comp} presents the 10 Myr SFR vs. stellar mass, i.e., the star-forming main sequence (SFMS) for our sample. Each point is colored by redshift. We only include galaxies whose SED-fitting was reliable, with at least three photometric filters, from here on out. Imposing a more strict criterion, such as requiring a minimum of four filters, would reduce our sample size significantly without any changes to our qualitative results. The median uncertainties on our measurements
are indicated on the lower right. 
Note that some galaxies in this figure are likely AGN-hosts, which we remove before compiling our final sample (\S~\ref{sec:finalsamp}).

To place our results in context of the literature, we also include the parameterization of the SFMS from the comprehensive compilation of \citet{Popesso23} at $z=2.5$, the mean redshift of our sample. The solid magenta curve indicates the best-fit SFMS over the mass range that they performed the fitting on and the dashed portion of the curve is an extrapolation to ${\logms}<8.7$. The shaded region represents the $\pm1\sigma$ confidence interval on the best-fit SFMS.
We find that on average, our data are in good agreement with the best-fit SFMS from the literature across several orders of magnitude in mass and SFR. However, there is fairly significant scatter around the median, typically $\approx$0.5 dex. At low masses (${\logms}\lesssim8$), very few galaxies are found to be below the expected SFMS, which could be due to emission-line selection biases finding more highly star-forming low-mass galaxies. As expected, our higher-$z$ galaxies appear to have higher SFR compared to lower-$z$ galaxies at the same masses.

\subsection{Morphologies}
\label{sec:morph}

We characterize the morphologies of galaxies using the multi-band imaging  compiled for each field. 
We perform this characterization in two ways: (1) quantitative metrics and (2) visual classifications of morphologies.

\subsubsection{Quantitative Morphology Measurements}
\label{sec:morphquan}

We measure galaxy structural properties using {\sc Statmorph} \citep{RG19,statmorphsoft}, which computes both non-parametric morphological diagnostics and 2D S{\'e}rsic-profile fits from images. 
{\sc Statmorph} uses a single-filter science image, segmentation map labeling individual sources, a ``weight map'' representing the 1$\sigma$ uncertainty, and a PSF for the filter. 
We apply {\sc Statmorph} on the highest-quality imaging available for each field, prioritizing a combination of spatial resolution, depth, and wavelength coverage. For the COSMOS-Web fields (6), we used JWST/NIRCam F150W imaging. In fields overlapping with the COSMOS footprint only (4), we use HST/ACS F814W, and likewise for the fields with HST follow-up imaging (7). For the lone field lacking ancillary HST or NIRCam imaging, we use NIRISS F150W. 
More details of our approach are provided in \S~\ref{app:sm}.

Among other quantities, {\sc Statmorph} returns the elliptical half-light radius along the semi-major axis, {\rh}, the S{\'e}rsic index, {\nsers}, and the apparent minor-to-major axis ratio, $B/A$.
It also measures non-parametric quantities including Concentration, $C$, Asymmetry, $A$, and Smoothness, $S$, statistics commonly used in the CAS morphological system \citep[e.g.,][]{Conselice03}, and the Gini coefficient $G$ and $M_{20}$ statistic, following \citet{Lotz04}. 
All scientific analysis below is limited to reliable measurements with a ``flag'' $\leq1$ or ``flag\_sersic'' $\leq1$ for S{\'e}rsic-related quantities.
We visually verify that the fits with these flags are satisfactory.
We exclude from our analyses any object with ${\nsers}<0.1$ as this is unphysical and likely due to poor fits \citep{Nedkova21}. We also require ${\rh}>0.2$~kpc in order to exclude objects which are smaller than half the full-width-at-half-maximum (FWHM) of our worst PSF (NIRISS F150W) at our maximum redshift \citep{Hausler13}.

\begin{figure}[t]
\centering
\gridline{
\fig{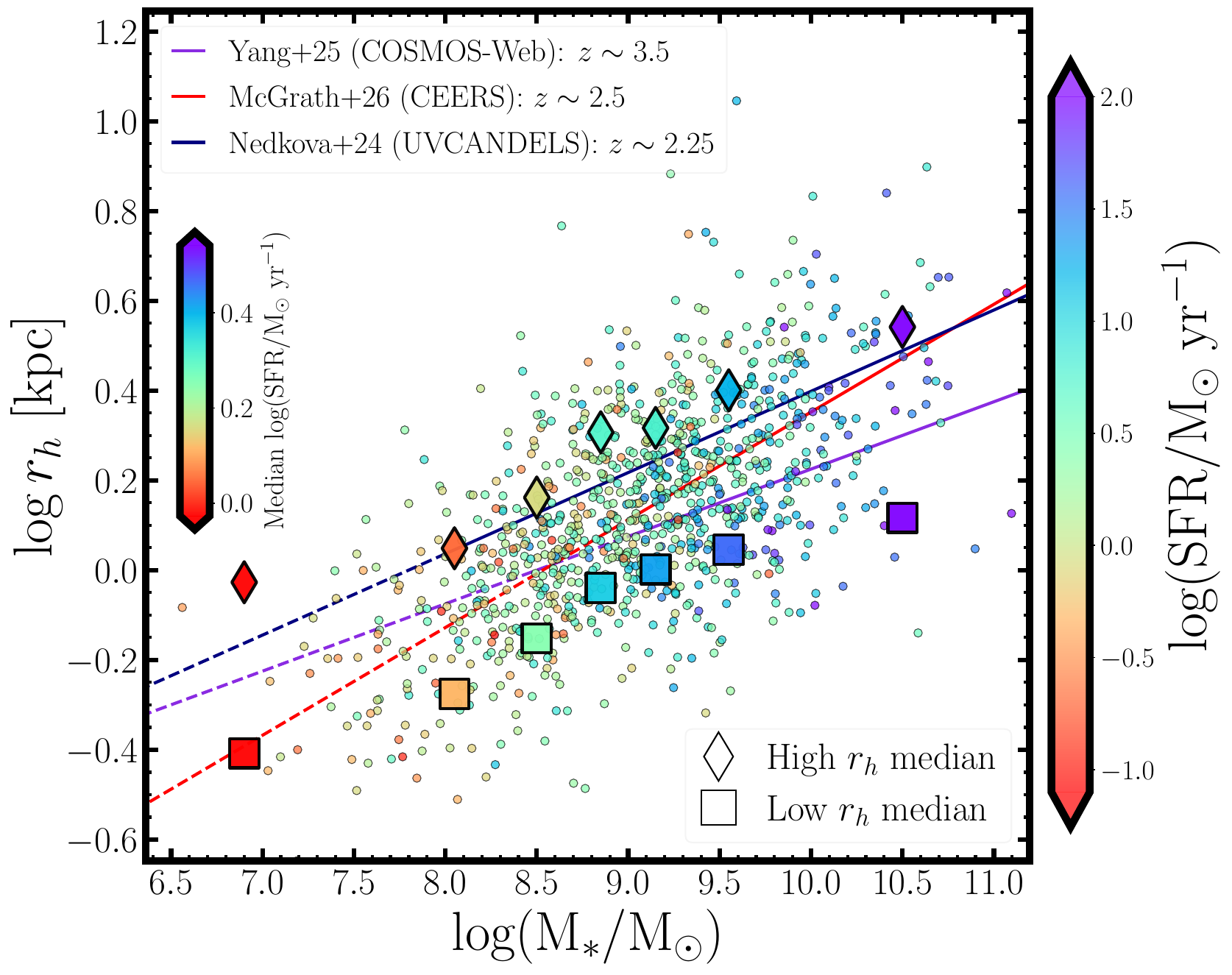}{0.46\textwidth}{}
}
\vspace{-25pt}
\caption{
Stellar mass vs. half-light radius for our galaxies. Individual points are colored by SFR, whose color-bar is on the right.
The median SFR for larger and smaller galaxies in different mass bins are indicated by shown by diamonds and squares, respectively, and the color-bars for these is on the left.
The lines represent best-fit relations from the literature at redshifts $z\!\sim\!2\!-\!3.5$. 
In general, larger galaxies tend to be less star-forming than the smaller ones.
}
\label{fig:mass-size}
\end{figure}

The rightmost panel of Fig.~\ref{fig:hists_sample} shows the histogram of the half-light radius {\rh}, with the median and middle 50\% of the distribution indicated by dashed and dotted lines, respectively.
The half-light radii are converted to physical kpc from arcseconds, using the measured redshifts.
The median {\rh} is $\sim$1.3~kpc, with the distribution of values typically $\sim~1-3$~kpc, comparable to the sizes measured for $z\sim1.5-3.5$ star-forming galaxies in many studies \citep{vanderwel14,Nedkova21,Nedkova24,Yang25}.

In Fig.~\ref{fig:mass-size}. we present the mass-size relation, i.e., {\mstar} vs. {\rh}, for our galaxies at $z\sim1.75-3.5$. Individual galaxies are color-coded by their SFR. We also show the median SFR of galaxies with large  (diamonds) and small (squares) sizes in eight bins of stellar mass.
For these, we consider the largest and smallest third of the {\rh} measurements in each {\ms} bin as the high and low {\rh} values, respectively. This binning approach is identical to that we adopt in \S~\ref{sec:mzr-size} when considering the effect of size on metallicity. 
Note that the small color-bar on the left side of the figure indicates the median {\logsfr} in each of these bins, separate from the larger color-bar which indicates {\logsfr} for the individual data points.
Galaxies become larger in size as mass increases, but interestingly, the larger galaxies in a given mass bin tend to have lower SFR than the smaller ones. We will return to this observation later (\S~\ref{sec:mzr-size}).

Best-fit empirical relations from the literature are also presented in Fig.~\ref{fig:mass-size} for comparison. In particular, we show the mass-size relation from UVCANDELS at $z\sim2-2.5$ \citep{Nedkova24}, 
CEERS at $z\sim2.5$ \citep{McGrath26}, and COSMOS-Web at $z\sim3.5$ \citep{Yang25}. The relations shown are for star-forming galaxies only, as our emission line-selected sample, by design, does not include a representative population of quenched galaxies. The dashed part of each line indicates the mass range at which we extrapolate the fitted relation. Furthermore, the samples and methods used in these measurements are heterogeneous. Despite these caveats, 
our goal is not to perform a rigorous analysis of the mass-size relation (which we defer to future work), but instead to provide additional validation for our size measurements. Therefore, we are encouraged by the broad agreement in the mass-size relation across our entire sample with the literature at overlapping redshift intervals.

We show further statistics of quantities we derive using {\sc Statmorph} in Fig.~\ref{fig:sm-stats} (Appendix~\S~\ref{app:sm}). 
As these plots show, distributions of structural parameters other than size are also broadly consistent with those expected for star-forming galaxies at cosmic noon. 
Fig.~\ref{fig:sm-stats} also shows the Gini coefficient, $G$, against the $M_{20}$ statistic, which we use to define mergers following the definition proposed by \citet{Lotz08}.

\subsubsection{Visual Morphology Classifications}
\label{sec:morphvis}

Visual classification is a longstanding approach to understanding galaxy morphologies, going back a century \citep[e.g.,][]{Hubble26,deV59,Buta13}. While this approach is highly subjective, with the high resolution of most HST and JWST imaging, vital salient features of galaxy structure can be identified \citep[e.g.,][] {Elmegreen2005}. 
We undertake a detailed visual classification approach where two expert reviewers classified the following morphological features: number of clumps, tadpole, chain, interacting morphology, presence of bulge, spiral features, and classification of apparent shape into elongated, spheroidal, or irregular.
More details of our approach are provided \S~\ref{app:vis}.

We present some of the results of our visual classification in Fig.~\ref{fig:vis-stats} (Appendix~\S~\ref{app:vis}). More than half of our galaxies have at least one clump (i.e., they are clumpy) and the clumpy fraction is in broad agreement with the literature. Furthermore, about a quarter of our galaxies have a clear central bulge, $\sim$1/6 show signs of interactions, and $\lesssim$~10\% show spiral features.

\subsection{Final Sample}
\label{sec:finalsamp}

We check for possible contamination from galaxies that host AGN, as these have harder ionizing spectra than stars that would render them unsuitable for the strong-line metallicity calibrations we use.
We use the ``mass-excitation diagnostic'' (MEx), which compares the stellar mass with the line ratio {\OIIIf}$\lambda5007$/{\Hb} \citep[e.g.,][]{Henry2013ApJ, Juneau2014ApJ, Coil2015ApJ}.
We remove 94 galaxies (or $\sim$8\% of our sample) which are likely AGN-ionized based on the MEx diagnostic. 
We only perform the MEx diagnostic experiment for sources with $S/N\geq5$ {\OIIIf} and $S/N\geq3$ {\Hb} detections. 
Further details of the AGN removal process are provided in Nedkova et al., (in prep.).


Our final sample of galaxies satisfies the following criteria: (1) redshift $1.74 \leq z \leq 3.45$; (2) {\OIIIf} $S/N\geq5$; (3) reliable SED-fits with at least three filters; (4) AGN removed via MEx diagnostic. (2) is necessary for ensuring that galaxies with robust redshifts are selected and they will contribute meaningfully to stacked spectra, with low noise levels.
We end up with 1015 sources in total that satisfy all these criteria.

As seen in Fig.~\ref{fig:hists_sample} (second from left),
our overall mass range spans nearly five dex, where the median is around ${\logms}\sim9$. Half of the galaxies have ${\logms}\lesssim8.2$ and ${\logms}\gtrsim9.7$. Our sample spans 3~dex in SFR (Fig.~\ref{fig:hists_sample}, third from left), with a median around ${\logsfr}\sim0.5$. Half of the sample has ${\logsfr}\lesssim0$ and ${\logsfr}\gtrsim1$. 
The masses and SFRs we reach are highly comparable to those of COSMOS-Web, which mostly includes photometric redshifts, at these epochs \citep{Shuntov25}.

\section{Analysis and Results} \label{sec:analysis}

\subsection{Stacked spectra}  \label{sec:stack}

Ideally, metallicities would be measured from each individual galaxy spectrum with high $S/N$. In reality, however, it is necessary to impose a $S/N$ limit on all the individual emission lines necessary to measure metallicity in each galaxy's spectrum. This can bias the sample toward galaxies with stronger emission lines, or hence higher SFRs, or higher metallicity-sensitive line ratios \citep[e.g.,][]{Sanders15,Curti20,Henry2021ApJ}.

Stacking spectra of galaxies sharing similar properties (particularly mass) has the advantage of increasing the $S/N$ of faint lines and enabling measurements of metallicity for galaxies whose individual spectra do not have sufficient $S/N$ for reliably quantifying line ratios that are used in these measurements \citep[e.g.,][]{Henry2013ApJ,Henry2021ApJ,Revalski2024ApJ}. 
At high $z$, this is particularly relevant, where individual line detections are often limited to the brightest emission-line galaxies \citep[e.g.,][]{AM13,Steidel2014ApJ,Strom2017ApJ}.
Our primary metallicity analysis is therefore limited to stacked spectra, which we leverage to understand galaxies as an ensemble \citep[see, e.g.,][for more details on the benefits of stacked over individual measurements]{Henry2021ApJ}.
We describe our stacking procedure briefly below and point interested readers to \citet{Henry2021ApJ} and \citet{Battisti22} for further details.


\begin{figure}[t]
\centering
\gridline{
\fig{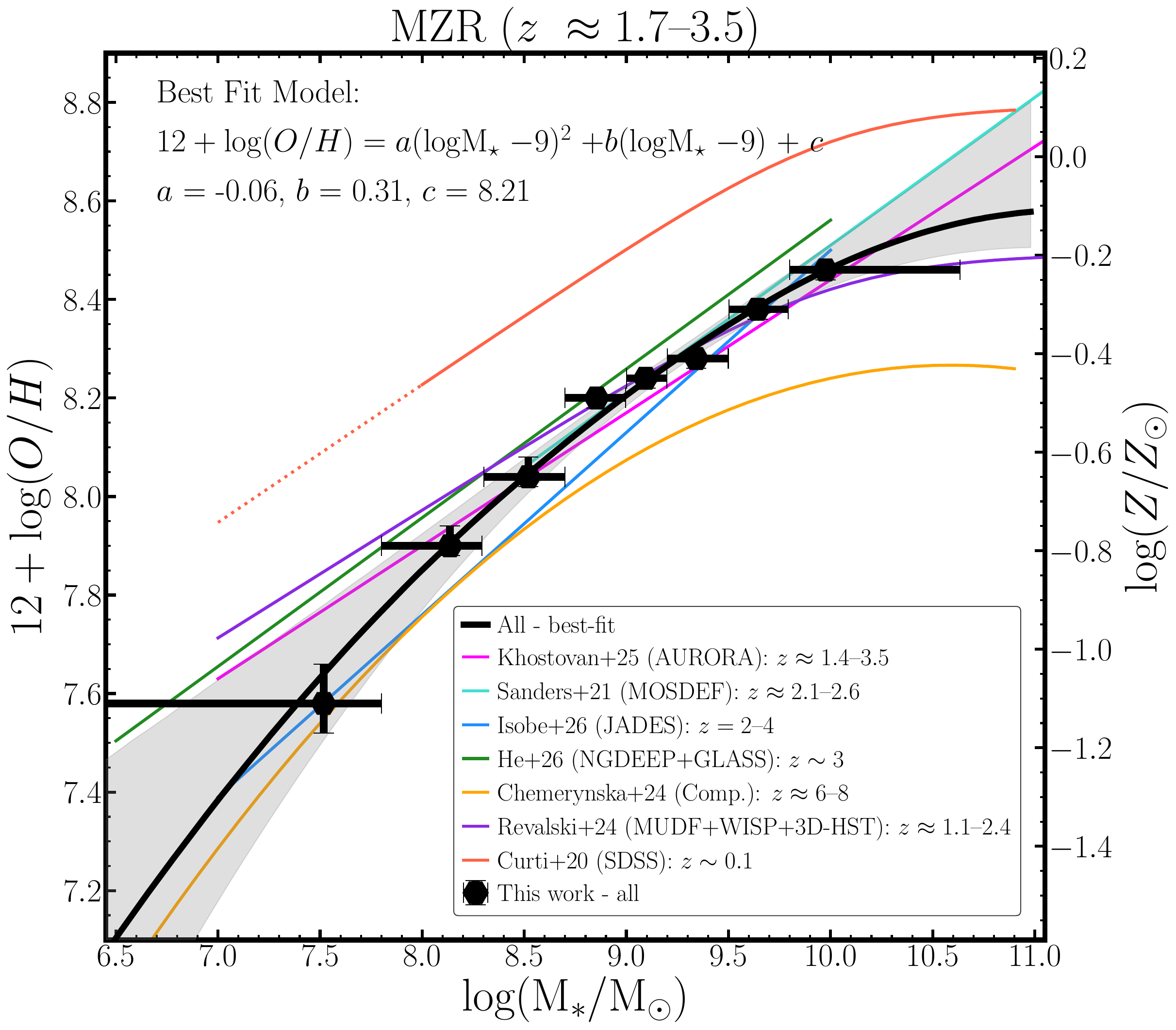}{0.48\textwidth}{}
}
\vspace{-25pt}
\caption{
The MZR of all galaxies.
The black data points and error bars show our binned MZR measurements, and the thick black curve and grey shaded region represent the best-fit quadratic equation (Eq.~\ref{eq:mzr}) and its $\pm1\sigma$ confidence interval, respectively. The best-fit parameters to our data are indicated on the top left.
The colored curves represent various MZR measurements from the literature using a variety of methods at different redshifts, from $z\sim0.1$ to $\sim$8.
A steep decline is seen at the lowest masses.
}
\label{fig:mzr_full}
\end{figure}

We stack continuum-subtracted, {\OIIIf}-normalized spectra in the rest frame, excluding masked pixels and adopting the median flux at each wavelength to avoid bias toward the brightest galaxies. Uncertainties are estimated from 1000 bootstrap resamplings, while emission-line fluxes are measured using constrained two-component Gaussian fits \citep[see][]{Henry2021ApJ}. 
We restrict stacking to subsamples with at least 15 galaxies (corresponding to Poisson error $\sim$25\%) to ensure statistically robust results.
We present our stacked spectra in the Appendix (\S~\ref{app:stack}) -- Figs.~\ref{fig:ms-stacks}, \ref{fig:sfr-stacks}, and \ref{fig:morph-stacks}. 
The results below are all derived from these.


\subsection{The MZR at $z\sim1.7-3.5$} \label{sec:mzr-full}

We apply our Bayesian metallicity measurement framework on spectra stacked in stellar mass. We choose 8 bins such that there are at least 50 galaxies in each bin for reliable statistics and that the most populated bin is only about twice as populated as the least populated bin. These choices help ensure relatively similar levels of statistical uncertainty in each bin. We verify that changing the number of bins by one or two or decreasing the minimum number of galaxies per bin by 20 or so does not affect our conclusions. 
We measure the maximum likelihood metallicity, which we denote the median metallicity, and the $\pm1\sigma$ uncertainty, in each stack. 

We model the MZR using a second-order polynomial,
\begin{equation} \label{eq:mzr}
    {\logoh} =a\left[\log M_{\ast}-9 \right]^2
      +b\left[\log M_{\ast}-9\right]+c,
\end{equation} 
where 
${\logms}=9$ is the median stellar mass of our sample and applied as a pivot to the fitting for numerical convenience.
The coefficient $c$ corresponds to the best-fit metallicity at ${\logms}=9$ -- a measure of the normalization. $b$ is the local slope at the pivot mass and $a$ is the curvature of the MZR.
This functional form is motivated by the need to fit the MZR over a fairly large range of masses \citep[e.g.,][]{Chemerynska24}.

\begin{deluxetable*}{cccc}
\tabletypesize{\footnotesize}
\setlength{\tabcolsep}{2pt} 
\tablecaption{Best-fit MZR parameters (Eq.~\ref{eq:mzr})}
\tablehead{
\colhead{Sample} &
\colhead{$a$} &
\colhead{$b$} &
\colhead{$c$}
}
\startdata
\tableline All & $-0.06^{+0.06}_{-0.02}$ & $0.31^{+0.05}_{-0.04}$ & $8.21^{+0.02}_{-0.02}$ \\ \tableline 
Low SFR & $0.03^{+0.01}_{-0.02}$ & $0.18^{+0.05}_{-0.04}$ & $8.27^{+0.02}_{-0.02}$ \\
High SFR & $-0.08^{+0.04}_{-0.02}$ & $0.30^{+0.04}_{-0.04}$ & $8.21^{+0.03}_{-0.03}$ \\ \tableline 
Large & $-0.04^{+0.05}_{-0.03}$ & $0.20^{+0.04}_{-0.04}$ & $8.27^{+0.02}_{-0.02}$ \\
Small & $-0.07^{+0.07}_{-0.05}$ & $0.35^{+0.06}_{-0.06}$ & $8.16^{+0.03}_{-0.03}$ \\ \tableline 
Low SFR surf. dens. & $0.00^{+0.03}_{-0.04}$ & $0.13^{+0.04}_{-0.03}$ & $8.31^{+0.02}_{-0.02}$ \\
High SFR surf. dens. & $-0.04^{+0.07}_{-0.06}$ & $0.34^{+0.05}_{-0.05}$ & $8.15^{+0.03}_{-0.03}$  \\
\tableline 
Bulge & $0.04^{+0.02}_{-0.03}$ & $0.12^{+0.04}_{-0.04}$ & $8.29^{+0.02}_{-0.02}$ \\
No bulge & $-0.05^{+0.06}_{-0.04}$ & $0.28^{+0.05}_{-0.04}$ & $8.20^{+0.02}_{-0.02}$ \\
\tableline 
Spiral features & $0.02^{+0.07}_{-0.04}$ & $-0.00^{+0.09}_{-0.00}$ & $8.41^{+0.02}_{-0.02}$ \\
No spiral features & $-0.06^{+0.06}_{-0.04}$ & $0.29^{+0.05}_{-0.05}$ & $8.19^{+0.03}_{-0.03}$ 
\\
\tableline 
\enddata
\tablecomments{
We report the median and $\pm1\sigma$ intervals of the Monte Carlo fit realizations.}
\label{tab:mzr-params}
\vspace{-15pt}
\end{deluxetable*}

We fit the MZR to this functional form using a Monte Carlo fitting procedure with 20,000 realizations. In each realization, the binned metallicity measurements were perturbed according to the metallicity uncertainties, while the median stellar mass of each bin was adopted as the representative mass.
To account for the finite spread in stellar mass within each bin, we propagated the mass uncertainty -- as the $\pm1\sigma$ width of the mass bin -- into the metallicity direction 
considering the local slope of the MZR.
Because the effective uncertainty depends on the local slope of the model, it was evaluated self-consistently during the fitting procedure. This treatment reduces the statistical leverage of bins spanning a broad range in mass while preserving the median mass.



We additionally imposed the physically motivated requirement that the MZR should not decrease with mass over the adopted mass range.
For a quadratic relation, this derivative is linear in stellar mass, so enforcing a non-negative derivative at the lower and upper boundaries guarantees that our best-fit MZR is monotonic throughout the entire mass range and allows it to flatten, but not turn over, at high masses.
All stellar mass bins were retained in every Monte Carlo realization; we did not bootstrap-resample bins with replacement. The median of the ensemble of 20,000 MZRs was adopted as the fiducial MZR, while the 16th and 84th percentiles at each stellar mass were used to define the $\pm1\sigma$  confidence intervals on the MZR.

In Fig.~\ref{fig:mzr_full}, we present our measured binned MZR and uncertainties with black data points and error bars, and our best-fit MZR and its $\pm1\sigma$ confidence interval with the thick black curve and gray shaded region, respectively.
The best-fit parameters and associated uncertainties of Eq.~\ref{eq:mzr} to our binned MZR are $a=-0.06^{+0.06}_{-0.02}$, $b=0.31^{+0.05}_{-0.04}$, and $c=8.21^{+0.02}_{-0.02}$, which are given in Table~\ref{tab:mzr-params}.
Note that these results are identical to those in Nedkova et al. (in prep).

All colored curves in Fig.~\ref{fig:mzr_full} represent estimates of the MZR at different redshifts compiled from the literature.
We note that these measurements are based on a heterogeneous mix of methods. Some estimates are based on direct auroral lines \citep[e.g.,][]{AA26}, while others use strong line ratios calibrated to direct metallicities \citep[e.g.,][]{Revalski2024ApJ}. Furthermore, different studies adopt different functional forms (e.g., linear or broken power-law) to describe the binned MZR.
Most existing cosmic noon measurements are for intermediate to high-masses (${\logms}\gtrsim8$, with only a handful of studies pushing the MZR determination to lower masses. In the range of masses where we can compare to a large number of existing MZR estimates at $1.5 \lesssim z \lesssim 3.5$, our own estimates are fairly consistent with literature, in terms of both the slope and normalization of the MZR.

\begin{figure*}[t]
\vspace{-10pt}
\centering
\gridline{
\fig{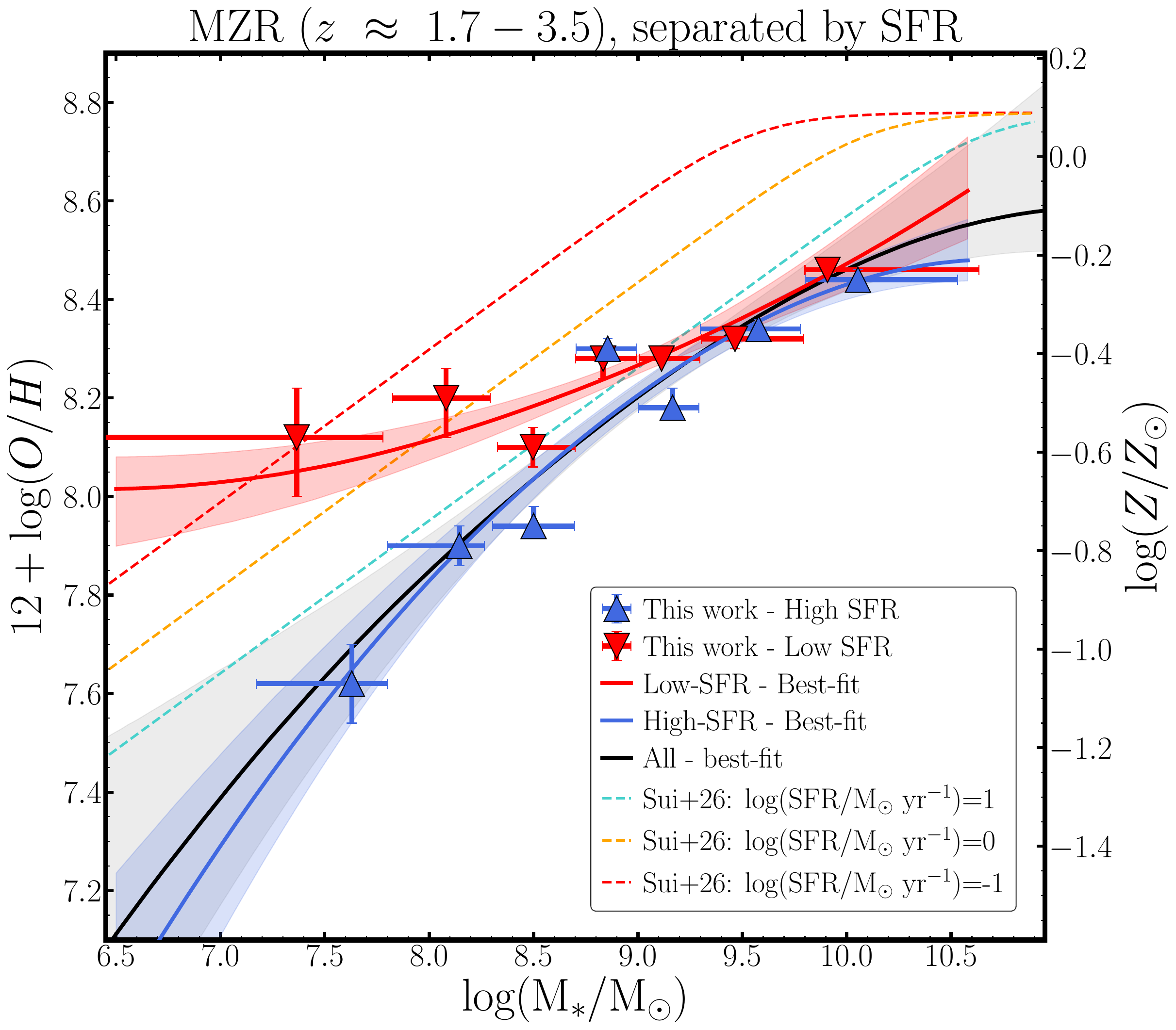}{0.525\textwidth}{}
\raisebox{5pt}{
\fig{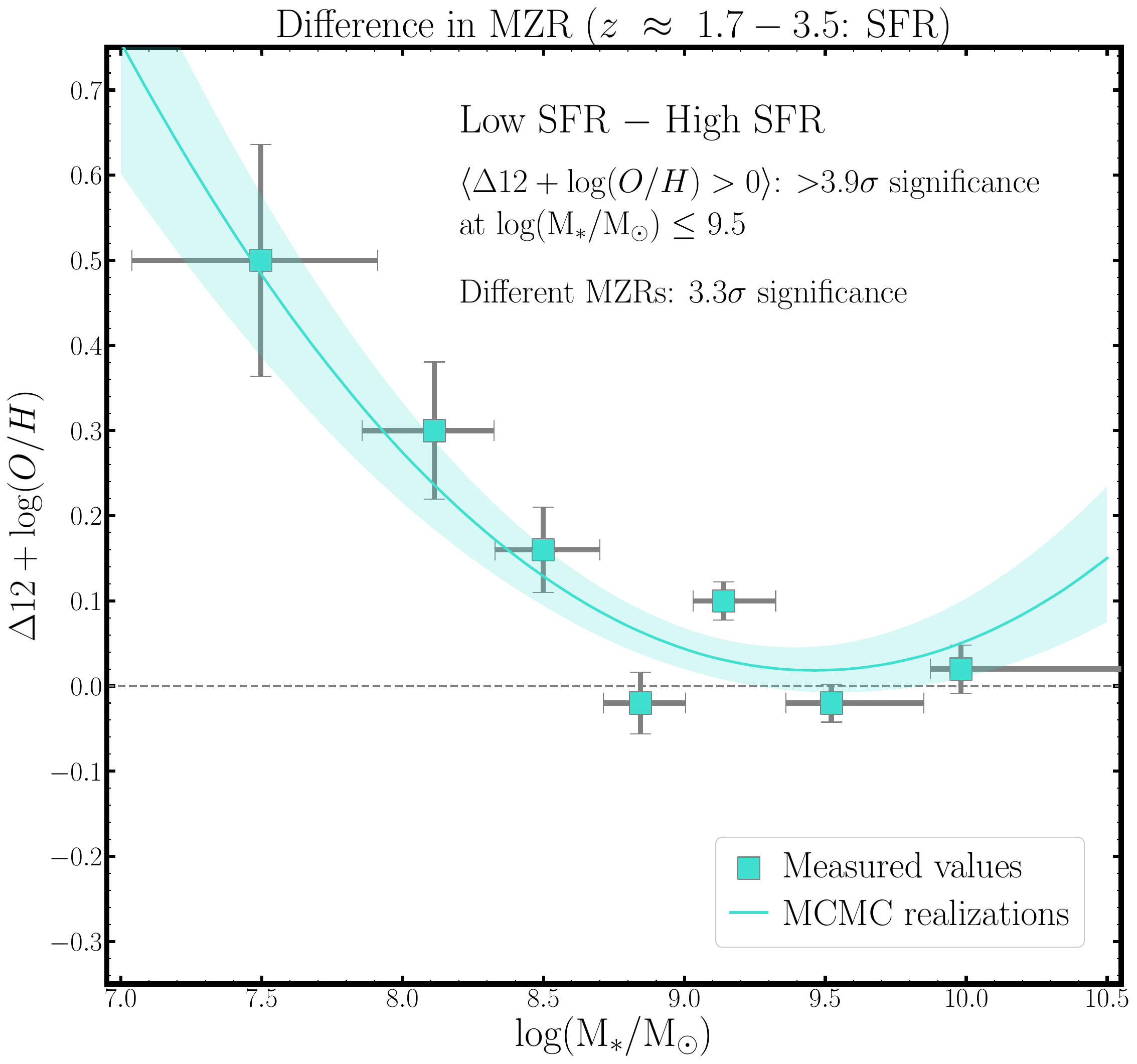}{0.47\textwidth}{}}
}
\vspace{-25pt}
\caption{
MZR separated by the SFR. 
{\it Left}: Our measured MZR for high-SFR (blue) and low-SFR (red) galaxies, as well as the MZR of the full population (black).
The data points and error bars represent the measured binned MZR and associated uncertainties of each population. 
The solid curves and shaded regions represent the best-fit MZR and $\pm1\sigma$ confidence interval of each population. 
The dashed curves show the SFR-dependence of the MZR in the local Universe \citep{Sui26}.
{\it Right}: The difference between the MZR of low and high SFR galaxies as a function of mass, also showing the statistical significance in the difference (see text).
We confirm, with high-confidence, that low-SFR galaxies are more metal-rich than high-SFR galaxies at ${\logms}\lesssim9.5$.
}
\label{fig:mzr-sfr}
\end{figure*}

At the less well-studied low-mass regime (${\logms}\lesssim8$), we find that the MZR drops fairly steeply. We find that on average, galaxies with ${\logms}\sim7.5$ have metallicity below 10\% solar. This is in agreement with the findings of \citet{Isobe26} based on JADES \citep{Eisenstein26} data with JWST/NIRSpec observations, but significantly lower than the estimates of \citet{He26} and \citet{Li23}, which are both based on NIRISS/WFSS observations. We note that these latter studies both have much smaller sample sizes than ours ($\lesssim$1 dex). We are statistically confident in our measurement in the lowest mass bin, which contains 76 galaxies in the stack.
For further discussion regarding the overall MZR, as well as any possible redshift evolution between $z\sim2$ to $z\sim3$ in our data, we refer the reader to Nedkova et al., (in prep).

\subsection{Dependence of MZR on SFR} \label{sec:mzr-sfr}

Next, we investigate the dependence of the MZR on the SFR, in other words, the FMR. 
We separate our sample into a high- and a low-SFR population in each mass bin and apply the same Monte Carlo framework for fitting the binned MZR to the quadratic form in Eq.~\ref{eq:mzr}.

For each mass bin, we first divide the sample into three quantiles ranked by SFR and stack the spectra of the upper third -- high-SFR -- and lower third -- low-SFR -- subsamples. We only consider the high and low SFR subsamples in our analysis, thereby ensuring that galaxies near the average star formation activity are not affecting our conclusions.
We remove extreme 3$\sigma$ outliers in SFR from each stack, as a way to ensure that a relatively small range of SFRs go into a given stack. We also experiment with the definition of high- and low-SFR subsamples by increasing and decreasing the percentage of galaxies going into the two stacks (e.g., changing to the top and bottom 40\% of the SFRs), finding that this does not affect our results.

We obtain two sets of stacked spectra for the high- and low-SFR populations.
Our approach is similar to those of \citet{Henry2021ApJ} and \citet{He26}.
Since we use just the top and bottom third of the SFRs, we 
now adopt seven bins so that we maintain at least 15 galaxies per stack.
Fewer bins would give increased statistics in each stack at the expense of a coarser ability to capture trends across mass. Nevertheless, our conclusions remain robust to any binning choice.
We apply our Bayesian framework to compute the median metallicity and $\pm1\sigma$ uncertainties in each stack and perform the same Monte Carlo realization procedure as above. 
Both populations are fit independently with separate best-fitting normalization, slope, and curvature.



The MZR separated by SFR is presented in the left panel of Figure~\ref{fig:mzr-sfr}. Both the measured MZRs in each bin (data points and error bars) and the best-fit relations and their $\pm1\sigma$ confidence intervals are shown. We represent the MZR of the low-SFR population in red and the high-SFR population in blue. 
The best-fit MZR and $\pm1\sigma$ confidence interval for the entire population obtained from these seven bins are represented by the black solid curve and shaded grey region. 
For context, we also show the SFR-dependence of the MZR in the local Universe based on DESI DR1 from 
\citet{Sui26},
with thin dashed curves showing three different SFR bins (red, orange, and turquoise).

Table~\ref{tab:mzr-sfr} presents the measurements in tabular form. Each row shows the number of galaxies, median, minimum, and maximum of stellar mass, median metallicity and $\pm1\sigma$ uncertainties, and median, minimum, and maximum {\logsfr} in a given stack.
Table~\ref{tab:mzr-params} presents best-fit parameters of both the low-SFR and high-SFR MZR.

\begin{deluxetable}{ccccccc}
\tabletypesize{\scriptsize}
\setlength{\tabcolsep}{1pt} 
\tablecaption{Measured values of the MZR separated by SFR}
\tablehead{
\colhead{Sample} & 
\colhead{$N_{\rm gal}$} & 
\colhead{$\log(M_\ast/M_\odot)$} & 
\colhead{\!$12+\log({\rm O/H})$\!} & 
\colhead{${\rm SFR}_{\rm med}$} & 
\colhead{${\rm SFR}_{\rm min}$} & 
\colhead{${\rm SFR}_{\rm max}$}
}
\startdata
Low SFR & 25 & $7.37^{+0.41}_{-1.11}$ & $8.12^{+0.10}_{-0.12}$ & -0.51 & -1.23 & -0.29 \\
High SFR & 26 & $7.63^{+0.17}_{-0.46}$ & $7.62^{+0.08}_{-0.08}$ & 0.28 & 0.05 & 0.63 \\
\tableline
Low SFR & 36 & $8.08^{+0.21}_{-0.26}$ & $8.20^{+0.06}_{-0.08}$ & -0.36 & -0.95 & -0.12 \\
High SFR & 37 & $8.14^{+0.12}_{-0.34}$ & $7.90^{+0.04}_{-0.04}$ & 0.53 & 0.27 & 1.15 \\
\tableline
Low SFR & 60 & $8.50^{+0.20}_{-0.17}$ & $8.10^{+0.04}_{-0.04}$ & -0.16 & -0.91 & 0.08 \\
High SFR & 61 & $8.50^{+0.20}_{-0.20}$ & $7.94^{+0.04}_{-0.02}$ & 0.71 & 0.42 & 1.36 \\
\tableline
Low SFR & 50 & $8.83^{+0.16}_{-0.13}$ & $8.28^{+0.02}_{-0.04}$ & -0.04 & -0.84 & 0.16 \\
High SFR & 52 & $8.86^{+0.14}_{-0.15}$ & $8.30^{+0.02}_{-0.02}$ & 0.73 & 0.53 & 1.47 \\
\tableline
Low SFR & 57 & $9.11^{+0.18}_{-0.11}$ & $8.28^{+0.02}_{-0.00}$ & 0.11 & -0.88 & 0.28 \\
High SFR & 58 & $9.17^{+0.12}_{-0.16}$ & $8.18^{+0.04}_{-0.00}$ & 0.99 & 0.68 & 1.62 \\
\tableline
Low SFR & 72 & $9.46^{+0.33}_{-0.16}$ & $8.32^{+0.02}_{-0.02}$ & 0.29 & -0.71 & 0.56 \\
High SFR & 72 & $9.58^{+0.20}_{-0.28}$ & $8.34^{+0.02}_{-0.00}$ & 1.19 & 0.91 & 1.75 \\
\tableline
Low SFR & 36 & $9.91^{+0.73}_{-0.11}$ & $8.46^{+0.02}_{-0.02}$ & 0.62 & -0.02 & 0.88 \\
High SFR & 35 & $10.05^{+0.48}_{-0.25}$ & $8.44^{+0.02}_{-0.02}$ & 1.46 & 1.31 & 1.86 \\ \enddata
\tablecomments{The third column presents the median, minimum and maximum stellar mass of each bin and fourth column presents the median metallicity and $\pm1\sigma$ uncertainties.
The last three columns show the median, minimum, and maximum values of {\logsfr} in each stack.
}
\label{tab:mzr-sfr}
\vspace{-20pt}
\end{deluxetable}

To rigorously assess the difference in the MZR of the low- and high-SFR populations, we quantify the difference between two MZRs using the metallicity offset at fixed stellar mass. 
The differential MZR was constructed directly from the Monte Carlo ensembles of the two independently fitted populations. For each of the 20,000 Monte Carlo realizations, the two best-fit MZRs were evaluated on the same stellar mass grid and subtracted point-by-point, producing an ensemble of $\Delta{\logoh} = [\logoh]_{\mathrm{Low~SFR}} - [\logoh]_{\mathrm{High~SFR}}$ curves. 
In this process, we propagate the uncertainties in both of the MZRs, including uncertainty in their normalization, slope, and curvature. In the right panel of Figure~\ref{fig:mzr-sfr}, we present the median of the $\Delta{\logoh}$ realizations with a solid turquoise curve and the 16th and 84th percentile of the realizations with the shaded turquoise region. The data points and error bars show the raw measured values of $\Delta{\logoh}$ in each mass bin.

The two MZRs appear almost indistinguishable at ${\logms}\gtrsim9.5$, but start to deviate from each other at lower masses. 
To quantify the difference in the MZR of high- and low-SFR galaxies, we performed two different statistical tests. First, we quantified the significance of an offset in the MZRs by calculating a metallicity offset $\Delta{\logoh}$ for each Monte Carlo realization integrated over a given mass interval. We used the ensemble of these offsets to determine the median offset in the MZRs, $\langle\Delta{\logoh}\rangle$, its $\pm1\sigma$ uncertainty, 
and the probability that $\langle\Delta{\logoh}\rangle>0$ across this mass interval.
We estimated a two-sided $p$-value by considering the fractions of Monte Carlo realizations with positive and negative $\Delta{\logoh}$,
and then converted this to an equivalent Gaussian significance in numbers of $\sigma$. This test measures how consistently one MZR lies above or below the other over the selected mass interval.

As a complementary test, we compared a model in which both galaxy populations share a single common MZR with a model in which the high- and low-SFR populations have their own separate MZRs. We estimated the $\chi^2$ goodness-of-fit statistic for both of these models and quantified their difference, $\Delta\chi^2 = \chi^2_{\rm common ~MZR} - \chi^2_{\rm separate~MZR}$. 
We then generated synthetic datasets assuming that both populations follow the same underlying MZR, refitted both models to each realization, and measured how often a $\Delta\chi^2$ as large as the observed value occurred by chance. 
This 
addresses the question: does the two populations having fully independent MZRs provide a statistically significant improvement over a common MZR?

From test 1, we find that the MZR of low-SFR galaxies is higher than that of high-SFR galaxies at ${\logms}\leq9.5$ at $>3.9\sigma$ significance, which we quote in Fig.~\ref{fig:mzr-sfr}. In this mass range, we find that the integrated median metallicity offset is $\langle \Delta{\logoh}\rangle=0.26^{+0.05}_{-0.05}$ dex. 
The median offset rises from $\Delta{\logoh}\lesssim0.1$ to $\Delta{\logoh}\gtrsim0.5$ between ${\logms}\sim9$ and ${\logms}\sim7$.
The difference in the MZRs is not significant at the $\geq3\sigma$ level at ${\logms}>9.5$.
Test 2 reveals that $\Delta\chi^2\approx18$, and that at 3.3$\sigma$ significance, two independent MZRs provide a statistically better description of the data than one common MZR for high- and low-SFR galaxies.

Finally, we repeat our experiments by considering only galaxies within a higher redshift window, $2.5\leq z \leq 3.45$. We find the same qualitative result as with our full redshift range, that low-SFR galaxies are more metal-rich than high-SFR galaxies of similar mass at ${\logms}\lesssim9.5$.

Overall, the results in this section indicate that, to a high degree of confidence, the MZR has a significant dependence on the SFR at redshifts $z\sim1.7-3.5$, such that at low and intermediate masses, galaxies with lower SFRs  are more metal-rich than more star-forming galaxies.

\subsection{Dependence of MZR on Morphological Properties}
\label{sec:mzr-morph}


Next, we apply the same methodology as \S~\ref{sec:mzr-sfr} to investigate the dependence of the MZR on morphological properties. 
For each mass bin, we separate galaxies into the largest and smallest third of any given property, e.g., half-light radius, and stack them. We then characterize the MZRs of the different subsamples.

\subsubsection{Size}
\label{sec:mzr-size}

\begin{figure*}[htbp]
\vspace{-10pt}
\centering
\gridline{
\fig{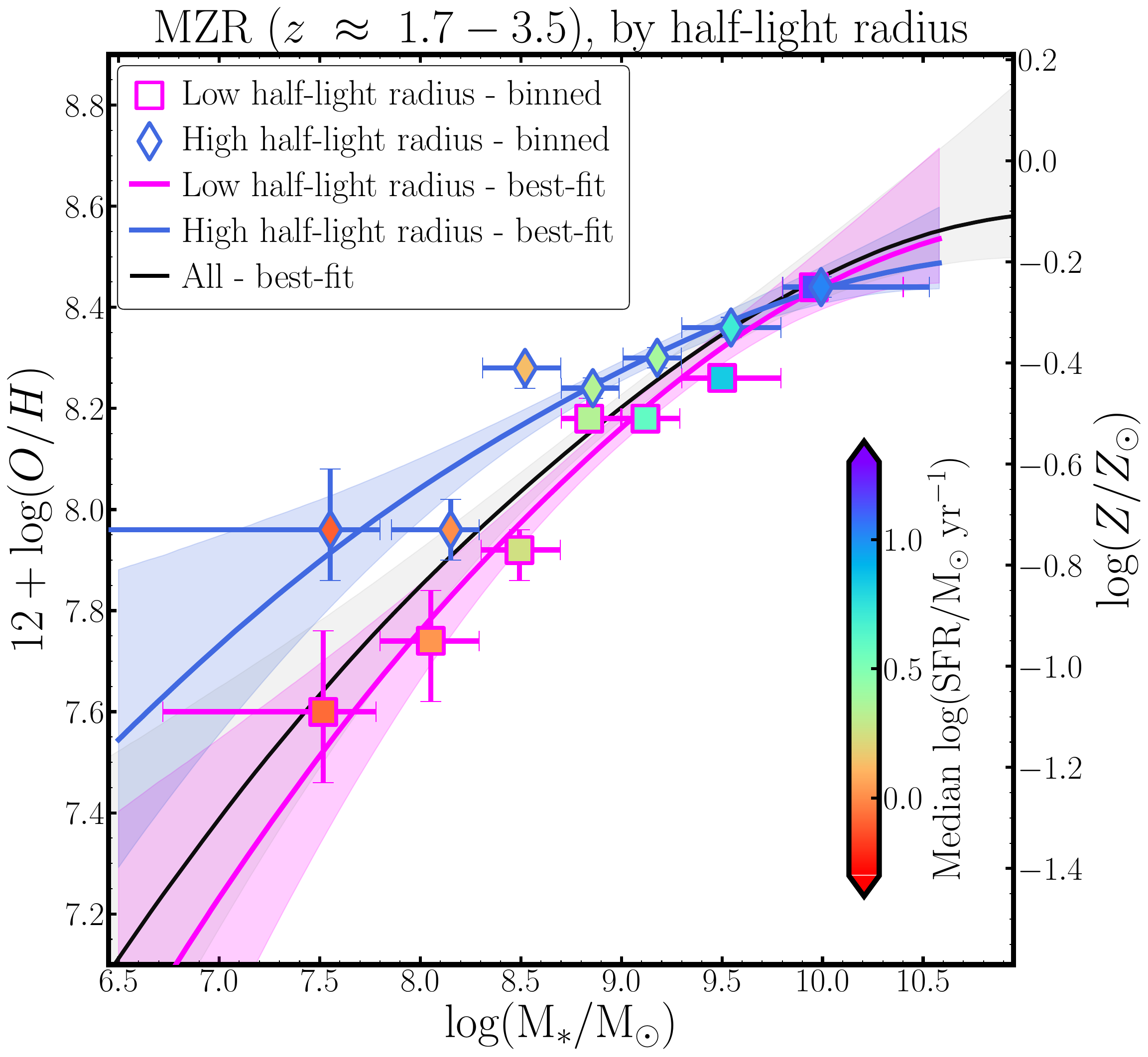}{0.515\textwidth}{}
\raisebox{7pt}{
\fig{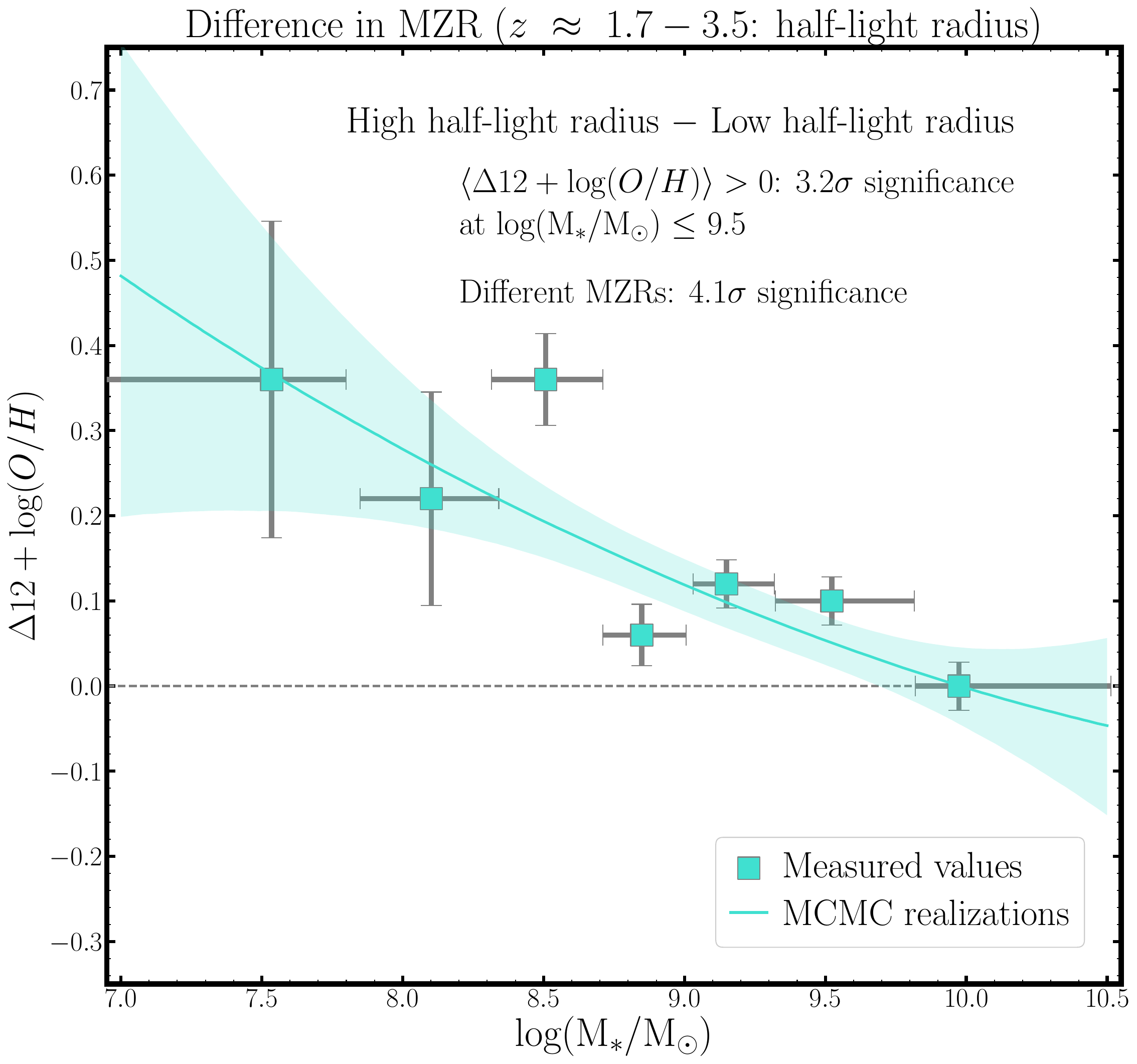}{0.48\textwidth}{}}
}
\vspace{-25pt}
\caption{
MZR divided by size. 
{\it Left}: our measured MZR for large (blue curve and error-bars and diamond markers) and small (magenta curve and error-bar and square markers) galaxies, as well as the MZR of the full population (black curve). Each individual marker is colored by the median of the SFR in that stack.
{\it Right}: the difference between the MZR of large and small galaxies as a function of mass.
At fixed mass, larger galaxies are more metal-rich than smaller galaxies at ${\logms}\lesssim9.5$.  
}
\label{fig:mzr-rh}
\end{figure*}

Similar to Fig.~\ref{fig:mzr-sfr},  Fig.~\ref{fig:mzr-rh} presents the MZR of galaxies separated by size (left panel) and the MZR offset between the large and small galaxies as a function of mass (right panel). 
In the left panel, the best-fit MZR of higher and lower half-light radius galaxies are represented by  blue and magenta curves, respectively. 
The binned measurements are represented by diamond markers and blue error-bars (large galaxies) and square markers and magenta error-bars (small galaxies). 
Each individual marker is also colored by the median SFR of galaxies in that particular stack (with the color-bar shown on the right side of the panel) in order to show the dependence of star formation activity on the trends.
As we have already seen SFR to have a secondary effect on the MZR, coloring binned data points by median SFR helps us obtain a more holistic picture of the secondary dependencies of the MZR.
Table~\ref{tab:mzr-rh} presents the measurements in tabular form (similar to Table~\ref{tab:mzr-sfr}), 
while the best-fit parameters of both the low- and high-half light radius subsamples are presented in Table~\ref{tab:mzr-params}.

At ${\logms}\lesssim9.5$, the two best-fit MZRs begin to diverge. In the regime ${\logms}\leq9.5$, the metallicity of larger galaxies is higher than that of smaller galaxies at a 3.2$\sigma$ significance level, with an integrated median metallicity offset $\langle \Delta{\logoh}\rangle=0.25^{+0.08}_{-0.08}$ dex.
The median offset rises from $\Delta{\logoh}\lesssim0.1$ at ${\logms}\sim9$ to $\Delta{\logoh}\gtrsim0.4$ at ${\logms}\sim7.5$.

We also find that two independent MZRs provide a statistically improved description of the data than one common MZR for large and small galaxies. This separate MZR model is preferred at a $4.1\sigma$ significance level.
These results suggest that extended and compact galaxies have separate MZRs at cosmic noon, with the former being more metal-rich than that latter at low masses.

At virtually all masses, the larger galaxies in the mass bin have somewhat smaller SFRs ($\sim$0.1 dex) than the smaller galaxies. This trend, also shown in Fig.~\ref{fig:mass-size}, suggests that recent star formation activity and size are interconnected phenomena and that we must consider both to fully capture chemical enrichment mechanisms. This is further discussed in \S~\ref{sec:discuss-morph}.

\begin{deluxetable}{ccccccc}
\tabletypesize{\footnotesize}
\setlength{\tabcolsep}{2pt} 
\tablecaption{Measured values of the MZR separated by size}
\tablehead{
\colhead{Sample} & \colhead{$N_{\rm gal}$} & \colhead{$\log(M_\ast/M_\odot)$} & \colhead{$12+\log({\rm O/H})$} & \colhead{${\rh}_{\rm ,med}$} & \colhead{${\rh}_{\rm ,min}$} & \colhead{${\rh}_{\rm ,max}$}
}
\startdata
Small & 16 & $7.52^{+0.26}_{-0.80}$ & $7.60^{+0.16}_{-0.14}$ & 0.40 & 0.23 & 0.49 \\
Large & 28 & $7.55^{+0.25}_{-1.30}$ & $7.96^{+0.12}_{-0.10}$ & 0.92 & 0.69 & 1.41 \\
\tableline
Small & 32 & $8.05^{+0.24}_{-0.25}$ & $7.74^{+0.10}_{-0.12}$ & 0.55 & 0.28 & 0.69 \\
Large & 33 & $8.15^{+0.14}_{-0.29}$ & $7.96^{+0.06}_{-0.06}$ & 1.09 & 0.91 & 1.76 \\
\tableline
Small & 55 & $8.49^{+0.20}_{-0.19}$ & $7.92^{+0.04}_{-0.06}$ & 0.70 & 0.41 & 0.83 \\
Large & 51 & $8.52^{+0.18}_{-0.21}$ & $8.28^{+0.00}_{-0.04}$ & 1.44 & 1.17 & 2.15 \\
\tableline
Small & 50 & $8.84^{+0.16}_{-0.14}$ & $8.18^{+0.04}_{-0.02}$ & 0.91 & 0.33 & 1.09 \\
Large & 43 & $8.86^{+0.13}_{-0.16}$ & $8.24^{+0.02}_{-0.02}$ & 1.99 & 1.60 & 2.91 \\
\tableline
Small & 54 & $9.12^{+0.17}_{-0.12}$ & $8.18^{+0.02}_{-0.02}$ & 1.01 & 0.60 & 1.26 \\
Large & 52 & $9.18^{+0.12}_{-0.17}$ & $8.30^{+0.02}_{-0.02}$ & 2.10 & 1.80 & 3.17 \\
\tableline
Small & 70 & $9.50^{+0.29}_{-0.20}$ & $8.26^{+0.02}_{-0.02}$ & 1.12 & 0.48 & 1.37 \\
Large & 71 & $9.54^{+0.25}_{-0.24}$ & $8.36^{+0.02}_{-0.02}$ & 2.49 & 2.02 & 3.84 \\
\tableline
Small & 36 & $9.96^{+0.45}_{-0.16}$ & $8.44^{+0.02}_{-0.02}$ & 1.26 & 0.54 & 1.72 \\
Large & 34 & $9.99^{+0.54}_{-0.19}$ & $8.44^{+0.02}_{-0.02}$ & 3.31 & 2.74 & 4.69 \\
\enddata
\tablecomments{
Similar to Table~\ref{tab:mzr-sfr}.
The last three columns show the median, minimum, and maximum values of {\rh} in kpc in each stack.
}
\label{tab:mzr-rh}
\vspace{-15pt}
\end{deluxetable}


\begin{figure*}[htbp]
\vspace{-10pt}
\centering
\gridline{
\fig{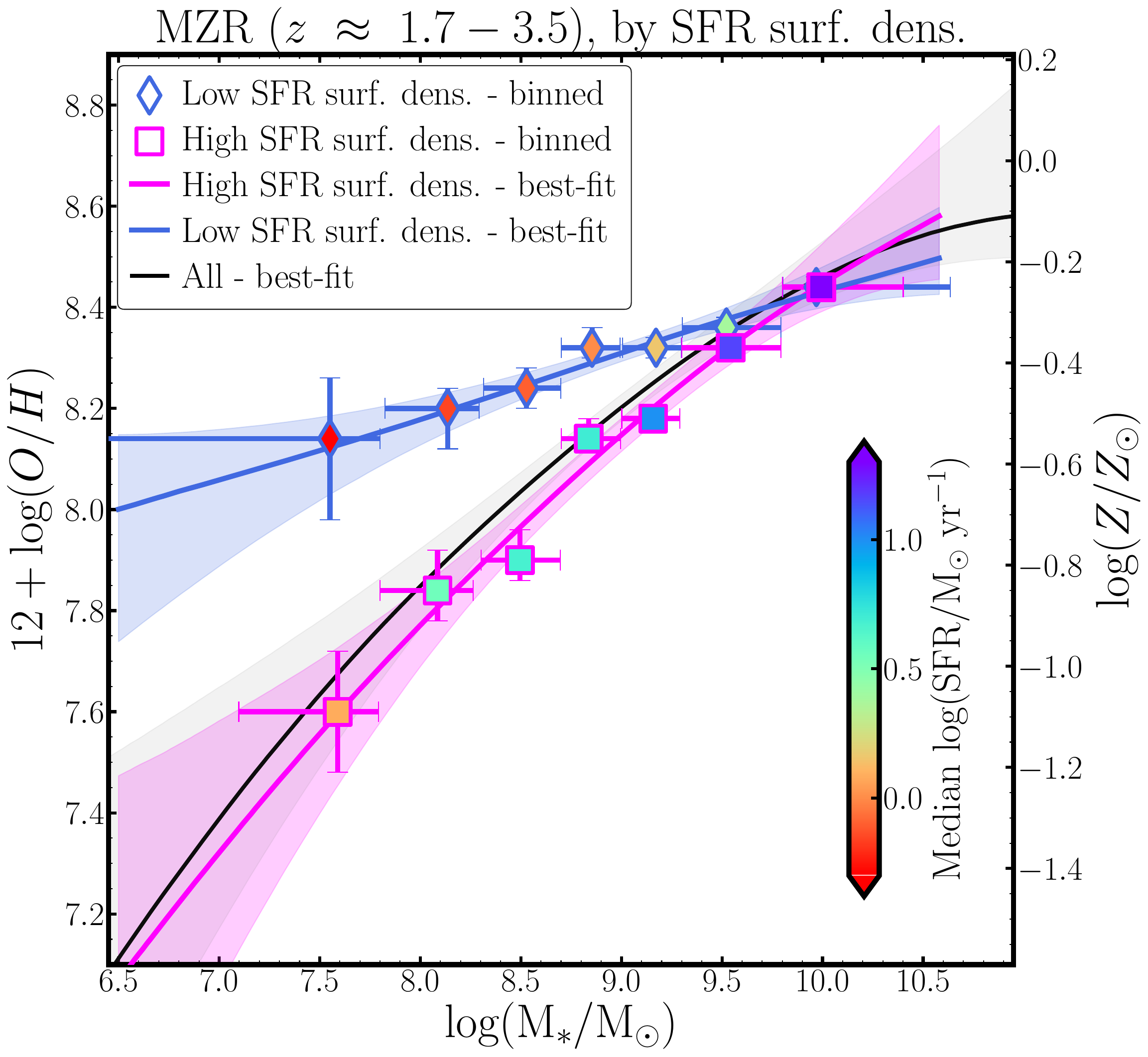}{0.515\textwidth}{}
\raisebox{7pt}{
\fig{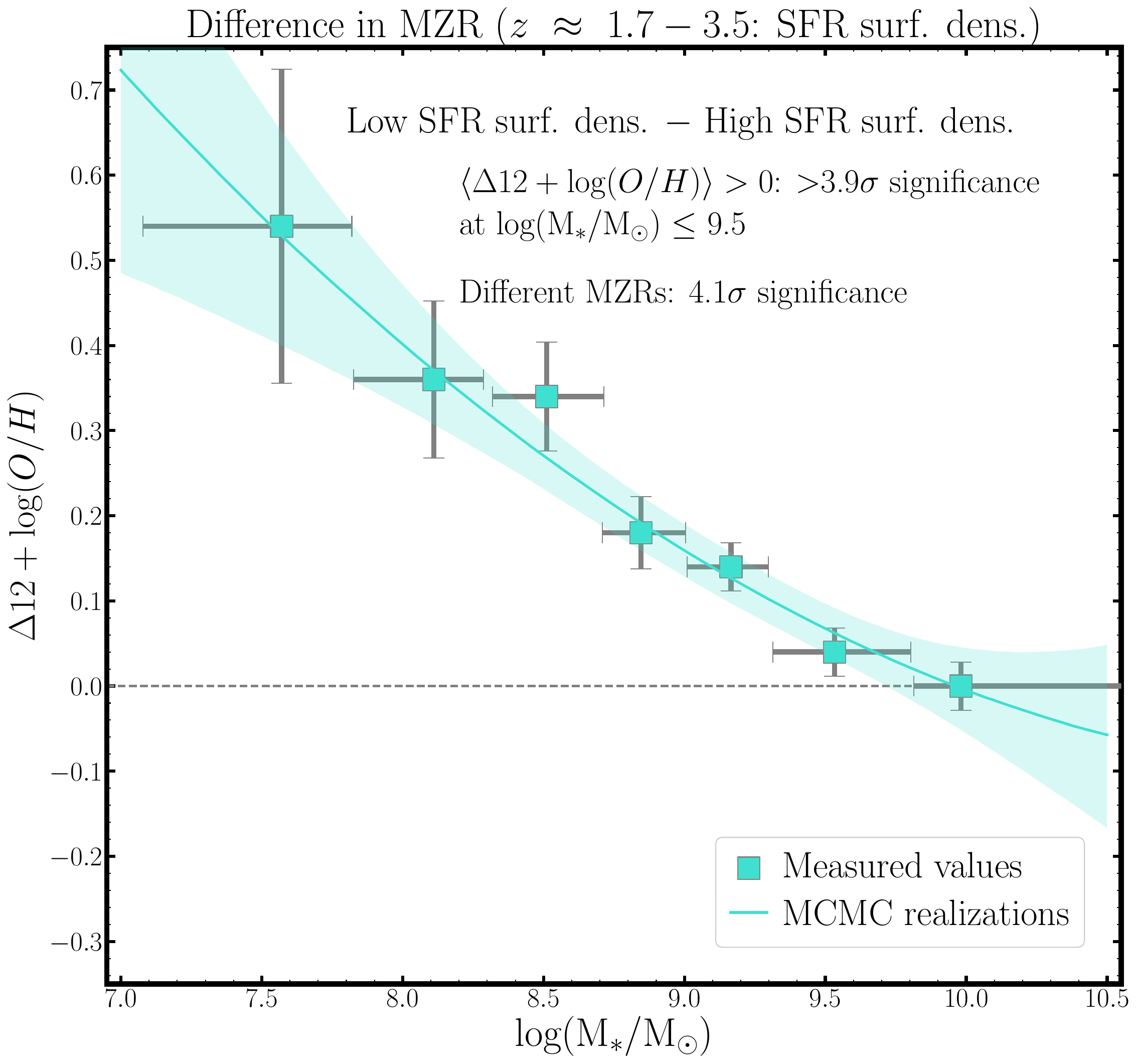}{0.48\textwidth}{}}
}
\vspace{-25pt}
\caption{
Similar to Figure~\ref{fig:mzr-rh}, but for SFR surface density. 
At fixed mass, galaxies with less concentrated SFR surface density are more metal-rich than those with more highly-concentrated SFR surface density at ${\logms}\lesssim9.5$.  
}
\label{fig:mzr-sfrdens}
\end{figure*}

\subsubsection{Morphological broadening and other observational biases}

In slitless spectroscopy, the galaxy itself effectively defines the spectral aperture:
the spatial light distribution projected along the dispersion direction is convolved with the intrinsic spectrum, such that spatially extended sources produce broader observed emission lines than compact sources at fixed intrinsic emission line width. This ``morphological broadening'' may lower the effective resolving power for extended galaxies relative to compact galaxies \citep[e.g.,][]{Kummel09,Momcheva16,Chen25_RAA}.

To assess whether morphological broadening of the NIRISS/WFSS line-spread function (LSF) introduces systematic biases in metallicities measured from stacked spectra, we construct a suite of synthetic spectra and passed them through our measurement pipeline. First, we verify the {\OIIIf} line width is proportional to the median observed half-light radius, {\rh} (in arcseconds), in our stacks of {\mstar} and {\rh} (see, e.g., Fig.~\ref{fig:morph-stacks}).
Based on the range of observed median metallicities (${\logoh}=7.6-8.4$) and rest-frame LSF width ($\sigma_{\rm{LSF}}=6-15$~{\AA}), we generate 100 independent realizations of a mock stacked spectrum using our emission-line forward model for each combination of {\logoh} and $\sigma_{\rm{LSF}}$. 
The line ratios are self-consistently derived from the  \citet{Cataldi25} strong-line calibrations at the input metallicity, and fluxes normalized such that the {\OIIIf}~$\lambda5007$ flux equals unity, as we do in our real stacked spectra. The line profiles share a common Gaussian width set by $\sigma_{\rm{LSF}}$. Gaussian noise was added in each realization at a similar level as our real stacked spectra ($\sim$0.1\% per pixel).

We measure the metallicity in these synthetic spectra using our Bayesian approach and compare the difference between the recovered and true metallicities, the metallicity ``bias''.
We quantify the degree to which line flux ratios bias the metallicity, via the difference between true and recovered ratios R$_2$, R$_3$, and O$_{32}$. 
We recover metallicities with a small bias that is {\it independent of} $\sigma_{\rm{LSF}}$ for ${\logoh}\geq8$. At this metallicity regime, none of the line ratios R$_2$, R$_3$, or O$_{32}$ are biased by $\sigma_{\rm{LSF}}$ either. However, at ${\logoh}\leq7.8$, we find a slight negative bias that increases with $\sigma_{\rm{LSF}}$, such that higher $\sigma_{\rm{LSF}}$ leads to an underestimation of metallicity by up to $\sim0.1$ dex. This is caused by a small overestimation of O$_{32}$ and underestimation of R$_2$ at large $\sigma_{\rm{LSF}}$,
due to a blending of {\OIIwave} with the continuum break.


Therefore, we can conclude that the observed trend of larger galaxies being metal-rich relative to smaller galaxies at fixed mass is not driven by morphological broadening at all but the lowest metallicities. At ${\logoh}\leq7.8$, instead, the median metallicity of larger galaxies (which have broader LSFs) is likely to be underestimated and should be treated with caution. This may actually cause the metallicity of the high-{\rh} galaxies in our lowest-mass bin to be higher in reality than what we show in Fig.~\ref{fig:mzr-rh}. 
Our findings are consistent with \citet{He24} who showed that despite significant flux reduction from morphological broadening in grism spectra, the flux ratios and metallicities do not change significantly.

We also investigate whether redshift evolution impacts our results on the dependence of the MZR on size. The average half-light radius of star-forming galaxies is known to decrease by $\approx$2-3 times between $z\sim1.7$ to 3.5 \citep[e.g.,][]{vanderwel14,Yang25}. In light of this, we repeat our MZR measurements separated by {\rh} in a narrower redshift bin of $z=1.74-2.3$. We find no statistical difference in the MZR of larger and smaller galaxies with respect to the case where we included all galaxies at $z=1.74-3.45$. Thus, our finding that larger galaxies are more metal-rich than smaller galaxies at fixed mass is not driven by systematically higher-$z$ galaxies in the larger size stacks at a given mass.

In addition, we consider the fact that we measure galaxy sizes in either rest frame-UV ($\sim$2500~{\AA} at our median $z$) or rest-frame optical ($\sim$5000~{\AA}), depending on the field. While there is a known systematic variation in sizes measured between rest-UV and rest-optical bands, this is typically only seen at higher masses -- ${\logms}\gtrsim9.5-10$ \citep[e.g.,][]{Nedkova24,Yang25,McGrath26}. Since we only see a significant difference in the MZR by size at ${\logms}<9.5$, the band-to-band variation in measured sizes are not likely to explain the difference. 

\subsubsection{SFR surface density}
\label{sec:mzr-sfrdens}

We measure the SFR surface density, $\mathrm{SFR}/(\pi r_h^2)$, 
and repeat our MZR measurements. 
Similar to Fig.~\ref{fig:mzr-rh}, we present the MZRs separated by SFR surface density in Fig.~\ref{fig:mzr-sfrdens}, along with the difference in the two MZRs. 
The best-fit parameters of the low- and high-SFR surface density MZRs are given in Table~\ref{tab:mzr-params}.
As with SFR and size, the MZRs of the low and high SFR density populations diverge at ${\logms}\lesssim9.5$. However, the difference in the MZR of the two populations is far more pronounced than with size or SFR at intermediate and low masses.

\begin{figure*}[htbp]
\vspace{-10pt}
\centering
\gridline{
\fig{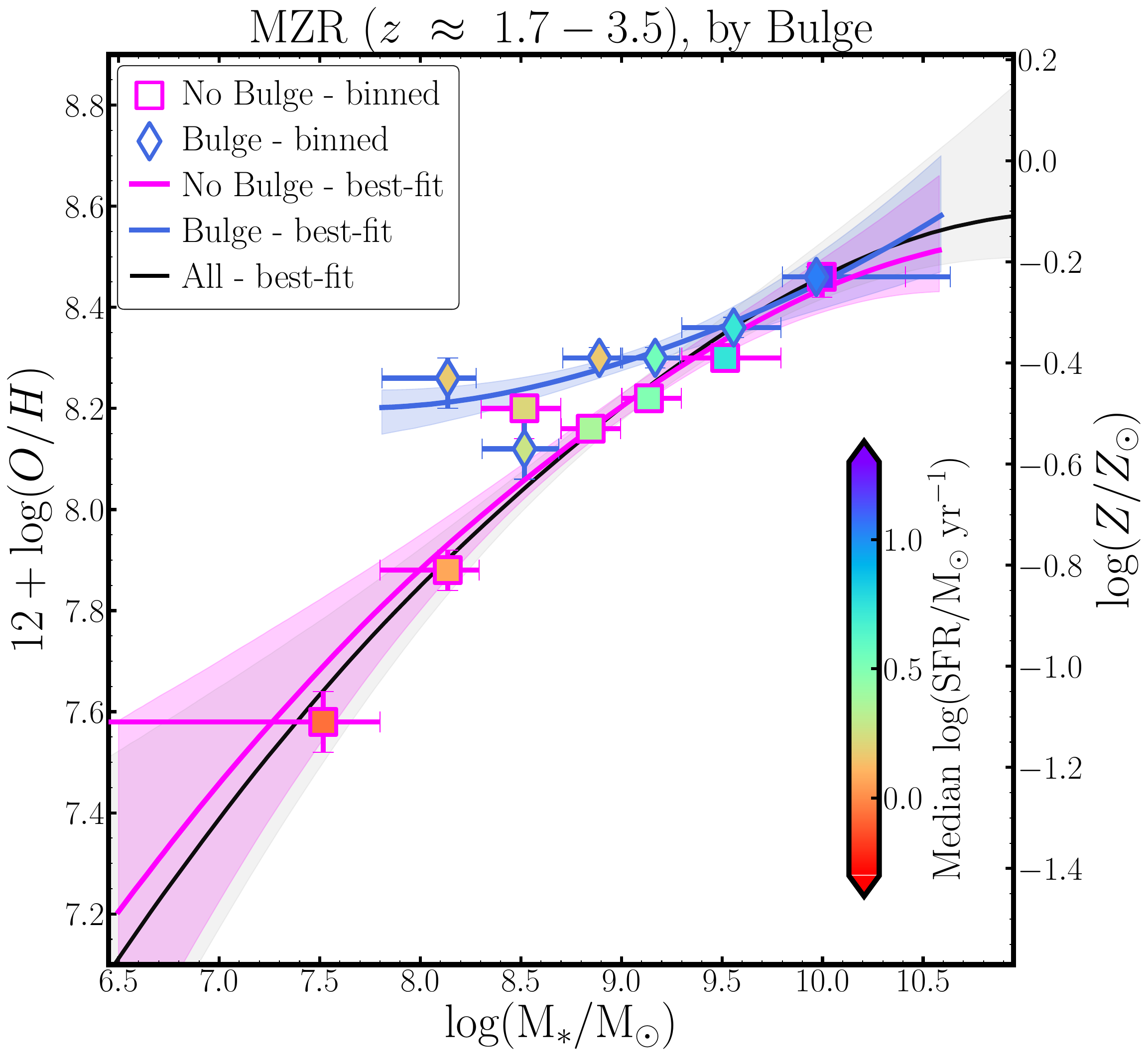}{0.45\textwidth}{}
\raisebox{7pt}{
\fig{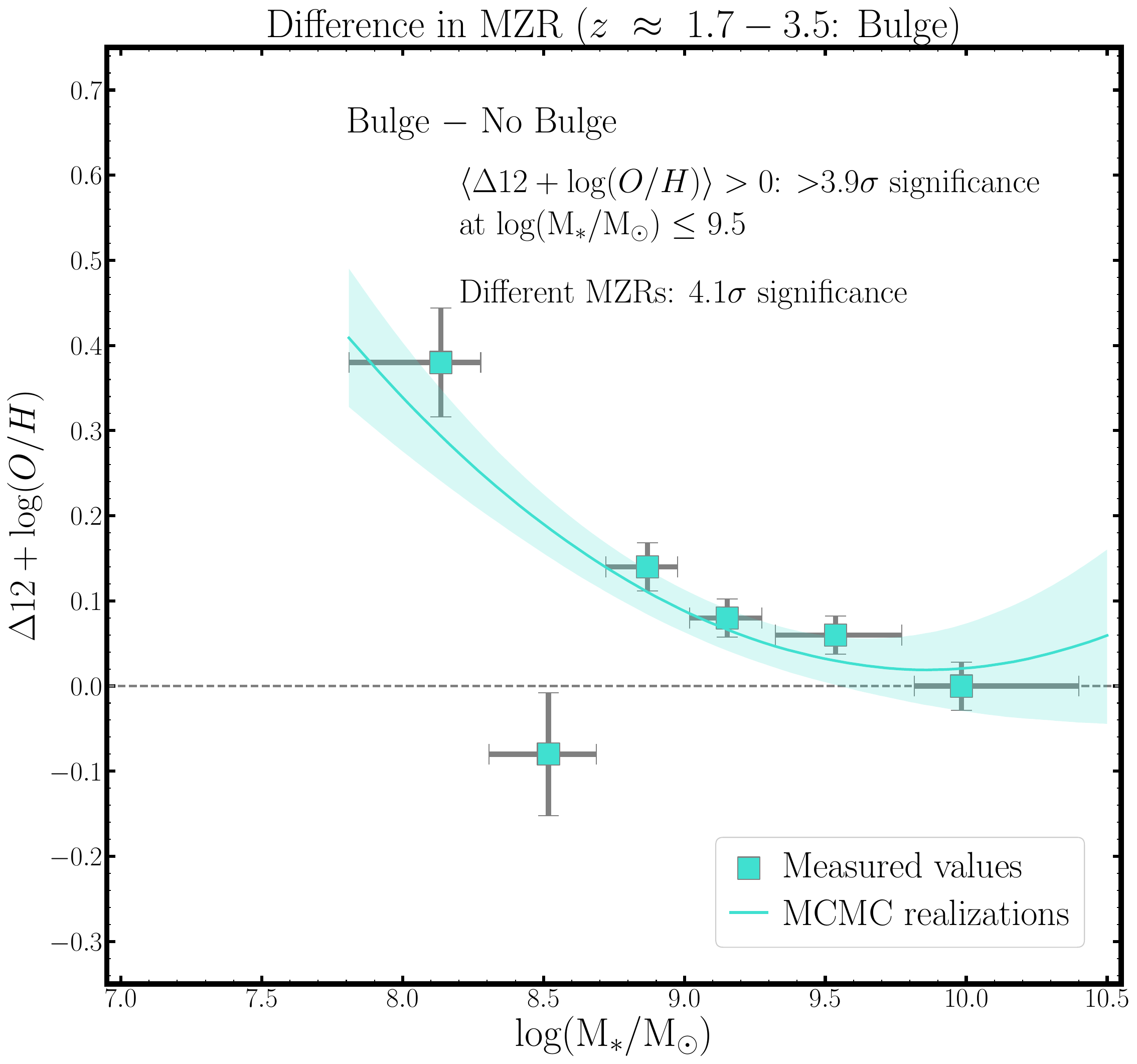}{0.42\textwidth}{}
}
}
\vspace{-25pt}
\gridline{
\fig{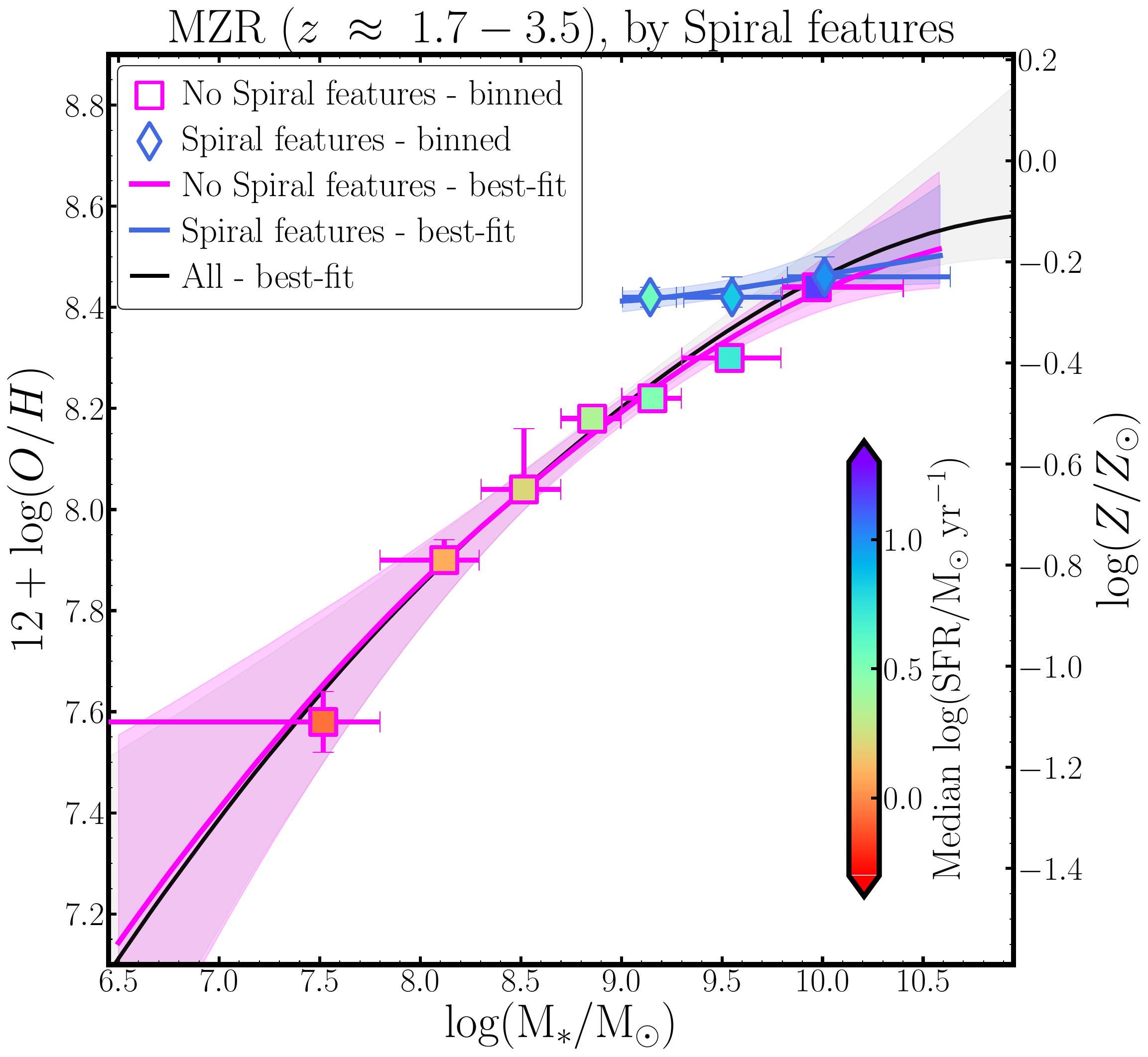}{0.45\textwidth}{}
\raisebox{7pt}{
\fig{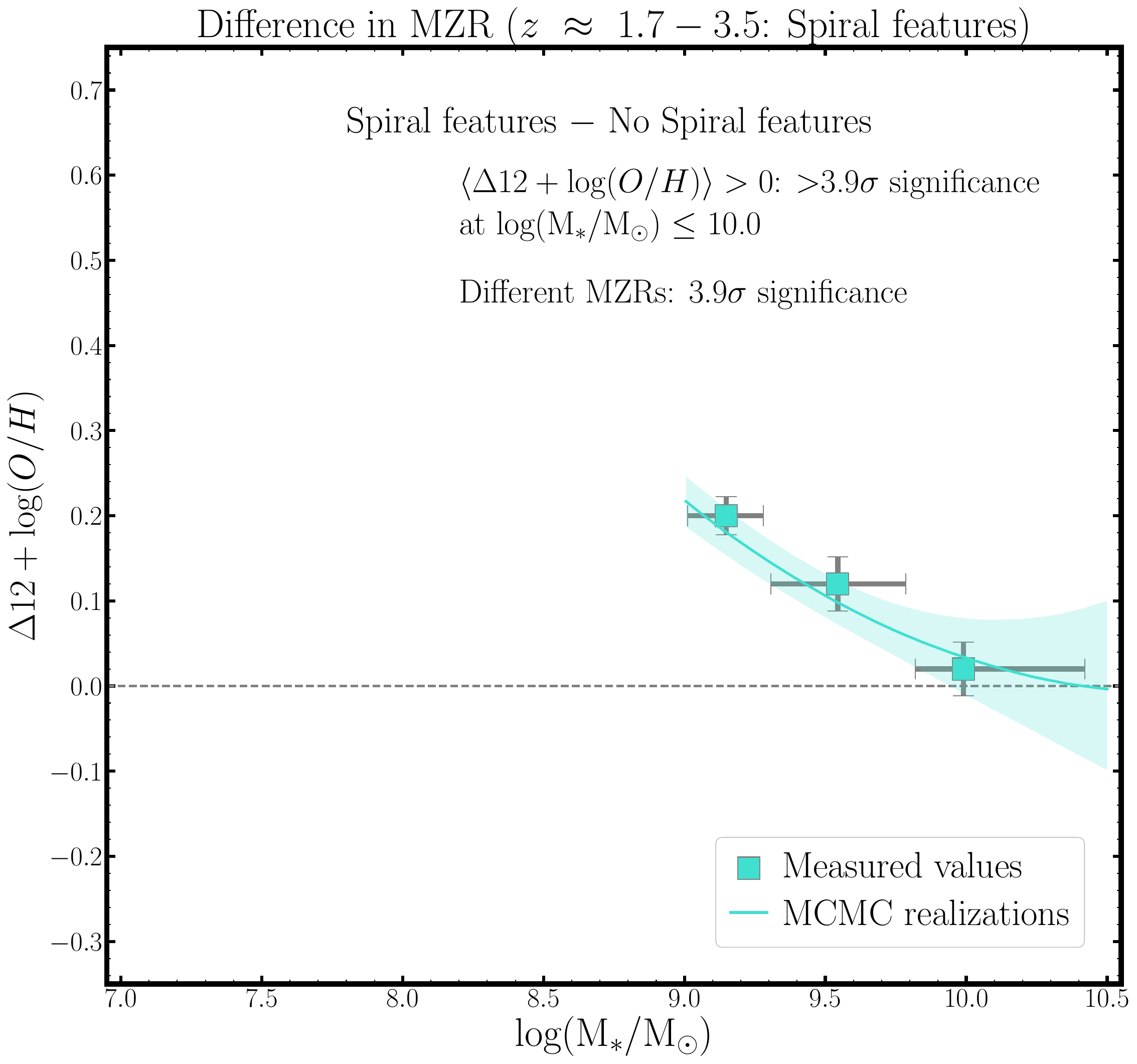}{0.42\textwidth}{}
}
}
\vspace{-25pt}
\caption{
Similar to Figure~\ref{fig:mzr-rh}, but for the presence of a bulge (top) and spiral features (bottom). At ${\logms}\sim8-9.5$ (${\logms}\sim9-10$), a central bulge (spiral features) are typically associated with higher metal-enrichment.
}
\label{fig:mzr-blg-spr}
\end{figure*}

At ${\logms}\leq9.5$, low-SFR density galaxies are more metal-rich than high-SFR density galaxies at a $>3.9\sigma$ significance, with an integrated median offset $\langle \Delta{\logoh}\rangle=0.35^{+0.06}_{-0.07}$ dex. The offset rises smoothly from $\Delta{\logoh}\lesssim0.05$ at ${\logms}\sim9.5$ to $\Delta{\logoh}\gtrsim0.7$ at ${\logms}\sim7$.
A separate MZR model is preferred to the common MZR model at 4.1$\sigma$ significance. Overall, the MZR depends on the SFR surface density with a very strong statistical significance.
The large magnitude of this difference is likely a composite effect of both the SFR and the half-light radius, which each have a statistically significant effect on the MZR. 
Fig.~\ref{fig:mzr-sfrdens} also underscores the importance of the SFR in determining the enrichment levels of our galaxies.

\subsubsection{Bulge and Spiral Features}
\label{sec:mzr-blg-spr}

We also see statistically significant dependence of the MZR on the presence of a central bulge and spiral features, albeit over smaller mass ranges.
The top row of Fig.~\ref{fig:mzr-blg-spr} presents the overall MZR and MZR offsets for galaxies with (blue) and without (magenta) a bulge. We find that galaxies with a bulge have a higher metallicity than those without one at ${\logms}\sim8-9.5$, with a $>3.9\sigma$ significance. The offset rises to $\Delta{\logoh}\approx0.4$ at ${\logms}\sim8$. 
However, galaxies with a bulge are only seen at ${\logms}>7.8$.

The bottom row shows Fig.~\ref{fig:mzr-blg-spr} shows the MZR separated by galaxies with and without spiral features. With the caveat that spiral features are only seen in the three highest-mass bins, ${\logms}\geq9$, the MZRs deviate significantly at $9\leq{\logms}\leq10$. The spiral galaxy MZR is higher than the non-spiral galaxy MZR at 3.9$\sigma$ significance. At ${\logms}\sim9$, spiral galaxies are more metal-rich than non-spiral galaxies by $\Delta{\logoh}\sim0.2$~dex.
Table~\ref{tab:mzr-params} presents best-fit parameters of the MZR for bulge, non-bulge, spiral, and non-spiral populations.


\subsubsection{Other morphological parameters}
\label{sec:mzr-other}

\begin{figure*}[htbp]
\vspace{-10pt}
\centering
\gridline{
\fig{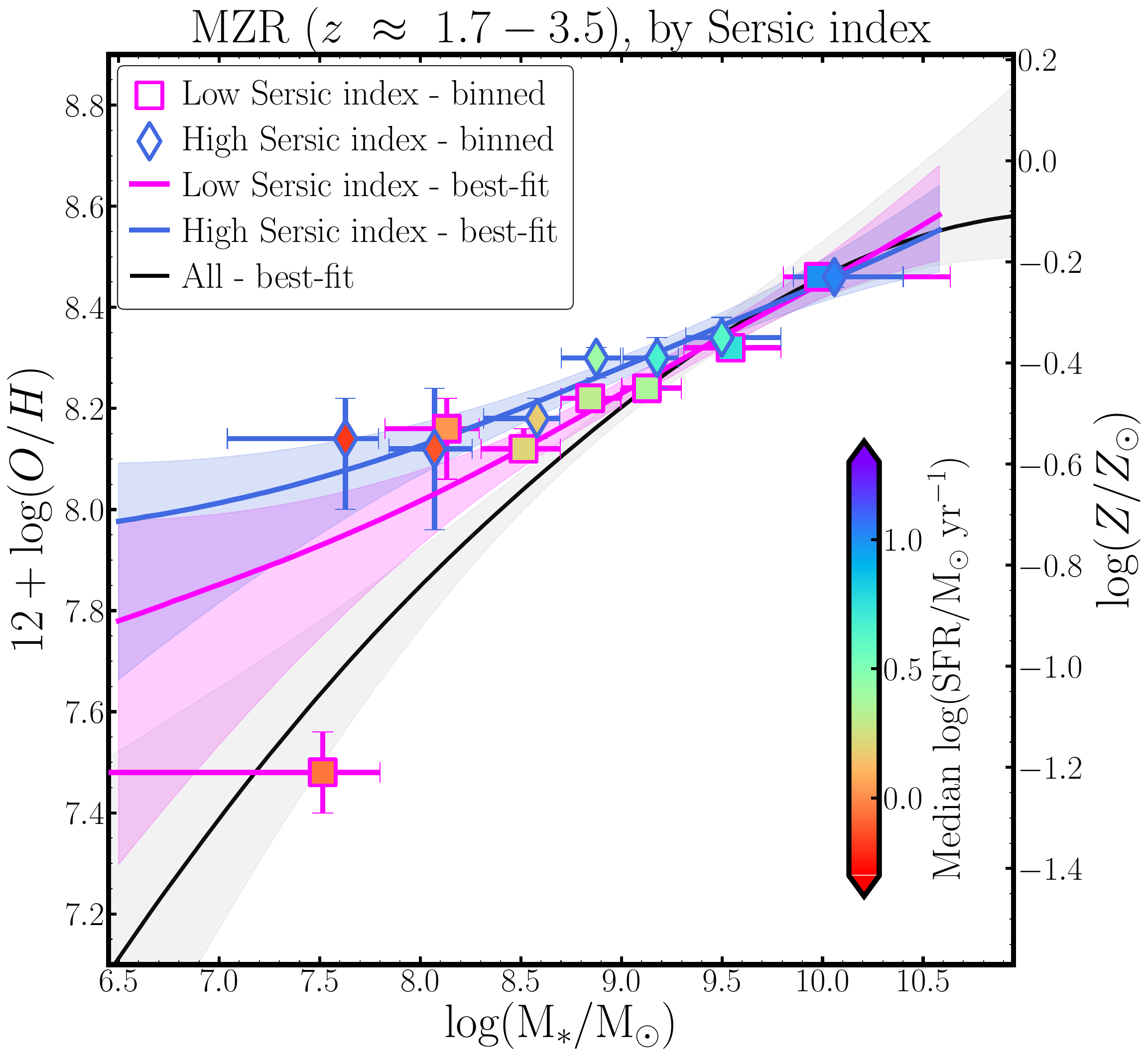}{0.33\textwidth}{}
\fig{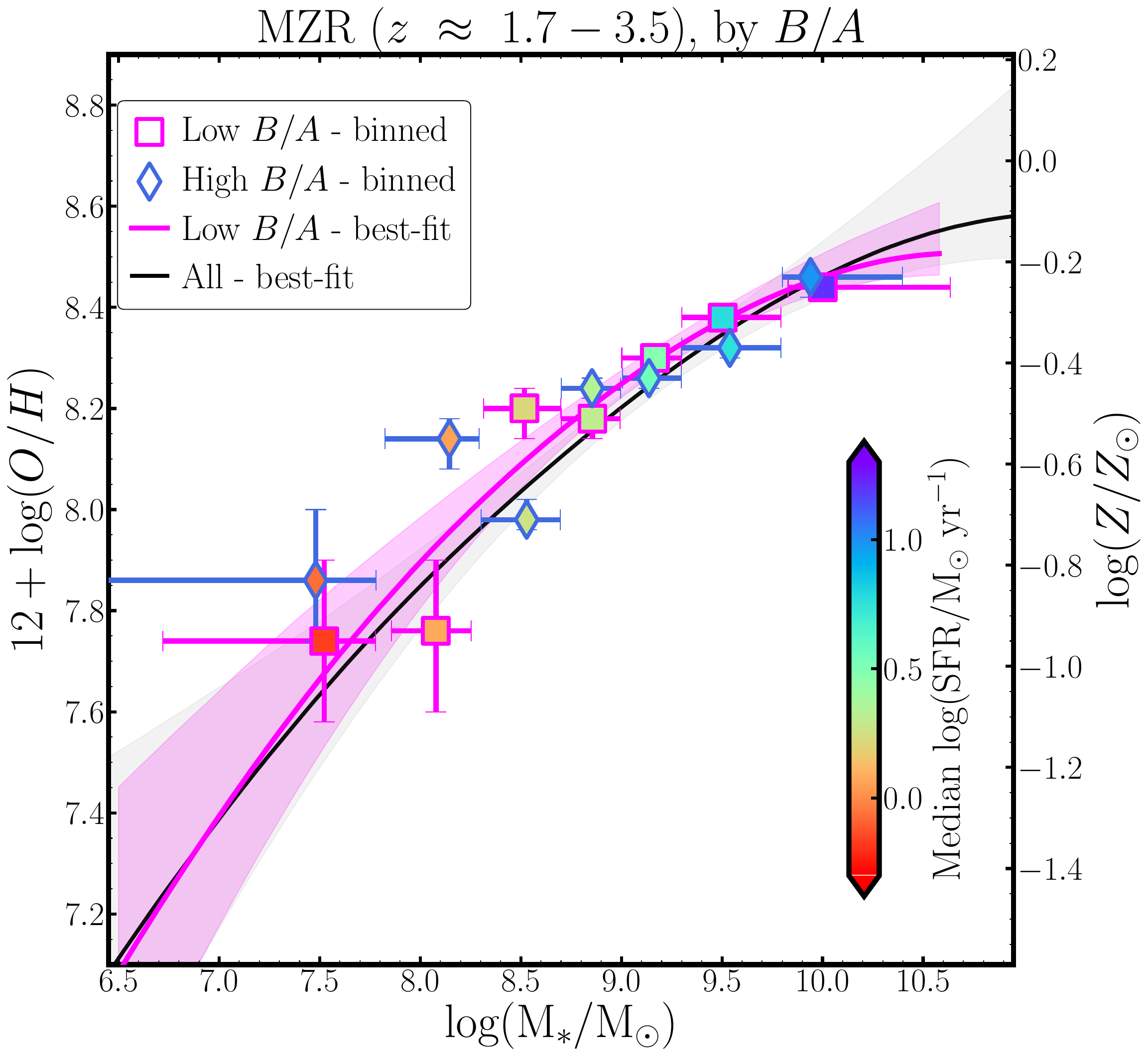}{0.33\textwidth}{}
\fig{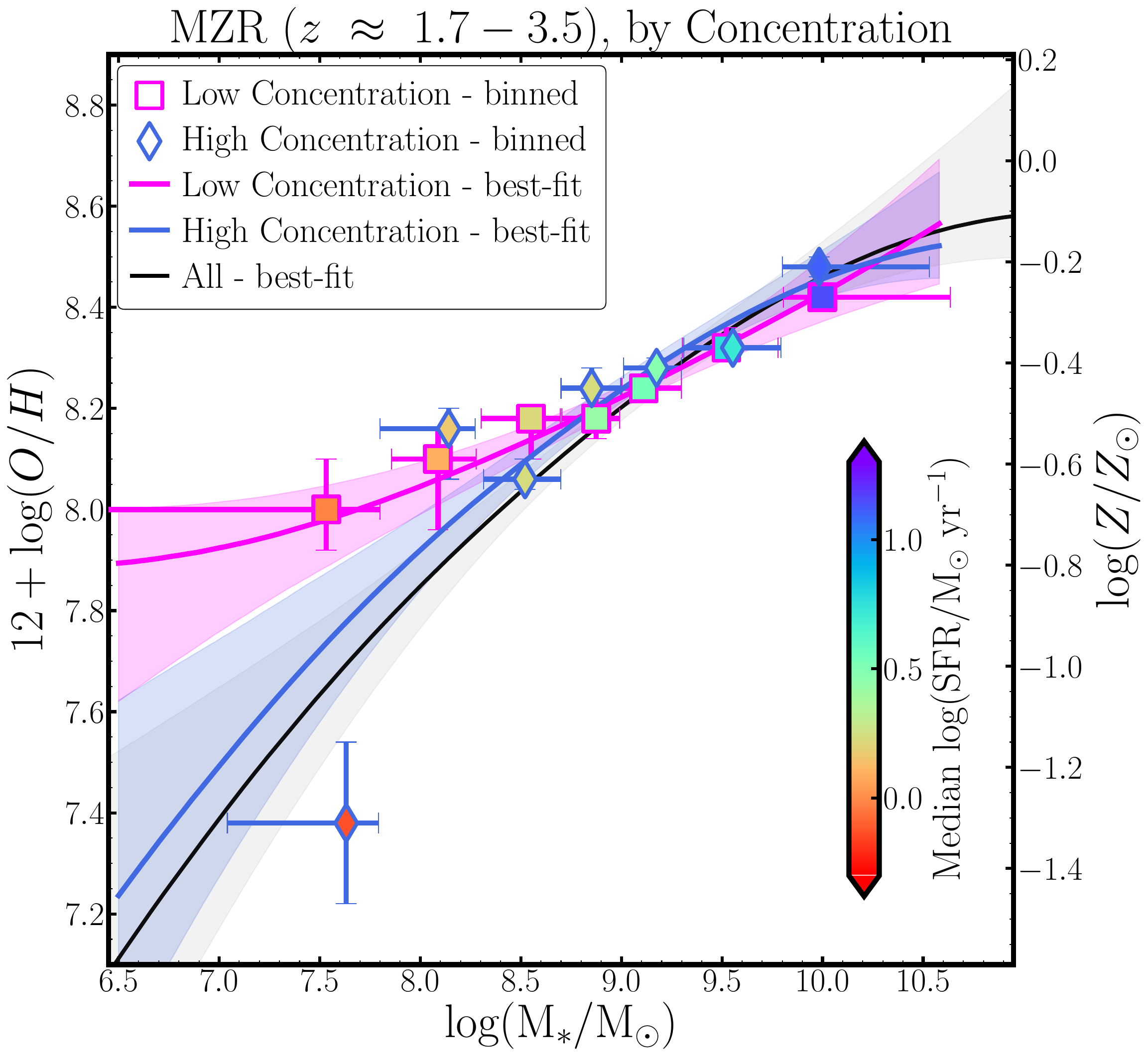}{0.33\textwidth}{}
}
\vspace{-25pt}
\gridline{
\fig{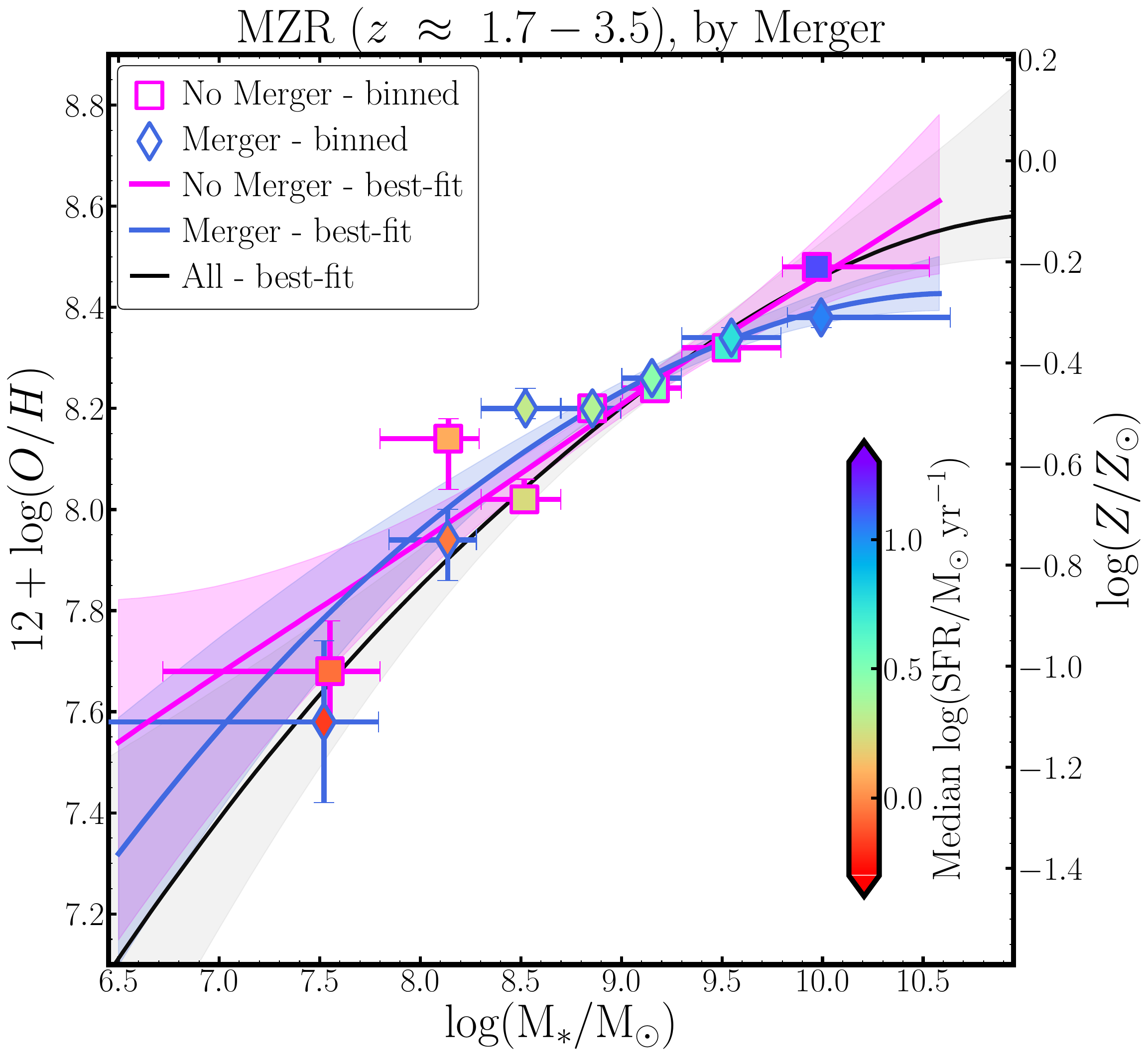}{0.33\textwidth}{}
\fig{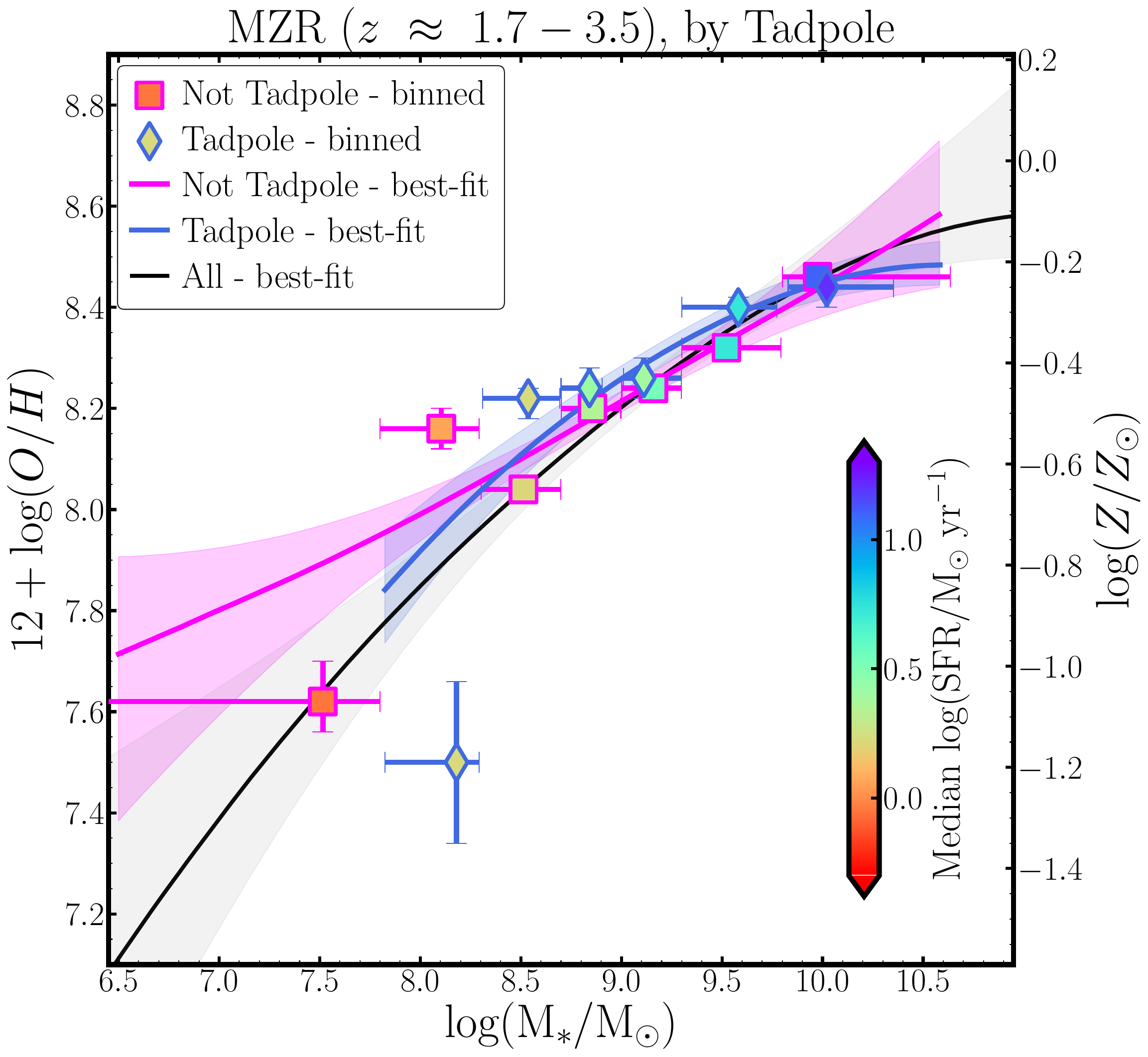}{0.33\textwidth}{}
\fig{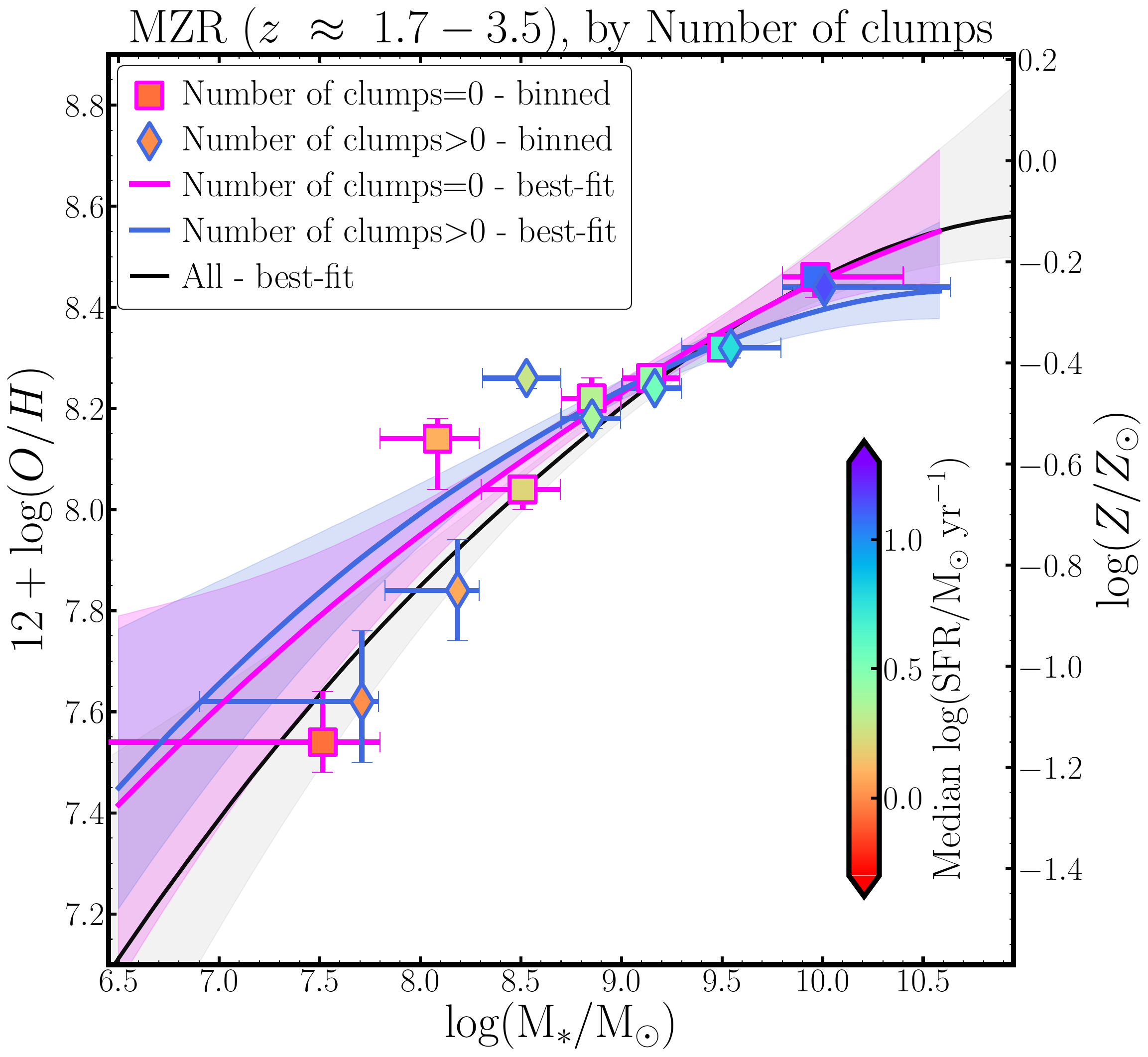}{0.33\textwidth}{}
}
\vspace{-25pt}
\caption{
Examples of the effect of some other morphological properties on the MZR.
\textit{Clockwise from top left}: S{\'e}rsic index, $B/A$ axis ratio, concentration, number of clumps, 
tadpole vs. non-tadpole galaxies,
and mergers vs. non-mergers (defined following \citet{Lotz08}). 
}
\label{fig:mzr-weak}
\end{figure*}

We test the effect of several other morphological parameters on the MZR. 
In Fig.~\ref{fig:mzr-weak}, we present the MZR separated by S{\'e}rsic index, minor-to-major axis ratio, central concentration, number of clumps, 
tadpole vs. non-tadpole galaxies, and mergers vs. non-mergers defined by the $G-M_{20}$ statistic  (\S~\ref{sec:morphquan}). 
We find small differences in some of these cases, but none of these properties show a statistically significant variation in the MZR. The S{\'e}rsic index shows the strongest effect, with high-{\nsers} galaxies being more metal-rich than low-{\nsers} galaxies at 2.3$\sigma$ significance. Thus, S{\'e}rsic index shows a marginally significant difference. Aside from this, only small variations are seen in a few bins in the other cases. 

\section{Discussion} \label{sec:discuss}



\subsection{SFR-dependence of the MZR}
\label{sec:discuss-fmr}

We uncover a strong dependence of the SFR on the MZR such that low-mass galaxies with ${\logms}<9.5$ have higher metallicity if they have lower SFR at fixed mass (\S~\ref{sec:mzr-sfr}). 
We have thus shown evidence that the FMR exists out to at least $z\sim3.45$, using one of the largest samples available to date for such studies at $z\gtrsim2$. We are able to probe the SFMS over $\sim$4 and $\sim$3 orders of magnitude in {\mstar} and SFR, respectively (Fig.~\ref{fig:sfr-comp}), allowing us to faithfully trace the fundamental plane of {\mstar}-SFR-{\logoh} at cosmic noon.

The approach we adopt to study the FMR is somewhat different from that commonplace in the literature.
The low-mass slope of the MZR is often fixed and the mass-metallicity-SFR 3D plane can be reduced to a 2D projection that produces the least scatter, in the form of ${\mu = {\logms} - \alpha{\logsfr}}$ \citep[e.g.,][]{Mannucci10,AM13,Sanders2021ApJ,Lam26}. This approach requires an $\alpha$ value that is parameterized by specific choices in the sample selection and the metallicity calibrations used. Consequently, different $\alpha$ values are used in different studies; for instance, strong-line and direct metallicities yield different values.

To determine a projection of least scatter, this approach can convolve and correlate the uncertainties in {\mstar} and SFR in a complex manner \citep[see, e.g.,][]{Revalski2024ApJ}.
In addition, this approach typically relies on measuring the residual scatter of individual galaxies around the best-fit MZR and SFMS relations \citep[e.g.,][]{Sanders2021ApJ,Stanton26}, which is not suitable for our stacking methodology.
For these reasons, we instead quantify the SFR-dependence of the MZR by stacking galaxies in low- and high-SFR subsamples in fixed mass bins, similar to the approach of \citet{Henry2021ApJ} for HST-based surveys at $z\sim1-2$. However, unlike their study, we do find a $>3\sigma$ difference in the MZR of low- and high-SFR galaxies. This could be due to lower uncertainties in our SED-derived SFRs, significantly improved statistics at the ${\logms}\lesssim8.5$ end, use of metallicity calibrations based on cosmic noon galaxies (instead of local analogs), or other reasons.

Qualitatively, the trends we find are consistent with a growing body of literature showing the existence of an FMR from the local Universe to the first $\sim1-2$~Gyr of the Universe. Depending on the use of direct metallicities or strong-line calibrations, ${\logoh}$ appears to decrease by $\approx$0.15--0.3 dex per dex rise in SFR at masses below ${\logms}\sim10$ \citep{Mannucci10,AM13,Curti20}.
While studies conducted with a variety of datasets and metallicity measurement methods have found a qualitatively similar result at higher redshifts, there is debate as to whether the {\it quantitative} SFR-dependence evolves with redshifts -- and if so, how. \citep[e.g.,][]{Sanders18,Topping21,Li23,Nakajima2023ApJS,Curti2024A&A,Lam26}.

This is one of the first studies to show the existence of an FMR down to ${\logms}\lesssim7$ at cosmic noon.
\citet{He26}, who studied the SFR-dependence of the MZR down to similar masses and redshifts as this study using stacked JWST/NIRISS data as well, reported no such evidence of an FMR. We posit that this difference could be due to systematics associated with the choice of metallicity calibrations, significantly smaller sample size ($\sim10\times$), biases due to the requirement of {\Hb} in stacks, or uncertainties in their SFR estimation, particularly as they did not have a direct estimate of dust extinction. 
In this work, we find that the difference in {\logoh} of low- and high-SFR galaxies grows significantly with decreasing {\mstar}, from $\Delta {\logoh}\lesssim0.1$ at ${\logms}\sim9$ to $\Delta {\logoh}\gtrsim0.6$ at ${\logms}\sim7.0$. In other words, the lower-SFR population at cosmic noon has a shallow MZR that almost flattens at very low masses compared to the highly star-forming population which exhibits a rapid steep MZR at low masses.


Our results are consistent with theoretical expectations as well. Standard equilibrium, gas-regulator, or ``bathtub'' models invoke the net balance of inflows and outflows of gas within galaxy ecosystems to explain this dependence \citep[e.g.,][]{Dave2012MNRAS,Dayal13,Lilly2013ApJ,PM14}. 
Gas inflows both increase the total gas fraction and decrease the gas-phase metallicity of the star-forming ISM. The former results in star formation, increasing the SFR of the galaxy, which eventually leads to an increase in metallicity. Feedback can eventually expel some of these metals out of the galaxy, marking the end of this enrichment cycle.
The FMR can therefore be interpreted as a consequence of galaxies being at different phases of this cycle.
Generally, both analytical models and cosmological simulations predict that star-forming galaxies grow in a state of quasi-equilibrium; gas inflows are {\it nearly} in equilibrium with consumption and outflows and metal-enrichment is {\it nearly} in equilibrium with metal-dilution \citep{Finlator17}.

To first order, the FMR is a result of the difference in the availability of gas between galaxies of similar stellar mass, which itself is due to a difference in the rate of gas accretion \citep[e.g.,][]{Bothwell13,Lagos16,DeRossi17,Wang26_ArXiV}.
Galaxies that have more recent gas accretion naturally have a higher rate of star formation and more recently diluted ISM.
The timescale on which the accreted gas is processed into stars, i.e., the gas consumption timescale, and the timescale on which gas is removed from the ISM by internal feedback determine how long it takes for the galaxy to return to its quasi-equilibrium state of being near the average MZR \citep[e.g.,][]{Lilly2013ApJ,Forbes14,PM14,Torrey18,Garcia2025MNRAS}. 

Observations indicate that gas depletion times vary strongly with deviation from the SFMS and that lower-mass galaxies generally have higher available gas fractions than higher-mass galaxies \citep[see][]{Tacconi20}.
In addition, outflows are observed to be stronger in lower mass galaxies across cosmic time \citep[e.g.,][]{Weiner09,Chisholm17,Chisholm18,XXu22,SL25}. This is in line with predictions of mass-loading factors of galactic winds being higher due to shallower gravitational potential wells in low-mass galaxies \citep[e.g.,][]{Muratov15,Mitchell20,Pandya21}.


Consequently, lower mass galaxies should naturally exhibit a stronger SFR-dependence in their MZR than higher mass galaxies. Low-mass galaxies are more sensitive to changes in their gas supply and feedback-driven outflows. Metal-poor accretion can quickly dilute the ISM, but strong outflows in shallow gravitational potential wells allow these galaxies to return to equilibrium quicker.
\citet{Forbes14} showed that the strength and duration of these departures from equilibrium is a competition between the stochasticity of gas accretion and the ability of outflows to expel gas from galaxies.
In higher-mass galaxies, this gas regulation is likely to be more stable, as the relative strength of outflows and available gas fractions are lower. 
The theoretical framework of \citet{Wang26_ArXiV} suggested
that the majority of star-forming galaxies at $z\gtrsim1$ reside in a so-called ``inflow-driven regime,'' which have not had the time to reach a state of equilibrium in the MZR. 
They showed that galaxies depart from this inflow-driven at the highest masses (${\logms}\gtrsim10$) at $z\sim2-3$, consistent with our interpretation of the observed FMR at cosmic noon.

\subsection{Morphology-dependence of the MZR}
\label{sec:discuss-morph}

Another key result of our study is the dependence of the MZR on morphological properties of galaxies, particularly at the ${\logms}<9.5$ end (\S~\ref{sec:mzr-morph}). 
In this regime, larger galaxies, lower SFR surface density galaxies, and galaxies with a clear central bulge are metal-enriched relative to compact, higher SFR surface density, and bulge-less galaxies of similar mass. At intermediate masses, ${\logms}\sim9-10$, galaxies with spiral features are more metal-rich compared to their non-spiral counterparts at fixed mass.
To our knowledge, this is the first study to statistically capture these trends linking chemical enrichment and structural assembly of galaxies at cosmic noon.

\subsubsection{Size and SFR surface density}

The positive correlation between stellar size and gas-phase metallicity is particularly noteworthy, as this is opposite to the trend commonly observed in the local Universe, where smaller galaxies are typically more metal-rich than larger galaxies at fixed mass \citep[e.g.,][]{Ellison08,BH12,DEugenio18,Ma24}. 
\citet{Yabe12} found a similar metallicity-size anti-correlation at $z\sim1.2-1.6$ using a sample of $\sim$70 star-forming galaxies, but only down to ${\logms}\sim9.5$. Compiling a larger sample down to lower masses and $z\sim0.6-1.8$, \citet{Gillman21} found only a moderate anti-correlation.
Using $\lesssim$100 JADES galaxies, \citet{Jia25} found that individual galaxies lying above the average mass-size relation tend to be below the MZR -- a conclusion that holds at $z\sim1-7$.

There are, however, some observational results qualitatively consistent with ours, i.e., smaller galaxies being metal-poor compared to larger galaxies.
\citet{Hoopes07} found that compact UV-luminous galaxies at $z\sim0-0.3$ are systematically metal-deficient compared to larger galaxies of similar mass. In the higher-$z$ universe, galaxies below the average mass-size relation were reported to be much more metal-poor than those above it by \citet{LH23}. Their study, using NIRSpec spectroscopy and NIRCam photometry, was based on $\sim$400 galaxies at $z\sim3-10$ down to ${\logms}\approx7$.  
Similarly, \citet{Tacchella23} found highly compact ${\logms}\approx8$ galaxies with extremely low metallicities ($\sim$6\%) at much earlier epochs, $z\sim7.5-8.5$.

\citet{SD18} proposed, using the EAGLE cosmological simulations \citep{Schaye15}, that the observed low-$z$ metallicity-size anti-correlation is due to differences in gas accretion histories of galaxies with similar masses \citep[see also][]{Boardman26}. 
Earlier accretion of gas would result in a galaxy assembling most of its stars at early times within a fairly compact region, compared to later accretion, which would allow it to form stars further from the center, ultimately increasing its half-light radius \citep[consistent with inside-out growth;][]{Bezanson2009ApJ,vd10}. Therefore, sustained late gas accretion can at once lower the ISM metallicity, raise the SFR, and grow the stellar size.
According to \citet{SD18}, EAGLE predicts that larger galaxies are metal-poor relative to smaller galaxies out to $z\sim8$, in clear tension with our observations.

A combination of the reduced metallicity in high-SFR and small galaxies leads to high SFR surface density galaxies being significantly metal-poor. 
Authors such as \citet{SM19} and \citet{Baker23}
have showed that the FMR exists on spatially resolved scales within galaxies, where the local metallicity is lower in regions of high SFR surface density 
\citep[as expected from analytical models; e.g.,][]{SA19}. Similar results were found at $z\sim0.3$ by \citet{Koller24}. 
Using spatially-resolved NIRISS observations, \citet{EC25} showed that individual star-forming regions with higher SFR surface density have lower metallicity at $z\sim0.6-1.4$. 
High SFR surface density has been associated with low metallicity in spatially-resolved observations at cosmic noon as well \citep[e.g.,][]{Cresci10,XinWang2020ApJ,Acharyya26}.

The SFR surface density is proportional to the gas surface density by the iconic Kennicutt-Schmidt law \citep{Kennicutt98}. A high SFR surface density indicates a highly-concentrated star-forming gas reservoir, which is likely to be built by fairly recent gas accretion. If this is the case, then highly concentrated recent star formation will be accompanied by a chemically diluted ISM. Furthermore, high SFR surface density may also drive strong outflows that expel metals produced by recent star formation, as seen in observations \citep[e.g.,][]{Weldon24,RC25}, and predicted when clustered SNe cause    ``super-bubble'' feedback \cite[e.g.,][]{Fielding18}.
Conversely, galaxies with less concentrated SFR activity likely experience less intense recent gas inflows and/or less powerful outflows, allowing them to enrich their ISM over longer timescales.

Taken together, our findings suggest that compact galaxies with highly concentrated star formation activity tend to be metal-poor at cosmic noon. 
We showed that, on average, the extended galaxies in our sample have lower SFRs than compact ones at the same mass (Fig.~\ref{fig:mzr-rh}), connecting the star formation mechanisms to structural evolution.
In contrast to the physical picture advanced by \citet{SD18} and others, we propose that at cosmic noon, recent gas accretion favors the formation of compact galaxies with intense, localized starbursts. Extended galaxies, on the other hand, may have undergone more sustained chemical evolution due to less concentrated star formation activity. 
This would imply that beyond gas accretion history, the  spatial distribution of the resulting star formation and structural assembly is a crucial aspect of this physical connection. Moreover, this may reflect differences in the physical mechanisms driving the baryon cycle $\sim$9--12 Gyr ago to today.

\subsubsection{Clumpiness}

Broadly, this picture is supported by the zoom-in simulations of \citet{Ceverino16}, who suggest that metal-poor inflows at $z\gtrsim2$ trigger localized starbursts in ${\logms}\sim9$ galaxies.  
Such scenarios often predict that pristine accretion results in lower metallicities in star-forming clumps than in the surrounding ISM outside the clumps. We find no significant difference in the metallicities of galaxies with and without visible clumps.
This may be because individual clumps (and therefore metallicity differences) are short-lived, on the order $\sim$20--50~Myr \citep{Ceverino16,Oklopcic17}.
Moreover, clumps may originate in gravitational instabilities from pre-existing enriched gas, rather than recently-accreted metal-poor gas and continued star formation and inward migration may progressively increase clump metallicities \citep[e.g.,][]{Ceverino2010MNRAS,Mandelker14,Mandelker2017MNRAS}.
Some observations prefer this scenario of secular clump formation over an external origin \citep[e.g.,][]{Bournaud08}.

Crucially, we do not measure metallicities in spatially-resolved clump vs. inter-clump regions, where these aforementioned metallicity differences are expected to appear. Indeed, most observational evidence linking clumps to low metallicity has been in the form of such spatial differences \citep[e.g.,][]{SA2014ApJ,Straugh15,EC25}. In this study, we measure the metallicity integrated over an entire galaxy, which can average out the metallicity differences between clump and inter-clump regions. However, from integrated metallicities in $\sim$300 $z\sim0.6-2.5$ galaxies, \citet{Sok26} found that galaxies with clumps are systematically metal-poor by $\lesssim$0.1~dex than those without clumps. The disagreement with our findings may be due to their use of individual measurements rather than stacking, uncertainties in the metallicity calibrations, differences due to identifying clumps by automated vs. visual methods, or some other systematic effect.

We stress that the absence of a systematic metallicity deficit in clumpy galaxies in our sample does not necessarily exclude accretion-driven clump formation, but that visible clumpiness alone is not a reliable indicator of recent metal-poor gas accretion. 
More broadly, our findings suggest that the metallicity differences between compact and extended galaxies may be associated with galaxy-wide variations in gas supply, star-formation activity, and metal retention, rather than localized chemical dilution within individual star-forming clumps.

\subsubsection{Bulges and spirals}

Further evidence of structural evolution being connected to chemical enrichment is provided by the observations of enriched ISM in bulge and spiral galaxies. 
Bulges can form through dissipative mergers, gravitational instabilities, and inward migration of massive star-forming clumps, all of which can cause a buildup of metallicity near the center \citep{Bournaud07,Bournaud2016ASSL,Ceverino15}. Therefore, at fixed mass, the presence of a prominent bulge may indicate that a larger fraction of the galaxy's stellar mass was assembled in early star formation episodes, allowing the galaxy to have more time to enrich its ISM. Bulge growth may also stabilize the gas disk against fragmentation, thereby reducing subsequent star formation activity \citep{Martig09}. We see some indication of lower SFR in bulge galaxies relative to bulge-less galaxies at ${\logms}\sim9-9.5$ in Fig.~\ref{fig:mzr-blg-spr} (left).
We note that previous studies have yielded mixed results on the effect of bulges on gas-phase metallicities  \citep[e.g.,][]{Ellison08,Pilyugin14b,Gillman21}.

The presence of spiral features suggests a fairly evolved dynamical state, including a more advanced stage of disk assembly.
Cosmological simulations predict that galaxies can evolve from compact, turbulent, and dispersion-dominated systems to extended, settled, rotationally-supported disks as they grow their stellar mass and gas metallicity \citep[e.g.,][]{SM94,Governato07,Ma17}. 
Spiral structure may also drive radial gas transport, and turbulent mixing, which can influence the spatial distribution of metals \citep[e.g.,][]{Orr23}.
Observations have lent support for the enrichment of spiral galaxies relative to galaxies without spiral structure and/or irregular structure \citep{Pilyugin13,Pilyugin14a,Pilyugin14b,Gillman21}.

Overall, the presence of bulge or spiral features may mark an evolutionary state in which a galaxy has undergone more sustained stellar assembly and chemical enrichment, with its current gas-phase metallicity less strongly dominated by recent, stochastic accretion and feedback.

\subsection{Towards a Unified Physical Picture}
\label{sec:discuss-phys}

Finally, we develop a consistent physical picture to explain our findings. We argue that secondary dependencies of the MZR on SFR and morphology are likely due to differences in the current (or recent) state of galaxy growth at cosmic noon, caused by several aspects of the baryon cycle. Most of these differences are only seen for intermediate- and low-mass galaxies (${\logms}\lesssim9.5$--10).

At fixed mass, galaxies that are more metal-rich appear to be more structurally evolved than their metal-poor counterparts. Metal-poor galaxies are undergoing more recent star formation activity, which is caused by fairly pristine gas accretion that dilutes the metallicity of the ISM. The star formation is concentrated in compact regions, leading to higher SFR surface density and smaller stellar sizes. These galaxies have not had much time to contaminate the ISM with metals synthesized by generations of star formation. Metal-rich galaxies, on the other hand, have experienced less recent star formation activity due to a lack of recent gas accretion that also helps them maintain high enrichment levels in their ISM. These galaxies have undergone more sustained chemical enrichment and developed highly extended structures, including bulges and/or spiral features such as disks and arms. 
Additional marginal evidence for this is in the slightly elevated metallicities of galaxies with higher S{\'e}rsic indices (Fig.~\ref{fig:mzr-weak}).

These two types of galaxies distinguished by their gas-phase metallicity may or may not represent an evolutionary sequence. The ``young'' and actively growing galaxies with low metallicity may eventually evolve into more dynamically settled and relaxed galaxies with higher metallicity as their accreted gas is processed into stars and their stellar structures develop. The two observed populations may instead originate from different evolutionary histories governed by a more complex interplay of gas cycling in and out of galaxies. The importance of both short and long-term accretion histories is intimately tied to this picture \citep[as also ascertained by, e.g.,][]{SD18,Boardman26,Wang26_ArXiV}.

Individual galaxies may rapidly oscillate between states of low and high metallicity following episodes of fresh accretion, rather than simply evolving monotonically from lower- to higher-metallicity systems. 
Simulations routinely predict (regardless of the model) short-timescale variability in gas accretion, depletion, and replenishment at high redshift -- leading to bursty star formation activity \citep[e.g,][]{Tacchella2016MNRAS,Sparre17,Iyer20}. 
Such variability is strongly supported by JWST observations which find that lower mass galaxies at higher redshifts have increasingly stochastic SFHs \citep[e.g.,][]{Tacchella22b,Clarke24,Cole25,Looser25,Mintz26,NC26}.
An illuminating future experiment would be to estimate the ages of stellar populations of galaxies of different enrichment levels, thus directly testing our picture of more evolved galaxies having higher metallicities.

Aside from SFR variations, stochasticity in gas accretion may also cause changes in some morphological properties, particularly among low-mass systems. Cosmological simulations predict cycles of gas compaction, depletion, and replenishment at high-$z$ which cause galaxies to transition between centrally concentrated, compact (``blue nugget'') phases and extended, disk-like/elongated configurations \citep{Zolotov2015MNRAS,Tacchella2016MNRAS.458}.
Stellar feedback can also cause rapid fluctuations in the gravitational potential, resulting in the stellar effective radius to vary by more than a factor of two over $\sim$200 Myr \citep{ElBadry16}. 
The light from young stars may also dominate the observed morphology of early galaxies, and changes in apparent size and structure may occur over mere tens of Myr \citep{Ma18}. 
Arguments of stochasticity could help explain why there is significant bin-to-bin variation in the MZR of galaxies of different morphological properties at ${\logms}\lesssim8.5$ (Fig.~\ref{fig:mzr-weak}), especially for galaxies with clumps which are expected to be fairly short-lived.

We can also imagine some galaxies evolving from a low- to high-metallicity state and after a fairly long time ($\gtrsim$100~Myr) experiencing intense accretion events that re-activate their star formation activities and lower their metallicities. Such an event would probably be instigated by mergers, which can drive low-metallicity gas toward galaxy centers, simultaneously increasing the SFR and diluting the ISM metallicity \citep[e.g.,][]{Rupke10,Bustamante2018MNRAS}.
However, we do not find statistically significant differences between the MZR of galaxies undergoing interactions and those that are not, whether this is classified by non-parametric $G-M_{20}$ statistics \citep{Lotz08} or visual inspection. This lends credence to the idea that gas accretion from the cosmic web and internal galactic phenomena play a bigger role in the chemical evolution of galaxies than mergers -- as least as far as our observations are concerned.

\section{Conclusion} \label{sec:conc}

Using JWST/NIRISS WFSS grism spectra and multi-wavelength imaging, we investigated the MZR at cosmic noon and explored its dependence on SFR and a variety of morphological properties.
We assembled one of the largest samples of galaxies at cosmic noon to date with JWST grism spectra and high-resolution JWST, HST, and ground-based imaging -- more than 1000 galaxies at redshifts $1.74 \leq z \leq 3.45$. 
We measured metallicities from emission-line fluxes and stellar masses and SFRs from SED-fitting. Moreover, quantitative morphological properties such as half-light radius, axis ratio, and S{\'e}rsic index are measured from single-band imaging and visual classifications were performed using multi-band imaging. 
By stacking spectra in mass, in SFR, and morphological properties, we measured the metallicities of galaxies as composite populations sharing similar characteristics. We investigated the MZR of galaxies in the mass range ${\logms}\simeq6.5-11$, its SFR-dependence, and its dependence on various morphological properties.
Our main findings are as follows.

\begin{enumerate}
\itemsep0em
    \item The general MZR is well fit by a second-order polynomial: ${\logoh}= a\left[\log{\mstar}-9\right]^2 + b\left[\log{\mstar}-9\right]+c$, where our best-fit parameters are $a=-0.06^{+0.06}_{-0.02}$, $b=0.31^{+0.05}_{-0.04}$, and $c=8.21^{+0.02}_{-0.02}$. (\S~\ref{sec:mzr-full}). 
    The MZR declines steeply at the low-mass end (${\logms}\leq8$), dropping to $\lesssim$10\% solar metallicity at ${\logms}\lesssim7.5$. At higher masses, our measured MZR is generally consistent with the literature at $z\sim1.5$--3.5 (see Nedkova et al., in prep).
    \item We show strong evidence that the FMR exists at ${1.74 \leq z \leq 3.45}$: galaxies with low-SFR have higher metallicity than those with high-SFR at ${\logms}\leq9.5$. The MZR of low-SFR galaxies in this mass range is higher than that of high-SFR galaxies at $>3.9\sigma$ significance, with the offset growing from $\Delta{\logoh}\lesssim0.1$ to $\Delta{\logoh}\gtrsim0.5$ dex from ${\logms}\sim9$ to ${\logms}\sim7$ (\S~\ref{sec:mzr-sfr}).
    \item At fixed mass, galaxies with larger half-light radius are more metal-rich than those with smaller half-light radius, at $>3\sigma$ significance over ${\logms}\leq9.5$. Larger galaxies have $\approx$0.1 ($\approx$0.4~dex) higher metallicity than smaller galaxies at ${\logms}\sim9$ (${\logms}\sim7.5$) (\S~\ref{sec:mzr-size}).
    \item The MZR also shows a statistically significant dependence on the SFR surface density at ${\logms}\leq9.5$: high-SFR surface density galaxies are metal-deficient compared to low-SFR surface density galaxies at fixed mass. The average offset grows to $\Delta{\logoh}>0.5$~dex at ${\logms}\sim7$ (\S~\ref{sec:mzr-sfrdens}).
    \item The presence of a central bulge and spiral features also show statistically significant effects on the MZR. Galaxies with a bulge are enhanced in metals, by $\sim$0.1-0.3~dex at ${\logms}\sim8-9.5$, and galaxies with spiral features are metal-enhanced by $\sim$0.1-0.2~dex at ${\logms}\sim9-10$ (\S~\ref{sec:mzr-blg-spr}). Most other morphological parameters either show no or partial statistical significance in terms of affecting the MZR (\S~\ref{sec:mzr-other}).
\end{enumerate}

Our results paint a picture in which galaxies in a more evolved state are more metal-rich than those in a undergoing more active formation and evolution. 
Low- and intermediate-mass galaxies that are more extended and have less recent star formation activity tend to be metal-enriched. 
Sometimes these also have a prominent central bulge and/or spiral structure. 
Compact and highly star-forming (and typically bulge-less and non-spiral) galaxies are much more metal-poor.
It is unclear if galaxies evolve monotonically from the low to high-metallicity state or if they oscillate between these states over fairly short or even longer timescales.
Regardless, our results highlight the complexities in the galactic baryon cycle and how they affect several inter-connected aspects of galaxy evolution. They also place novel constraints for theoretical models of galaxy formation.

We stress the importance of both time-resolved analyses of cosmological simulations and spatially-resolved observations at cosmic noon and earlier times in order to test the physical picture we propose in detail. 
While beyond the scope of this work, it would be very insightful to perform a detailed investigation tracking the evolution of gas-phase metallicity, SFR, and morphological characteristics of individual galaxies in high-spatial and -temporal resolution simulations. Simulations such as {\sc Vela} \citep{Ceverino2023MNRAS}, TNG50 \citep{Nelson19}, or FIRE-3 \citep{Hopkins23} could reveal the timescales of chemical evolution and better constrain the effect of physical prescriptions of inflows, feedback, and recycling. 
Simulations which can resolve the ISM on $\lesssim$10~pc scales over very fine time-steps (few Myr) will yield important insights into the physics controlling the mechanisms that give rise to the observed trends.


Spatially-resolved observations are key to mapping variations in the SFR surface density, metallicity, and other properties across large samples of galaxies at high redshifts. For example, Integral Field Spectroscopy with NIRSpec could be leveraged to test our picture of ramped-up star formation activity in compact regions leading to dilution of their metallicity. Such observations would also enable the measurement of metallicities in individual clumps and distinguish them from the globally-averaged metallicities we measure here.
Additionally, large NIRCam/WFSS programs such as POPPIES (GO-5398; PIs: Kartlatepe \& Rafelski), SAPPHIRES (GO-6334; PI: Egami), and COSMOS-3D (GO-5893; PI: Kakichi) will allow us to extend investigations of the SFR and morphology-dependence of the MZR to higher redshifts, when galaxies were less polluted by star formation activity and in principle we could trace earlier and more pristine gas accretion.
Moreover, molecular gas measurements with ALMA combined with JWST-based spectroscopy can provide powerful constraints on both feedback-driven outflows and gas inflows at low-mass galaxies in cosmic web.


One aspect of the MZR which we did not discuss is the effect of large-scale environment. For instance, protocluster environments have been shown to affect the MZR at cosmic noon, owing to a variety of mechanisms related to gas accretion, shock-heating, etc. \citep[e.g.,][]{Sattari21,XinWang2022ApJ,YYang26} 
The lack of contiguous sky coverage combined with the small FOV of HST and JWST prevent 
these observatories 
from robustly constraining the MZR as a function of environment. 
However, the wide-field capabilities and deep grism surveys of both Euclid 
and Roman will enable an unprecedented mapping of the cosmic web and galactic environments at $z\geq1$ \citep[e.g.,][]{ROTAC25,Euclid26_Kraljic,Hasan26}. This will open up new avenues to assemble  statistically significant samples of galaxies across various large-scale environments and directly map the effect of the baryon cycle (particularly gas inflows) on the chemical evolution of galaxies.

\appendix

\section{Stacked Spectra}
\label{app:stack}

In this section, we present our stacked spectra. Fig.~\ref{fig:ms-stacks} shows the  we present the overall stacked spectrum for all galaxies (black), followed by spectra stacked in bins of decreasing stellar mass (different colors).  The locations of common emission lines and the number of galaxies going into each stack are indicated in the figure. Bootstrapped errors on the stacked fluxes are shown in light grey. As we mentioned in \S~\ref{sec:stack}, the stacks are all normalized to the {\OIIIf} flux.

\begin{figure*}[htbp]
\vspace{-10pt}
\centering
\gridline{
\fig{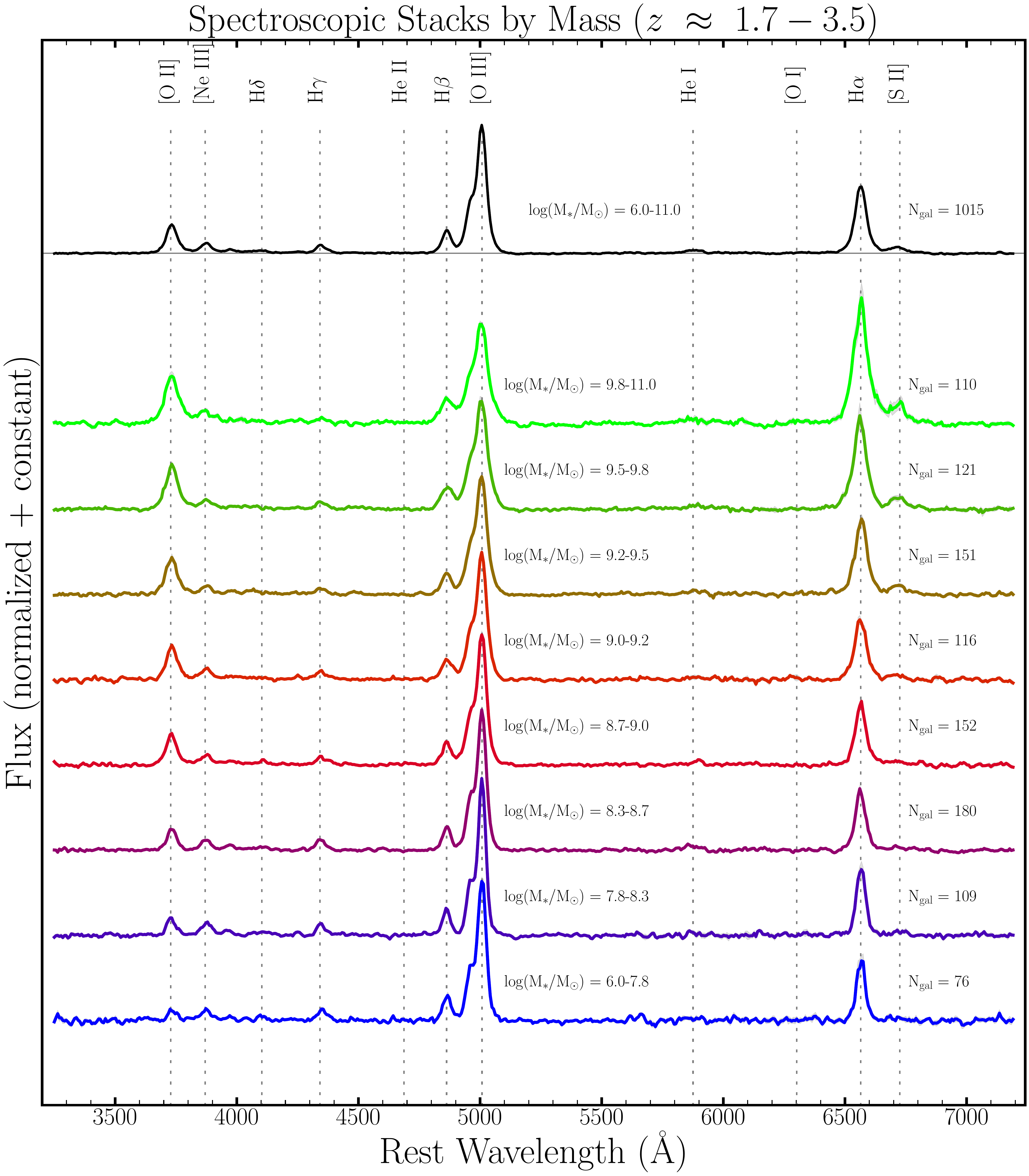}{0.9\textwidth}{}
}
\vspace{-25pt}
\caption{
Spectroscopic stacks by stellar mass, with the overall stack of 1015 galaxies shown at the top. 
The spectra of individual galaxies are normalized by {\OIIIf}  flux prior to stacking.
The number of galaxies going into each stack are  shown on the right.
}
\label{fig:ms-stacks}
\end{figure*}

We present the spectra stacked by {\mstar} and SFR in Fig.~\ref{fig:sfr-stacks}. The left (right) column shows spectral stacks of low-SFR (high-SFR) galaxies. 
We present two examples of spectra stacked by mass and morphology in Fig.~\ref{fig:morph-stacks}. The top row shows spectral stacks in mass and half-light radius, the bottom row in mass and SFR surface density.
The low {\rh} and SFR surface density stacks are shown on the left, while the high {\rh} and SFR density stacks are shown on the right. 


\begin{figure*}[htbp]
\vspace{-10pt}
\centering
\gridline{
\fig{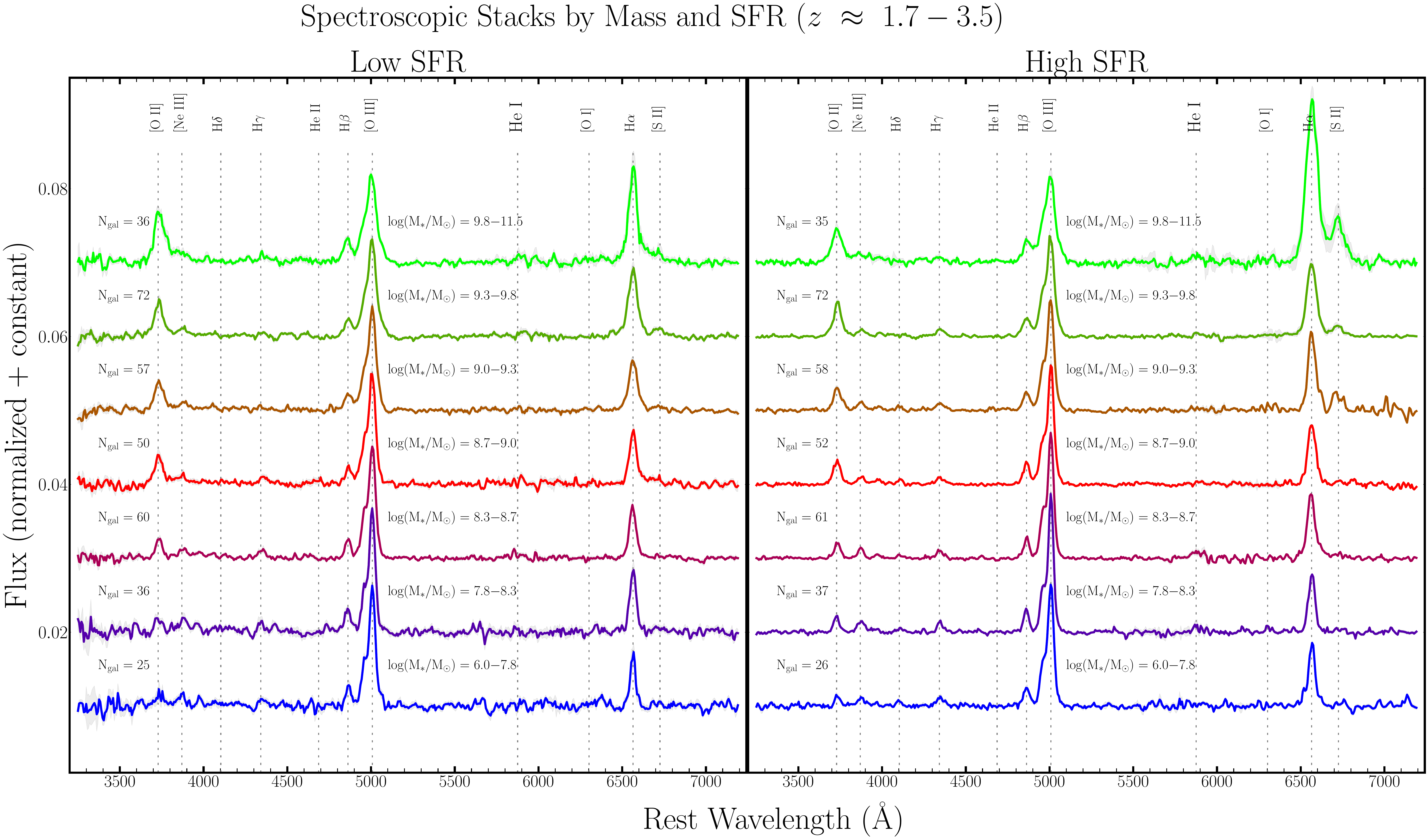}{0.99\textwidth}{}
}
\vspace{-25pt}
\caption{
Spectroscopic stacks by stellar mass and SFR, with the left column showing low-SFR stacks and right column showing high-SFR stacks. The spectra are normalized by {\OIIIf}.
}
\label{fig:sfr-stacks}
\end{figure*}


\begin{figure*}[htbp]
\vspace{-10pt}
\centering
\gridline{
\fig{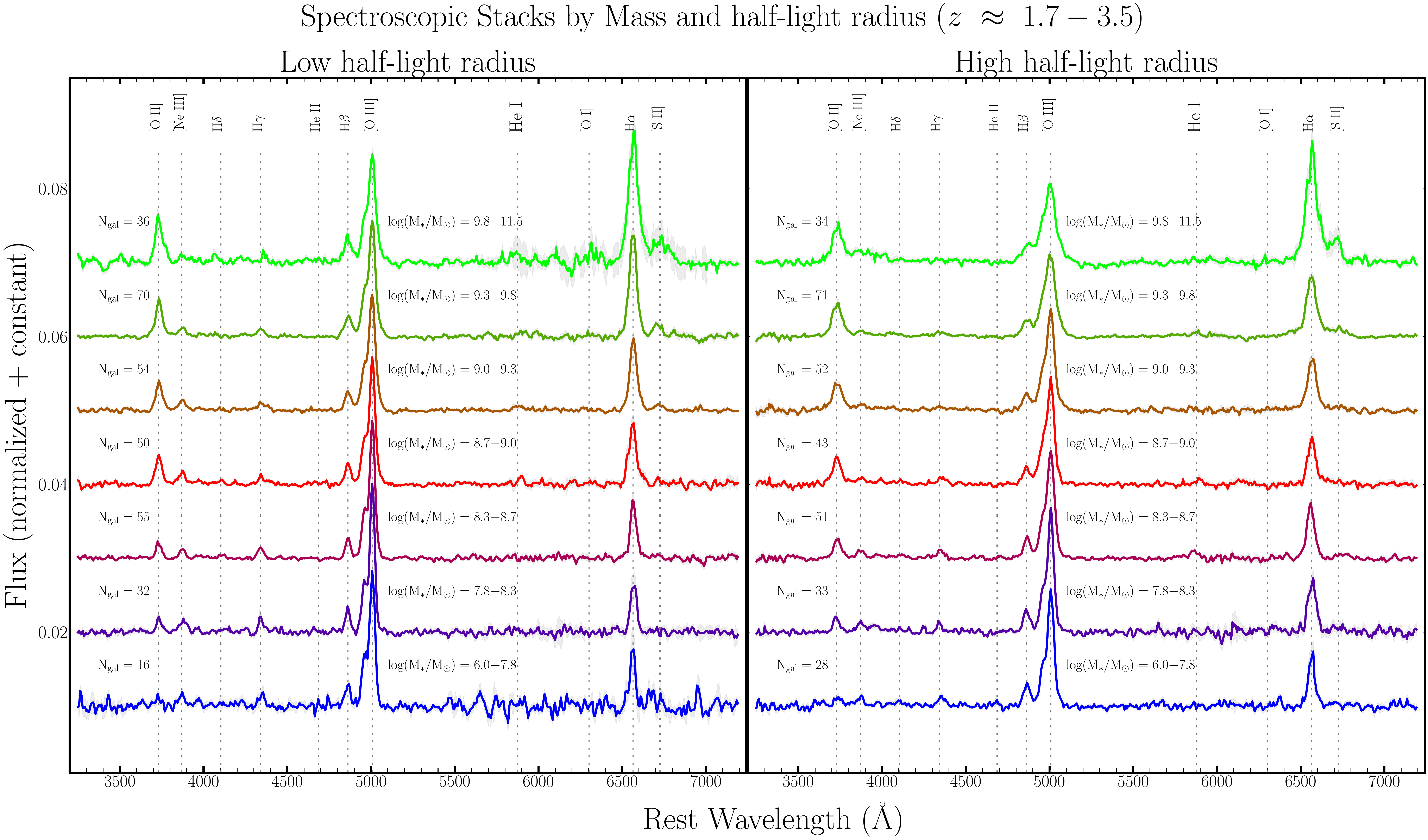}{0.99\textwidth}{}
}
\vspace{-25pt}
\gridline{
\fig{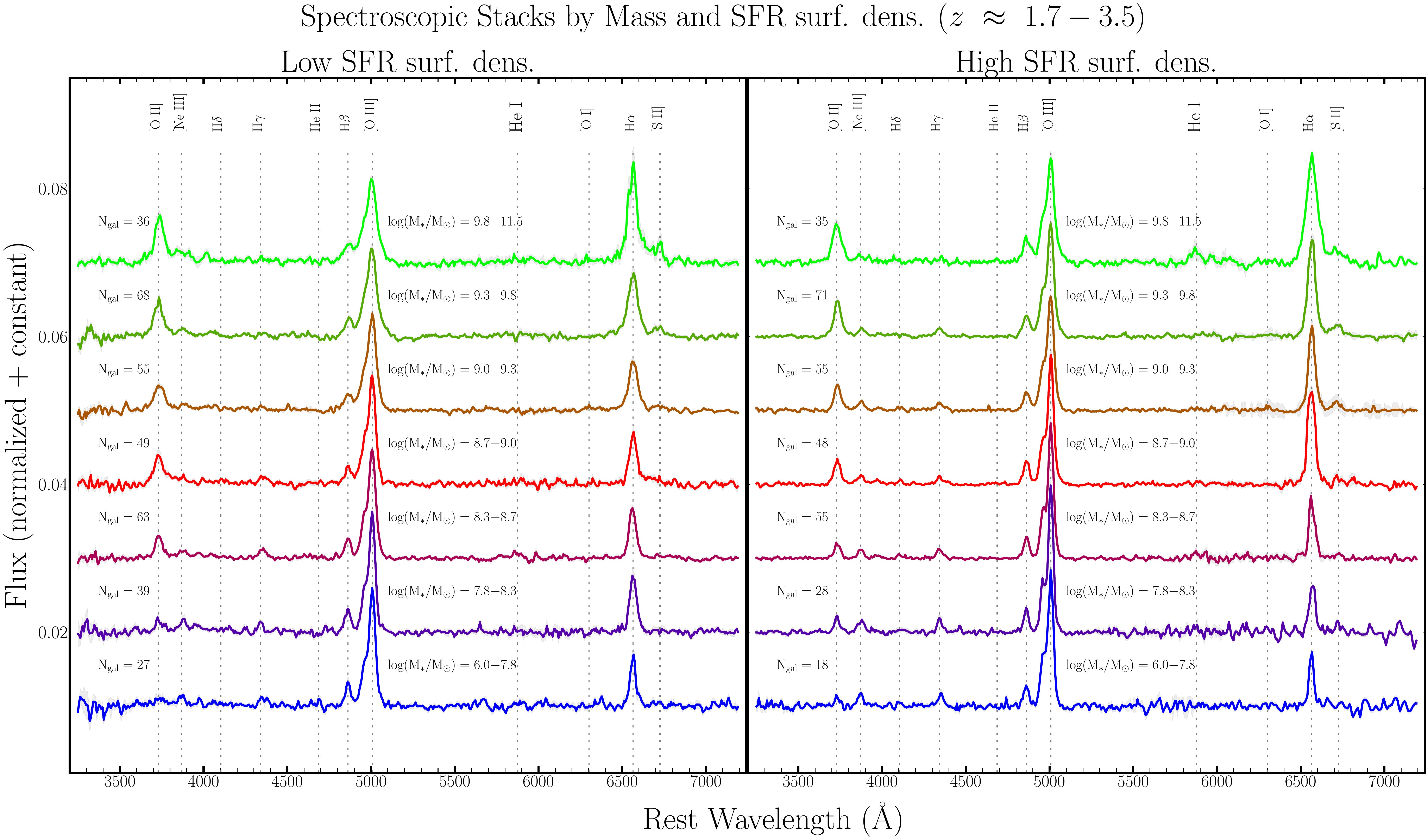}{0.99\textwidth}{}
}
\vspace{-25pt}
\caption{
Spectra stacked by mass and half-light radius (upper row) and mass and SFR surface density (lower row), normalized by {\OIIIf}.
}
\label{fig:morph-stacks}
\end{figure*}

\section{Morphology measurements and classifications}

\subsection{{\sc Statmorph} measurements}
\label{app:sm}

We select F150W as the preferred JWST filter in {\sc Statmorph} because it provides a useful balance between tracing the underlying stellar structure and preserving high spatial resolution and sensitivity. 
Over the redshifts we probe in this work, F150W samples rest-frame wavelengths of roughly $0.33-0.55$~{\micron}, spanning the near-UV to optical and therefore tracing a larger fraction of the older stellar population than shorter-wavelength imaging while remaining sensitive to the star-forming structure of the galaxies.
In comparison, F115W probes bluer rest-frame wavelengths and is more strongly weighted toward young stellar populations and dust-sensitive structure, in addition to having lower throughput. 
The F277W and F444W filters have fairly high throughput and probe the red rest-frame optical and NIR wavelengths better tracing the true stellar mass distribution, but at the cost of broader PSFs and poorer spatial resolution. 
Using F150W therefore provides an effective middle ground to reduce the strongest rest-frame UV morphological biases, while retaining a substantially narrower PSF than the redder NIRCam bands and generally high imaging depth. 
For Par682 morphology measurements, we adopt NIRISS F150W for similar reasons given that it is roughly at the same wavelength as NIRCam F150W (albeit with lower spatial resolution).

For fields with only ancillary HST coverage, we choose ACS F814W as it is the HST filter available to us that is closest in wavelength to F150W and has a high enough resolution and throughput. For this filter, the rest-frame wavelengths are $\sim0.17-0.3$~{\micron}. 
The use of these different filters inevitably introduces a wavelength-dependent morphological $k$-correction effect \citep[e.g.,][]{Conselice03,Lotz04}. The F814W measurements can consequently be more sensitive to patchy star formation, young stellar populations, and dust attenuation, while the longer-wavelength F150W images provide a closer representation of the underlying stellar-mass distribution, particularly toward the lower-redshift end of the sample.
We therefore account for the imaging band used when interpreting field-to-field differences in the measured morphological distributions and regard comparisons between the F814W- and F150W-based measurements with appropriate caution.

Furthermore, the dither pattern of the NIRISS observations often resulted in insufficiently subsampling the PSF and therefore an irregular pattern is introduced for some galaxies. This is an additional reason the NIRISS F150W is not chosen as the default filter.

We generate segmentation maps from NIRCam/NIRISS F150W and ACS F814W images using {\sc photutils} \citep{photutilssoft} and identify sources above a 1.5$\sigma$ threshold after background subtraction and Gaussian smoothing, with at least 10 connected pixels. 
We
perform {\sc Statmorph} measurements on cutouts extending to 2.5 times the elliptical Kron radius, and only consider sources with $\geq10$ segmented pixels and photometric $S/N>5$. 
We adopt empirical NIRCam and ACS PSFs from COSMOS-Web \citep{Shuntov25} and 3D-HST \citep{Brammer12}, respectively. We generate NIRISS PSFs using {\sc stpsf} \citep{Perrin12,Perrin14}, accounting for PSF effects in S{'e}rsic fitting \citep{RG19}.



The left three panels of Fig.~\ref{fig:sm-stats} present histograms of quantities measured with {\sc Statmorph}: S{\'e}rsic index {\nsers}, Concentration $C$, and minor-to-major axis ratio $B/A$ 
The S{\'e}rsic index distribution is strongly peaked near ${\nsers}\sim1$, with $\sim$80\% of galaxies having ${\nsers}<2$, indicative of predominantly exponential/disk-like light profiles \citep[e.g.,][]{Chang13,Martorano25}. 
The concentration distribution peaks at $C\sim2.3-2.7$, consistent with the typical $\sim2.4-3$ for high-$z$ galaxies \citep{Law12,Gillman20}. 
The projected axis-ratio distribution is broad and centered near $B/A\sim0.6-0.7$, in agreement with previous measurements at $z\sim2$ \citep{Chang13,vanderwel14b,Zhang19,Pandya24}. 

The right panel of Fig.~\ref{fig:sm-stats} presents the Gini coefficient, $G$, against $M_{20}$ color-coded by $C$. The dashed lines indicate the criteria used to define mergers, early-types and late-type galaxies by \citet{Lotz08}, based on rest-frame optical morphologies of $z<1.2$ galaxies with HST. We find that the vast majority of our galaxies are late-type galaxies according to this classification scheme, and only a handful are early-types. A fairly significant fraction are mergers. We caution, however, that rest-frame wavelength and spatial resolution can shift galaxies within the $G-M_{20}$ plane, and some recent JWST studies have found somewhat modified criteria to better define mergers based on visual inspection at $z\gtrsim1.2$ \citep[e.g.,][]{Yao23,Li26}. However, we did not find a noticeable improvement between the older and newer definitions of mergers in visual inspection of our galaxies, for which we retain the standard \citet{Lotz08} definition.

\subsection{Visual classifications}
\label{app:vis}


We visually inspect multi-filter cutouts from JWST/NIRCam, JWST/NIRISS, HST/ACS, and HST/WFC3, prioritizing spatial resolution and broad wavelength coverage to identify morphological features such as star-forming clumps \citep[e.g.,][]{Guo15}. Cutouts are generated at 2.5$\times$ the Kron radius and a fixed 4{\arcsec} scale, with multiple image stretches and available redshifts of neighboring sources to facilitate classification. We also construct RGB composites using a modified version of {\sc Trilogy}\footnote{\url{https://github.com/dancoe/Trilogy}}, combining an arbitrary number of filters with wavelength-dependent color assignments and independently optimized intensity scaling. For four COSMOS fields, HST/ACS images are reprojected onto the JWST WCS to ensure consistent alignment. Given the differing pixel scales of ACS ($\sim$0.03{\arcsec}/px) and NIRISS ($\sim$0.065{\arcsec}/px), RGB composites are inspected alongside individual-filter images to mitigate resolution-related artifacts. We exclude $\lesssim$8\% of sources that are too compact or noisy for reliable morphological classification.

In Fig.~\ref{fig:vis-stats}, we present some statistics of our visual morphological classifications. The left panel is a bar chart showing the number of galaxies with salient morphological features identified visually. Over 300 ($\approx$25\% of the sample) galaxies show a central bulge, and about 100 have possible spiral features, including spiral arms or (face-on or edge-on) disk-like structures. Less than 200 show signs of interactions or mergers, and over 300 are either tadpoles or chains. 

The middle panel is a bar chart showing the number of galaxies as a function of number of clumps. More than half of our galaxies have at least one clump, with a non-negligible fraction assessed to have many more -- up to fifteen clumps. The right panel presents the fraction of clumpy galaxies -- defined as galaxies with at least 1 clump (green) or at least 2 clumps (pink) -- in bins of stellar mass. The clumpy fraction increases fairly monotonically with mass, rising to $\sim$70\% at ${\logms}>9.5$. The clumpy fraction (with $\geq1$ clumps) rises above 50\% at ${\logms}>8.5$. These results are qualitatively consistent with past results with HST and JWST in the literature \citep[e.g.,][]{Guo15,Sattari2023ApJ}, keeping in mind that automated methods were typically used to identify clumps instead of visual classification.

We present a gallery of RGB image cutouts of several galaxies in our sample in Fig.~\ref{fig:rgb_examples}, where we show examples of various morphologies, including both quantitative and qualitative morphological features.

\begin{figure*}[htbp]
\vspace{-10pt}
\centering
\gridline{
\fig{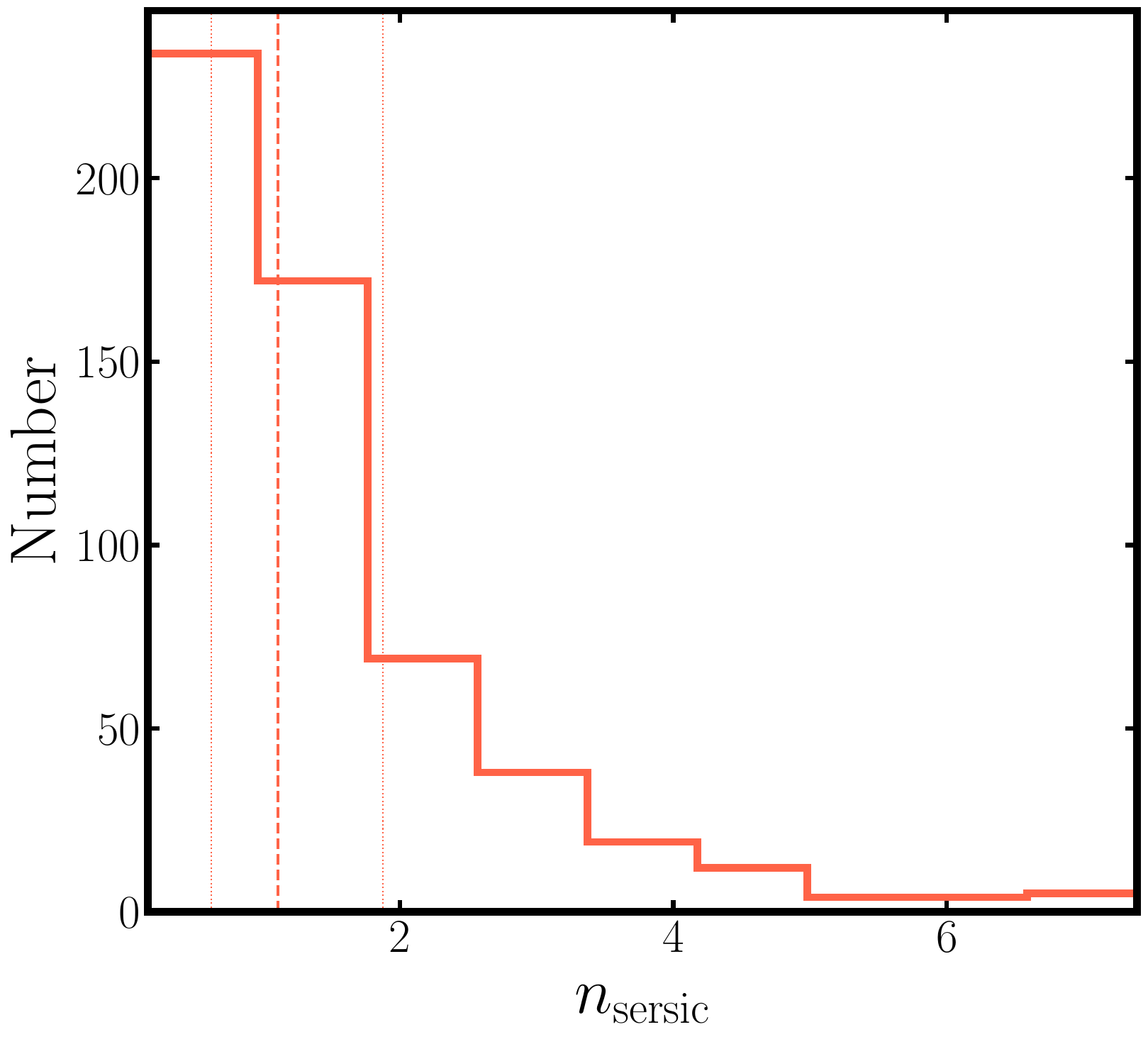}{0.24\textwidth}{}
\fig{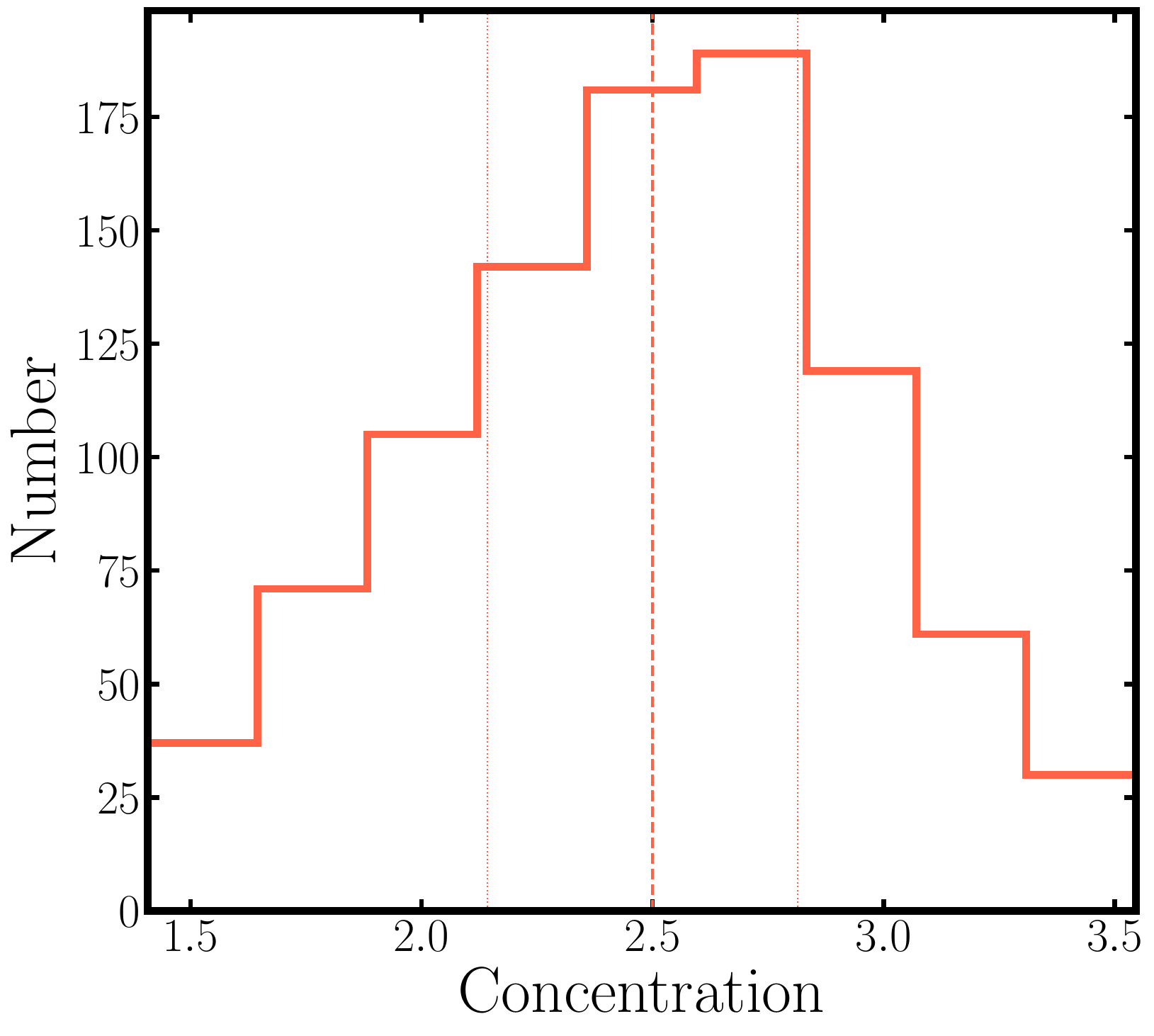}{0.24\textwidth}{}
\fig{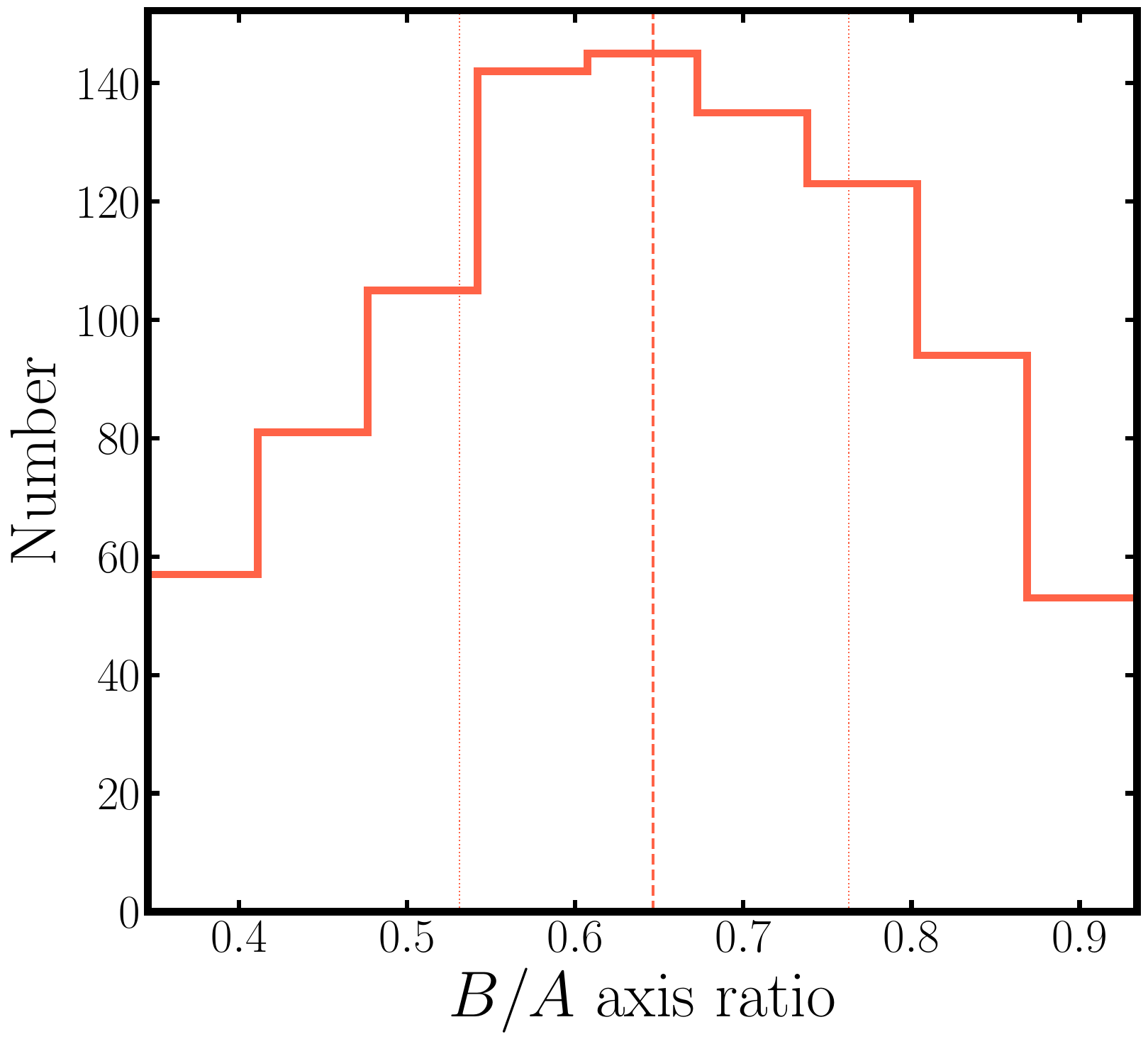}{0.24\textwidth}{}
\fig{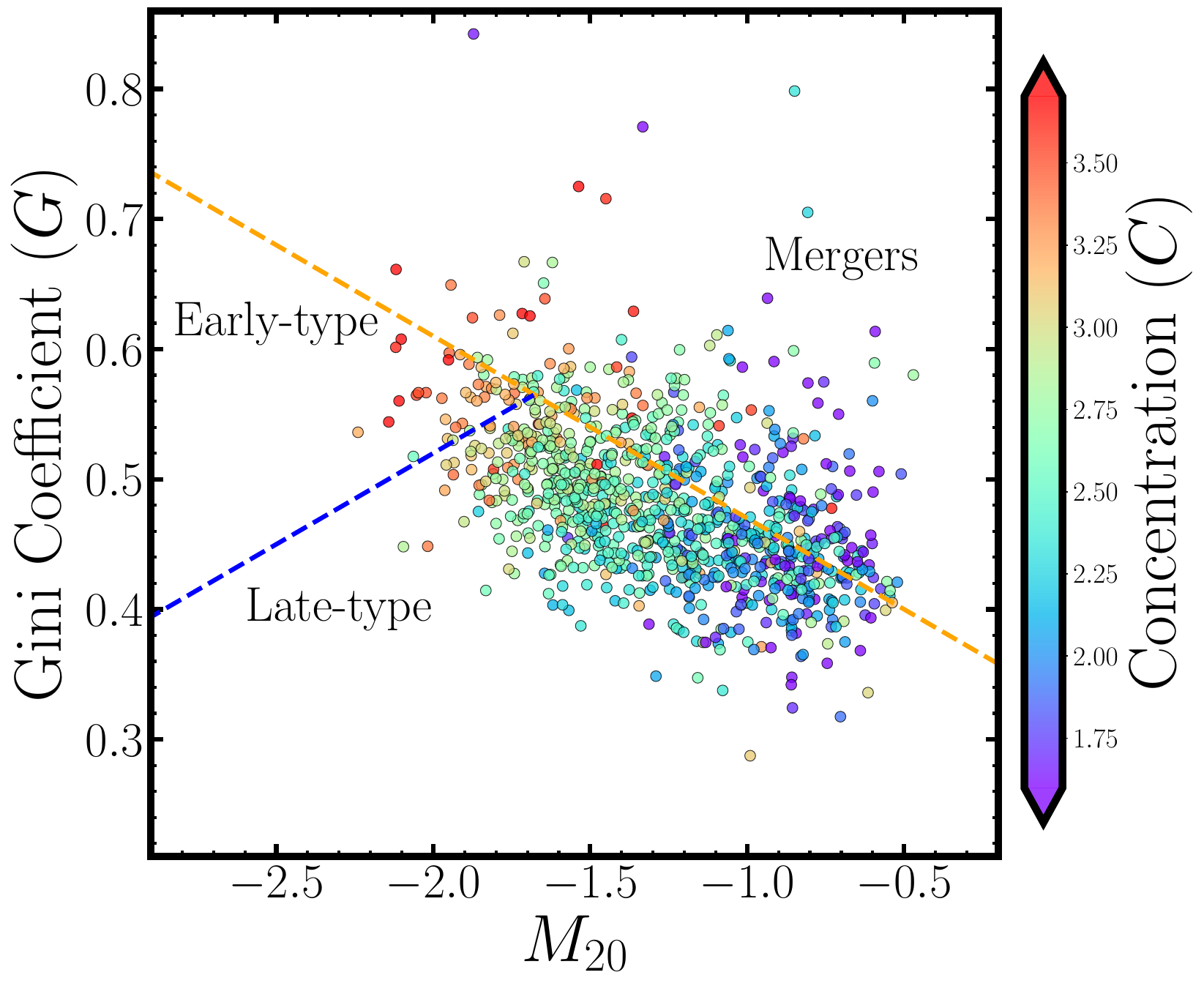}{0.26\textwidth}{}
 }
\vspace{-25pt}
\caption{
Statistics of quantities derived by {\sc Statmorph}.
The left three panels present
histograms of S{\'e}rsic index ({\nsers}), concentration ($C$), minor-to-major axis ratio $B/A$.
The dashed and dotted vertical lines represent the median and middle 50\% of each distribution in these panels.
The right panel shows Gini vs. $M_{20}$, colored by $C$. Lines separate early-type, late-type, and merger galaxies, following \citet{Lotz08}.
}
\label{fig:sm-stats}
\end{figure*}

\begin{figure*}[htbp]
\centering
\gridline{
\fig{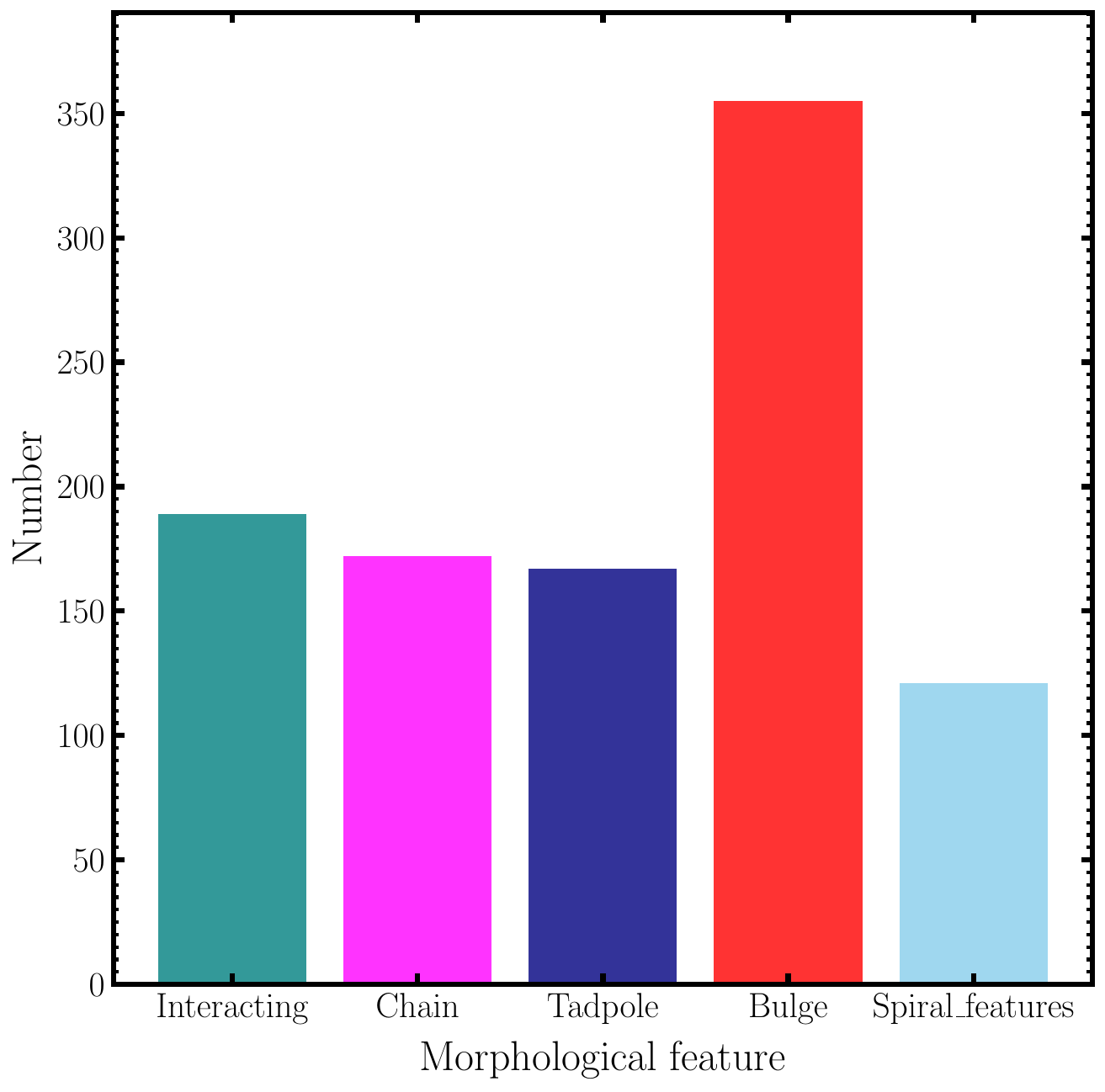}{0.26\textwidth}{}
\fig{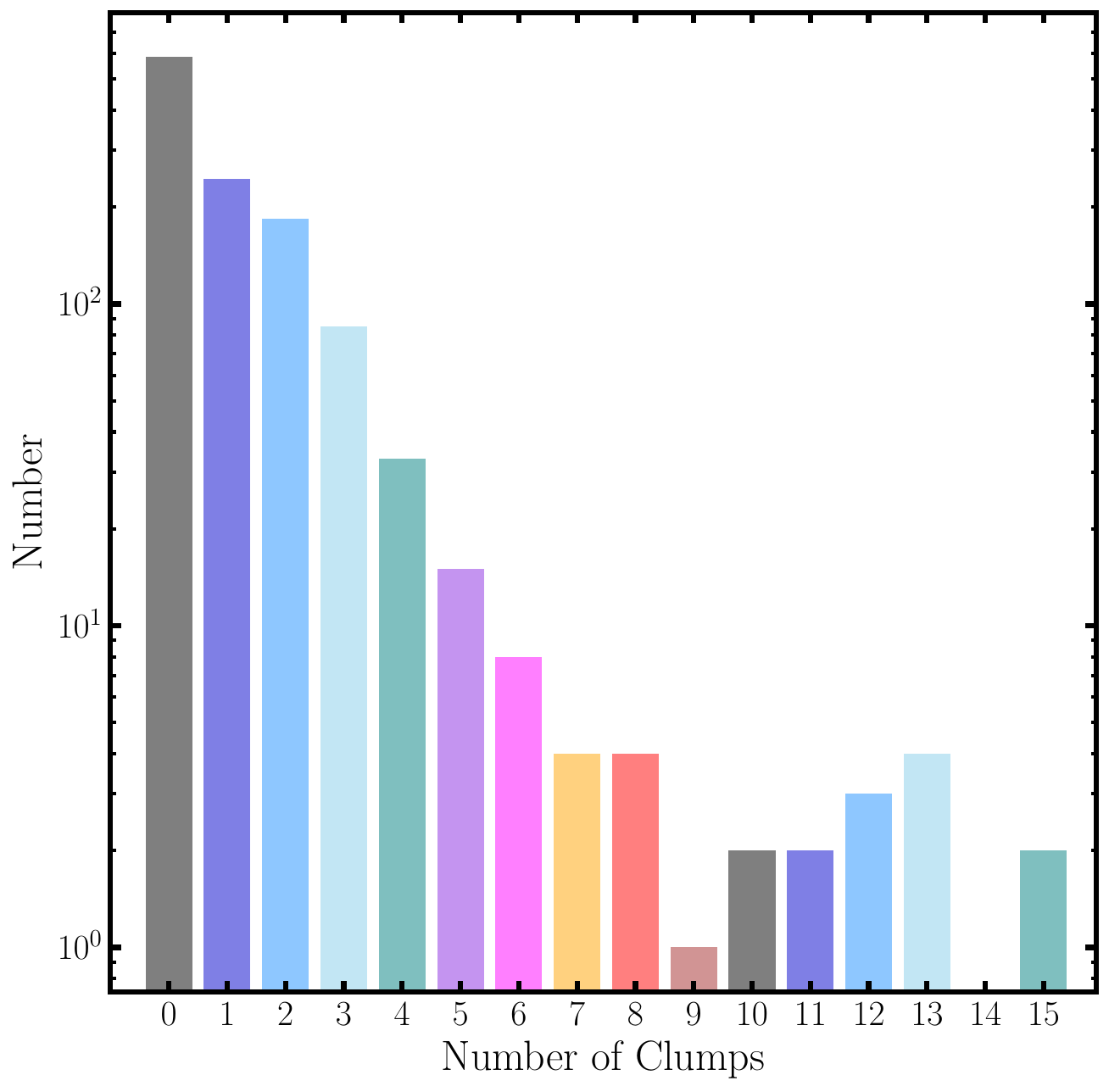}{0.26\textwidth}{}
\fig{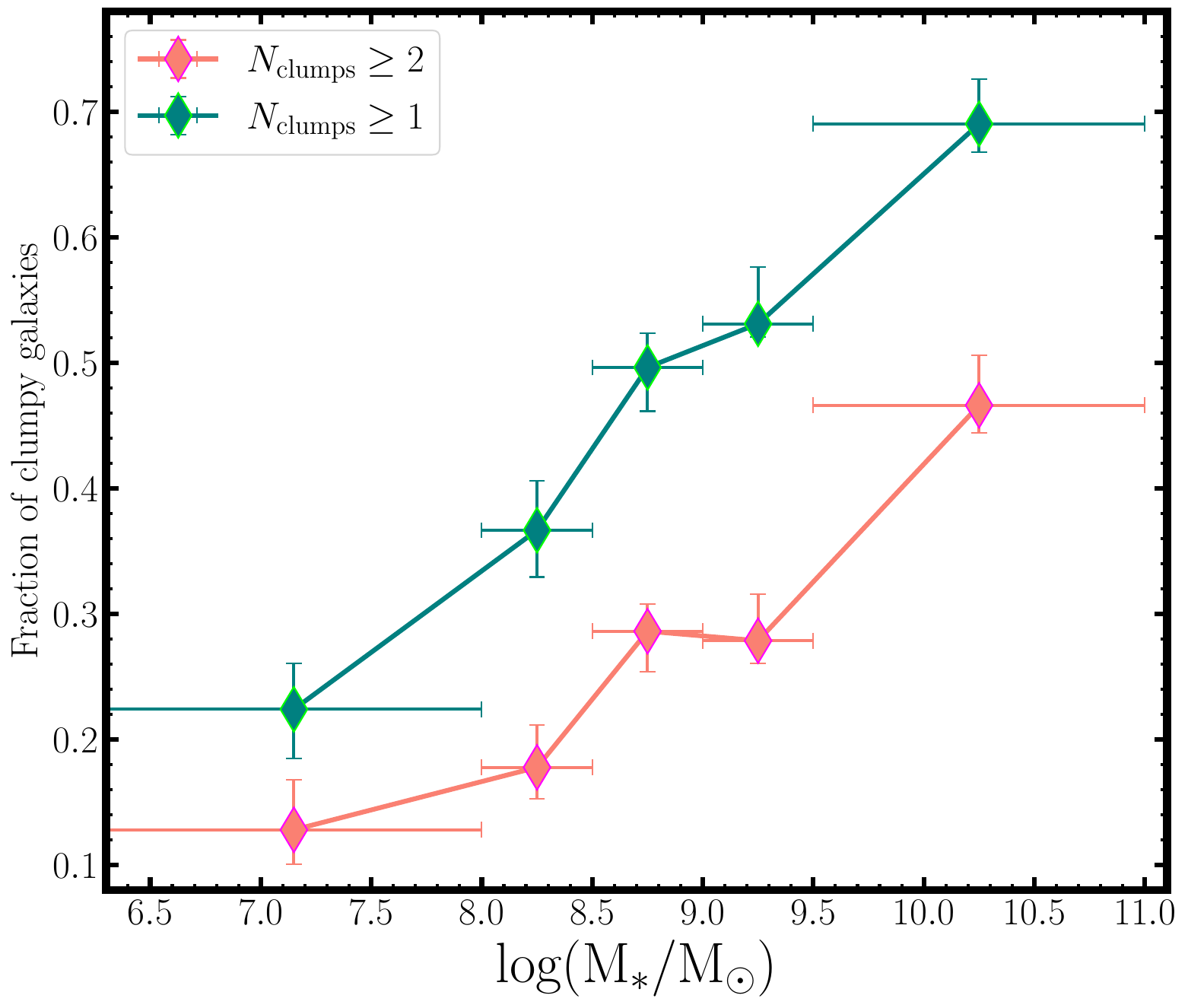}{0.3\textwidth}{}
}
\vspace{-25pt}
\caption{
Statistics of visual classification.
{\it Left:} Number of galaxies with different visual features;
{\it Middle:} Number of galaxies with different numbers of clumps;
{\it Right:} Fraction of galaxies with at least 1 (green) and at least 2 (pink) clumps as a function of mass.
}
\label{fig:vis-stats}
\end{figure*}

\begin{figure*}[t!]
\vspace{-10pt}
\centering
\gridline{
\fig{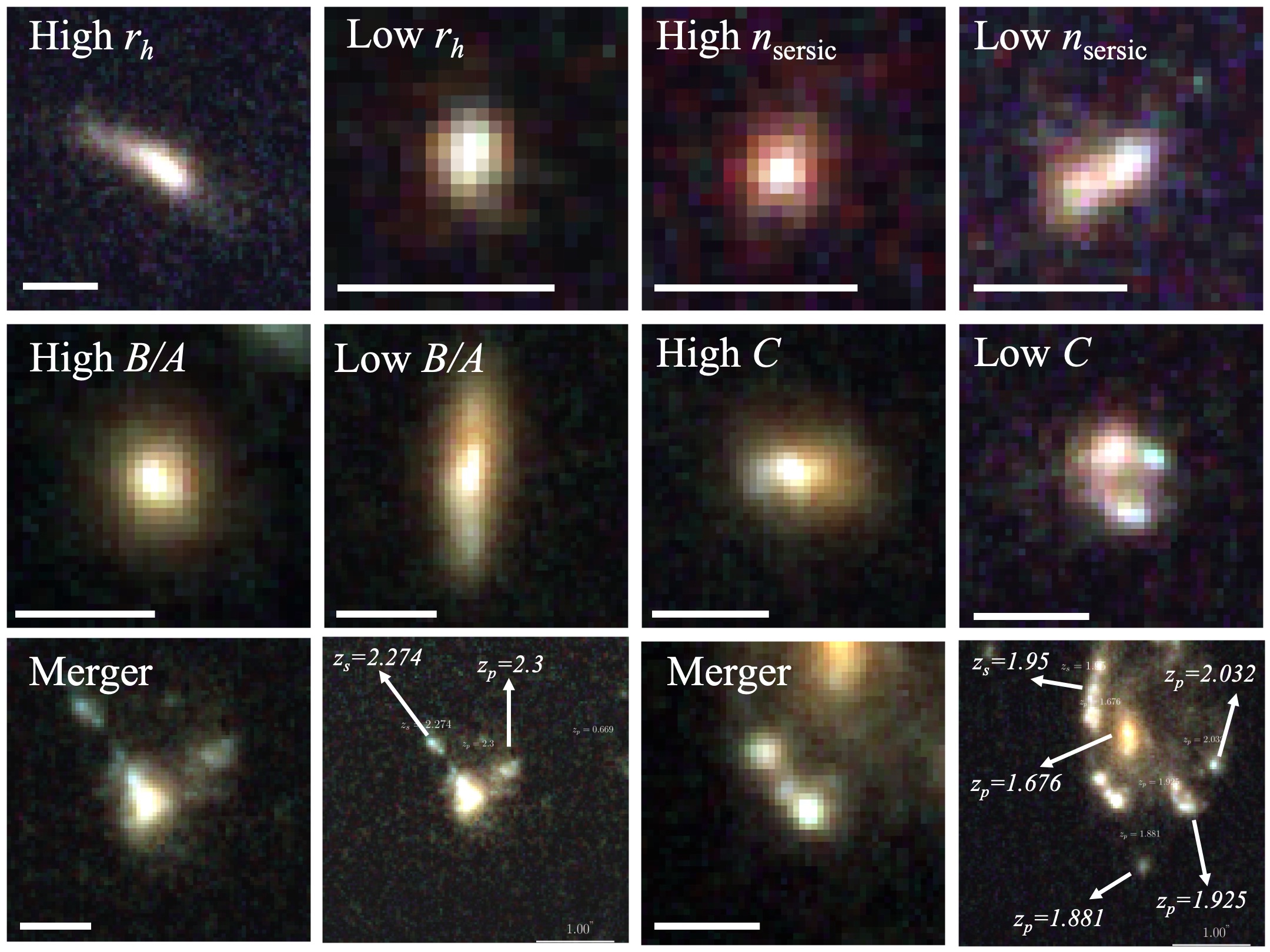}{0.82\textwidth}{}
}
\vspace{-25pt}
\gridline{
\fig{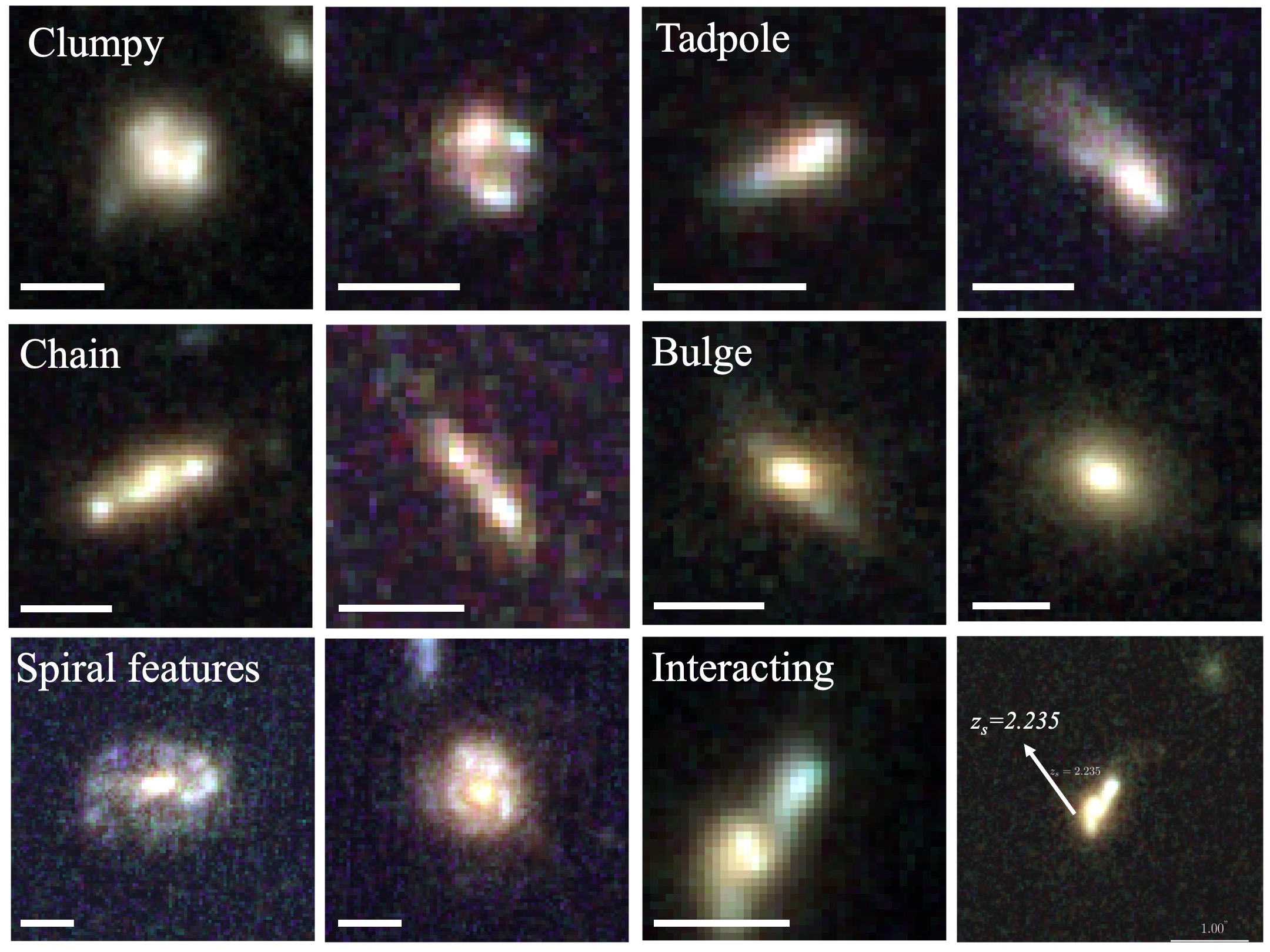}{0.82\textwidth}{}
}
\vspace{-25pt}
\caption{
RGB images showing examples of galaxies with different morphological characteristics, both quantitative and visual.
The thick white horizontal line show a physical scale of 0.5{\arcsec} in all but the two zoomed-out cutouts (which are 4{\arcsec} on a side). 
Some panels indicate spectroscopic or photometric redshifts of additional galaxies.
}
\label{fig:rgb_examples}
\end{figure*}

\begin{acknowledgments} 
Based on observations with the NASA/ESA/CSA James Webb Space Telescope and NASA/ESA Hubble Space Telescope obtained at the Space Telescope Science Institute, which is operated by the Association of Universities for
Research in Astronomy, Incorporated, under NASA contract NAS5-03127 and NAS5-26555. Support for JWST-GO-01571 was provided through a grant from the STScI under NASA contract NAS5-03127. Support for Program number HST-GO-18022 was provided through a grant from the STScI under NASA contract NAS5-26555. AJB acknowledges funding from the ``FirstGalaxies" Advanced Grant from the European Research Council (ERC) under the European Union’s Horizon 2020 research and innovation program (Grant agreement No. 789056).
PJW acknowledges support from the Agence Nationale de la Recherche (ANR), grant ANR-25-CE31-0192-01 (CIGALE-2D).
This research was supported by the International Space Science Institute (ISSI) in Bern, through ISSI International Team project \#24-624. 
MR also acknowledges support from the STScI's Director's Discretionary Research Funding (PID D0101.82527).
We acknowledge insightful discussions with G, Noirot, V. Estrada-Carpenter, J. Favaro, A. Fox, S. Faber, F. Jiang, Y. Tang, R. Sanders, R. Trainor, C. Carreira, C. Steel, C. Tremonti, D. Setton, N. Pirzkal.

AI tools (Claude Sonnet 4.6, ChatGPT GPT-6) were used to assist with refinement and optimization of code used in the research process and reported in the manuscript, as well as to edit and improve clarity of the text. The authors independently verified all AI-generated content.

\end{acknowledgments}

\bibliography{Refs}{}
\bibliographystyle{aasjournal_nikki}



\end{document}